\documentclass[twocolumn]{aastex63}
\usepackage{amsmath}
\usepackage{xspace}

\defcitealias{SO_2019}{SO19}
\newcommand{\soforecast}{\citetalias{SO_2019}}
\newcommand{\soforecastp}{\citepalias{SO_2019}}
\defcitealias{Planck_2013_XIII}{P13}
\newcommand{\planck}{Planck\xspace}
\newcommand{\gaia}{Gaia\xspace}
\newcommand{\pysm}{\texttt{PySM}\xspace}
\newcommand{\pco}{\Pi_{\rm CO}\xspace}
\newcommand{\cothreetwo}{CO$(3\hbox{--}2)$\xspace}
\newcommand{\cotwoone}{CO$(2\hbox{--}1)$\xspace}

\providecommand{\sorthelp}[1]{}

\begin{document}

\title{The Simons Observatory: Galactic Science Goals and Forecasts}

\author[0000-0001-7449-4638]{Brandon~S.~Hensley}
\email{bhensley@astro.princeton.edu}
\affiliation{Department of Astrophysical Sciences,  Princeton University, Princeton, NJ 08544, USA}

\author[0000-0002-7633-3376]{Susan E.~Clark}
\email{seclark1@stanford.edu}
\affil{Department of Physics, Stanford University, Stanford, CA 94305, USA}
\affil{Kavli Institute for Particle Astrophysics \& Cosmology (KIPAC), Stanford University, Stanford, CA 94305, USA}

\author{Valentina~Fanfani}
\affiliation{Department of Physics, University of Milano-Bicocca, Piazza della Scienza 3, 20126 Milano, Italy}

\author[0000-0002-5501-8449]{Nicoletta Krachmalnicoff}
\affiliation{International School for Advanced Studies (SISSA), Via Bonomea 265, 34136 Trieste, Italy}
\affiliation{Institute for Fundamental Physics of the Universe (IFPU), Via Beirut 2, 34151 Grignano (TS), Italy}
\affiliation{National Institute for Nuclear Physics (INFN), Sezione di Trieste, Via Valerio 2, I-34127 Trieste, Italy}

\author[0000-0002-3255-4695]{Giulio~Fabbian}
\affiliation{Center for Computational Astrophysics, Flatiron Institute, 162 5th Ave, New York, NY 10010, USA}
\affiliation{School of Physics and Astronomy, Cardiff University, The Parade, Cardiff, Wales CF24 3AA, UK}

\author[0000-0001-9807-3758]{Davide~Poletti}
\affiliation{Department of Physics, University of Milano-Bicocca, Piazza della Scienza 3, 20126 Milano, Italy}

\author[0000-0002-0689-4290]{Giuseppe~Puglisi}
\affiliation{Universit\`a di Roma - Tor Vergata, Via della Ricerca Scientifica 1, 00133 Roma, Italy}
\affiliation{Computational Cosmology Center, Lawrence Berkeley National Laboratory, Berkeley, CA 94720, USA}
\affiliation{Space Sciences Laboratory, University of California, Berkeley, CA 94720, USA}
\affiliation{Department of Physics, University of California, Berkeley, CA 94720, USA}

\author[0000-0002-6362-6524]{Gabriele~Coppi}
\affiliation{Department of Physics, University of Milano-Bicocca, Piazza della Scienza 3, 20126 Milano, Italy}
\affiliation{Department of Physics and Astronomy, University of Pennsylvania, 209 South 33rd Street, Philadelphia, PA 19104, USA}
\affiliation{National Institute for Nuclear Physics (INFN), Sezione di Milano-Bicocca, Piazza della Scienza 3, 20126 Milano, Italy}

\author[0000-0001-8042-5794]{Jacob Nibauer}
\affiliation{Department of Astrophysical Sciences, Princeton University, Princeton, NJ 08544, USA}
\affiliation{Center for Particle Cosmology, Department of Physics and Astronomy, University of Pennsylvania, Philadelphia, PA 19104, USA}

\author[0000-0003-0398-639X]{Roman Gerasimov}
\affiliation{Center for Astrophysics and Space Sciences, University of California, San Diego, La Jolla, California 92093, USA}

\author[0000-0001-7225-6679]{Nicholas Galitzki}
\affiliation{Department of Physics, University of California San Diego, La Jolla, CA 92093, USA}

\author[0000-0002-9113-7058]{Steve~K.~Choi}
\affiliation{Department of Physics, Cornell University, Ithaca, NY 14853, USA}
\affiliation{Department of Astronomy, Cornell University, Ithaca, NY 14853, USA}

\author{Peter~C.~Ashton}
\affiliation{SOFIA-USRA, NASA Ames Research Center, MS N232-12, Moffett Field, CA 94035, USA}
\affiliation{Physics Division, Lawrence Berkeley National Laboratory, 1 Cyclotron Rd. Berkeley, CA 94720, USA}

\author[0000-0002-8211-1630]{Carlo Baccigalupi}
\affiliation{International School for Advanced Studies (SISSA), Via Bonomea 265, 34136 Trieste, Italy}
\affiliation{Institute for Fundamental Physics of the Universe (IFPU), Via Beirut 2, 34151 Grignano (TS), Italy} 
\affiliation{National Institute for Nuclear Physics (INFN), Sezione di Trieste, Via Valerio 2, I-34127 Trieste, Italy}

\author{Eric Baxter}
\affiliation{Institute for Astronomy, University of Hawai’i, 2680 Woodlawn Drive, Honolulu, HI 96822, USA}

\author[0000-0001-5817-5944]{Blakesley Burkhart}
\affiliation{Department of Physics and Astronomy, Rutgers University,  136 Frelinghuysen Rd, Piscataway, NJ 08854, USA}

\author[0000-0003-0837-0068]{Erminia Calabrese}
\affiliation{School of Physics and Astronomy, Cardiff University, The Parade, Cardiff, Wales CF24 3AA, UK}

\author[0000-0003-3725-6096]{Jens Chluba}
\affiliation{Jodrell Bank Centre for Astrophysics, Department of Physics and Astronomy, The University of Manchester, Manchester, M13 9PL, UK}

\author[0000-0002-1419-0031]{Josquin Errard}
\affiliation{Universit\'{e} de Paris, CNRS, Astroparticule et Cosmologie, F-75013 Paris, France}

\author[0000-0002-1984-8234]{Andrei V. Frolov}
\affiliation{Simon Fraser University, Department of Physics, 8888 University Drive, Burnaby BC, Canada}

\author[0000-0002-4765-3426]{Carlos Hervías-Caimapo}
\affiliation{Department of Physics, Florida State University, Tallahassee, FL 32306, USA}

\author[0000-0001-7109-0099]{Kevin M. Huffenberger}
\affiliation{Department of Physics, Florida State University, Tallahassee, FL 32306, USA}

\author[0000-0002-6898-8938]{Bradley R. Johnson}
\affiliation{University of Virginia, Department of Astronomy,
Charlottesville, VA 22904, USA}

\author{Baptiste Jost}
\affiliation{Universit\'{e} de Paris, CNRS, Astroparticule et Cosmologie, F-75013 Paris, France}

\author[0000-0003-3118-5514]{Brian~Keating}
\affiliation{Department of Physics, University of California San Diego, La Jolla, CA 92093, USA}

\author{Heather McCarrick}
\affiliation{Joseph Henry Laboratories of Physics, Jadwin Hall, Princeton University, Princeton, NJ 08544, USA}

\author[0000-0002-8307-5088]{Federico~Nati}
\affiliation{Department of Physics, University of Milano-Bicocca, Piazza della Scienza 3, 20126 Milano, Italy}

\author[0000-0002-9761-3676]{Mayuri Sathyanarayana Rao}
\affiliation{Raman Research Institute, C V Raman Avenue, Sadashivanagar, Bengaluru 560097, India}

\author[0000-0002-3495-158X]{Alexander van Engelen}
\affiliation{School of Earth and Space Exploration, Arizona State University, Tempe, AZ 85287, USA}

\author[0000-0002-5855-4036]{Samantha Walker}
\affiliation{Department of Astrophysical and Planetary Sciences, University of Colorado Boulder, Boulder, CO 80309, USA}
\affiliation{Quantum Sensors Group, National Institute of Standards and Technology (NIST), Boulder, CO, 80305, USA}

\author{Kevin Wolz}
\affiliation{International School for Advanced Studies (SISSA), Via Bonomea 265, 34136 Trieste, Italy}

\author[0000-0001-5112-2567]{Zhilei Xu}
\affiliation{MIT Kavli Institute, Massachusetts Institute of Technology, 77 Massachusetts Avenue, Cambridge, MA 02139, USA}

\author[0000-0002-3037-2003]{Ningfeng Zhu}
\affiliation{Department of Physics and Astronomy, University of Pennsylvania, 209 South 33rd Street, Philadelphia, PA 19104, USA}

\author[0000-0001-6841-1058]{Andrea Zonca}
\affiliation{San Diego Supercomputer Center, University of California San Diego, La Jolla, CA 92093, USA}

\date{\today}

\begin{abstract}
Observing in six frequency bands from 27 to 280\,GHz over a large sky area, the Simons Observatory (SO) is poised to address many questions in Galactic astrophysics in addition to its principal cosmological goals. In this work, we provide quantitative forecasts on astrophysical parameters of interest for a range of Galactic science cases. We find that SO can: constrain the frequency spectrum of polarized dust emission at a level of $\Delta\beta_d \lesssim 0.01$ and thus test models of dust composition that predict that $\beta_d$ in polarization differs from that measured in total intensity; measure the correlation coefficient between polarized dust and synchrotron emission with a factor of two greater precision than current constraints; exclude the non-existence of exo-Oort clouds at roughly 2.9$\sigma$ if the true fraction is similar to the detection rate of giant planets; map more than 850 molecular clouds with at least 50 independent polarization measurements at 1\,pc resolution; detect or place upper limits on the polarization fractions of \cotwoone emission and anomalous microwave emission at the 0.1\% level in select regions; and measure the correlation coefficient between optical starlight polarization and microwave polarized dust emission in $1^\circ$ patches for all lines of sight with $N_{\rm H} \gtrsim 2\times10^{20}$\,cm$^{-2}$. The goals and forecasts outlined here provide a roadmap for other microwave polarization experiments to expand their scientific scope via Milky Way astrophysics.
\end{abstract}

\section{Introduction}
Observations of the cosmic microwave background (CMB) have yielded many of the tightest constraints to date on a number of cosmological parameters \citep[e.g.,][]{BICEP2018,PB2020,Aiola_2020,Choi2020,Planck_2018_VI,SPT3g-ext2021,BICEPSPT2021,SPT3g2021}. Current and next-generation CMB instruments offer significant additional science returns, particularly through measurement of the polarized light from the CMB \citep{CMBS42019,SO_2019}. The search for primordial B-mode polarization from inflationary gravitational waves necessitates unprecedented sensitivity on scales $\gtrsim 1^\circ$. Measurements at smaller angular scales that probe, e.g., the weak gravitational lensing of the CMB, the neutrino mass hierarchy, and light relics from the hot big bang all require observations at high angular resolution over large fractions of the sky ($\sim$50\%). 

Crucially, all of these science cases depend on the capability to measure and extract the polarized Galactic dust and synchrotron emission using channels at both higher and lower frequencies than the peak of the CMB emission at $\sim$160\,GHz. Thus while these experiments are built for cosmology, their combination of sensitivity, angular resolution, large sky area, and frequency coverage in the $\sim$30--280\,GHz range also furnish sensitive new probes of the structure and physics of the magnetic interstellar medium (ISM) of the Milky Way.

By virtue of all-sky observations, CMB satellite missions have a long legacy of informing our understanding of the Galaxy. For instance, the Diffuse InfraRed Background Explorer (DIRBE) aboard the Cosmic Background Explorer (COBE) provided the first all-sky measurement of 3.5--12\,$\mu$m emission from polycyclic aromatic hydrocarbons (PAHs) at $0.7^\circ$ angular resolution, attesting their ubiquity in the Galactic ISM \citep{Dwek:1997}. COBE's Far-InfraRed Absolute Spectrophotometer (FIRAS) made full-sky maps of interstellar [CII] and [NII] emission at $7^\circ$ angular resolution \citep{Fixsen:1999} and measured in detail the frequency dependence of Galactic dust emission \citep{Finkbeiner:1999}. COBE's Differential Microwave Radiometer (DMR) provided the initial evidence for the existence of the anomalous microwave emission \citep[AME;][]{Kogut_1996}. Wilkinson Microwave Anisotropy Probe (WMAP) observations of polarized synchrotron emission at $1^\circ$ angular resolution \citep{Gold:2011} serve as a primary input to 3D models of the Galactic magnetic field \citep{Jansson:2012,Unger:2017}.

The latest experiment in this tradition is the \planck satellite, which mapped the full sky in nine frequency bands, seven of which had sensitivity to polarization and angular resolution $<10^\prime$ \citep{Planck_Early_I}. These data have had an enormous impact on understanding a range of topics including interstellar turbulence \citep{Planck_Int_XX}, the role of magnetic fields in governing the structure of molecular clouds \citep{Planck_Int_XXXV}, the geometry of the Galactic magnetic field \citep{Planck_Int_XLII}, the composition of interstellar dust in the Milky Way \citep{Planck_Int_XXII} and the Magellanic Clouds \citep{Planck_Early_XVII}, the nature of AME and its spectral variations in the Galaxy \citep{Planck_Int_XV,Planck_2015_X}, grain alignment \citep{Planck_2018_XII}, the ubiquity of high density ``cold clumps'' \citep{Planck_2015_XXVIII}, the geometry of synchrotron-bright radio loops \citep{Planck_2015_XXV}, and the spectral energy distribution (SED) of synchrotron emission \citep{Planck_2015_X, Planck_2015_XXV}, among many others.

Ground-based CMB experiments have also made important discoveries in Milky Way astrophysics. Observations at 14.5 and 32\,GHz from the Owens Valley Radio Observatory as part of the RING5M experiment were key for establishing the existence of AME \citep{Leitch:1997}. More recently, the Magellanic Clouds have been mapped in total intensity by the South Pole Telescope \citep[SPT;][]{Crawford:2016}, while the Atacama Cosmology Telescope (ACT) has furnished a multi-frequency view of magnetic fields in the Galactic center at arcminute resolution \citep{Guan:2021}. Next-generation ground-based CMB experiments promise to expand greatly on these studies by virtue of enhanced sensitivity, sky area, frequency coverage, and angular resolution.

The Simons Observatory (SO) is a set of new telescopes optimized for CMB survey observations, now under construction in the Chilean Atacama Desert, that will measure the temperature and polarization of the sky, beginning in 2023 \citep[][hereafter \soforecast]{SO_2019}. SO will have three 0.5\,m telescopes and one 6\,m aperture telescope. The Large Aperture Telescope \citep[LAT;][]{SO_LAT,SO_LAT2} will map 40\% of the sky in six frequency bands (27, 39, 93, 145, 225, and 280\,GHz) and angular resolution from $\sim7.4$--$0.9^\prime$. The LAT 225\,GHz band has  a projected sensitivity improvement of over three in temperature and over four in polarization when compared to the \planck 217\,GHz band \citep[\soforecast,][]{Planck_2018_I}. The Small Aperture Telescopes \citep[SATs;][]{SO_SAT} will measure 10\% of the sky angular resolutions from $\sim$91--10$^\prime$ and achieve more than an order of magnitude higher polarization sensitivity than the \planck satellite. 

An overview of the science goals that these telescopes will pursue was presented by \soforecast. In this paper, we expand on \soforecast\ to illustrate myriad new investigations into the multi-scale physics of Galactic structure and the physics of Galactic emission to be undertaken by SO using data collected by its CMB surveys. We provide quantitative forecasts that assess how the capabilities of the instruments translate into constraints on models of Galactic emission. In star-forming regions, the scales probed by SO bridge the high-resolution measurements from ALMA and the large-scale measurements from \planck, connecting collapsing cold core regions to the larger environment. SO will have the polarization sensitivity to map magnetic fields in a statistical sample of molecular clouds, allowing analyses to marginalize over effects like inclination angle in assessing the dynamical importance of magnetic fields. On larger scales, SO observations can test the connection of the gas and dust to the Galactic magnetic field, illuminating mechanisms of magnetic hydrodynamic turbulence as they operate in the ISM, such as the dissipation scale. On even larger scales, both the polarized dust and synchrotron emission measured by SO will contribute to the on-going, multi-probe effort to map the global magnetic field of the Galaxy. With frequency coverage extending from 27 to 280\,GHz, SO will also enable detailed tests of physical models of the frequency dependence of Galactic emission mechanisms in both total and polarized intensity.

This paper is organized as follows: in Section~\ref{sec:framework} we describe the models of the SO LAT and SAT Surveys, as well as ancillary data, on which our forecasting is based. In Section~\ref{sec:galactic_emission}, we quantify how SO data will test models of Galactic emission, including the energetics of synchrotron emission, the composition of interstellar dust, and the nature of the observed spatial correlation of dust and synchrotron emission. In Section~\ref{sec:multiscaleISM}, we describe the use of SO observations to explore the multiscale physics of the ISM, from debris disks, to dust and CO line emission in molecular clouds, to dust emission and turbulence in the diffuse ISM. Our results are summarized in Section~\ref{sec:conclusions}.

\section{Survey Description and Noise Models}\label{sec:framework}
Throughout this work, we adopt the models of the SO instruments and sky surveys presented in \soforecast. In this section, we review the SO noise models, the SO survey footprints, and the ancillary data used in various forecasts presented in this work. The SO noise model is publicly available\footnote{\url{https://github.com/simonsobs/so_noise_models}}.

\subsection{Noise Models}
\label{subsect:noise}

\subsubsection{SO Polarization Noise Model}
\label{subsubsect:SOnoise}
\begin{deluxetable}{ccccc}
  \tablewidth{0pc}
      \tablecaption{SO Polarization Noise Model Parameters\label{table:so_noise_model}}
    \tablehead{\colhead{Frequency} & \colhead{Noise}
      & \colhead{$\alpha_{\rm knee}$} & \colhead{$\ell_{\rm knee}$} \\ 
      $\left[{\rm GHz}\right]$ & [$\mu$K-arcmin] & & }
    \startdata
    \multicolumn{4}{c}{LAT}\\
    \hline
    27 & 100.4 & -1.4 & 700 \\
    39 & 50.9 & -1.4 & 700 \\
    93 & 11.3 & -1.4 & 700 \\
    145 & 14.1 & -1.4 & 700 \\
    225 & 31.1 & -1.4 & 700 \\
    280 & 76.4 & -1.4 & 700 \\    
    \hline
    \multicolumn{4}{c}{SAT}\\
    \hline
    27 & 49.5 & -2.4 & 30 \\
    39 & 29.7 & -2.4 & 30 \\
    93 & 3.7 & -2.5 & 50 \\
    145 & 4.7 & -3.0 & 50 \\
    225 & 8.9 & -3.0 & 70 \\
    280 & 22.6 & -3.0 & 100 \\
    \enddata
    \tablecomments{The noise model is normalized such that $N_{\rm red} = N_{\rm white}$, and this is the value reported in the ``Noise'' column. All values are quoted for $Q$ and $U$ maps.}
\end{deluxetable}

We adopt the SO noise power spectrum described in \soforecast, which takes into account both the atmospheric and instrumental noise. At high frequencies and large angular scales, the atmospheric $1/f$ noise becomes significant. The noise model used has the form:

\begin{equation}
    N_{\ell} = N_{\rm red} \left( \frac{\ell}{\ell_{\rm knee}} \right)^{\alpha_{\rm knee}} + N_{\rm white}
\label{SO_noise_model}
\end{equation}
for both the LAT and the SAT, where $N_{\rm white}$ is the white noise component and $N_{\rm red}$, $\ell_{\rm knee}$ and $\alpha_{\rm knee}$ describe the contribution from correlated atmospheric noise. We use the parameter values corresponding to the ``baseline'' model for all forecasts presented here. Throughout, we assume nominal SO mission parameters for a 5-year survey with parameter values listed in Table~\ref{table:so_noise_model}. We note that we have adopted the ``pessimistic'' value of $\ell_{\rm knee}$ described in \soforecast.

For simplicity, we assume delta function bandpasses (i.e., detectors sensitive to emission only at the nominal frequency) for most of the analyses presented here. Accounting for bandpass uncertainties would slightly increase the forecasted uncertainties on parameter constraints, but we expect our assessments of the relative improvements afforded by SO compared to existing data to be robust to this assumption since it is also applied to all other datasets considered.

\subsubsection{SO Intensity Noise Model}
The LAT intensity noise model \soforecastp\ assumes a common $\ell_{\rm knee}=1000$ and $\alpha_{\rm knee}=-3.5$ for all frequencies. The parameter $N_{\rm red}=12000\,\mu$K$^2$\,s was calibrated on ACT polarization observations at 145\,GHz \citep{louis2017} and extrapolated to the SO frequency bands using the brightness temperature variance due to changes in the precipitable water vapor (PWV) level computed using the ATM code \citep{Pardo2001}.

The noise model developed in \soforecast\ did not include the SAT intensity noise. For the purpose of deriving the results of Section~\ref{subsec:co_pol} that rely on the SAT intensity data, we estimate the SAT intensity noise angular power spectrum adapting the noise power spectrum observed by the Atacama B-Mode Search (ABS) experiment \citep{abs} assuming the SAT survey scanning strategy specifications. To account for the increased sensitivity of the SO SAT compared to ABS, we match the white noise plateau of the typical ABS detector noise angular power spectrum to the white noise level in intensity derived from the public SAT $145$\,GHz noise model \citep{sat}. We estimate the correlated part of the SAT noise power spectrum (i.e., the $\ell$-dependent part of Equation~\eqref{SO_noise_model}) at 145\,GHz and rescale it to all the other SAT frequencies assuming the same relative rescaling between frequency adopted for the LAT intensity noise model.

The final SAT intensity noise power spectrum is the sum of the correlated noise component and the white noise level for each frequency channel. ABS is the most technologically similar experiment to the SO SAT and also operated in the Atacama desert. Thus, although there is uncertainty inherent to this extrapolation due to unknown averaging properties of the correlated part of the noise as a function of number of detectors, this procedure is reasonable given available data. Furthermore, we note that our procedure does not assume any specific analysis technique to reduce the correlated noise (e.g. common mode subtraction or data high-pass filtering), thus it can be considered as a conservative estimate. 

\subsubsection{Noise Models for Ancillary Data} \label{subsubsect:noise}
In addition to SO data, some forecasts employ ancillary data from the Planck and WMAP satellites, as well as low-frequency ground-based data from C-BASS and S-PASS. In simulating these data, we use the noise models described below.

For \planck frequency channels we use the same noise power spectrum model adopted for SO, with the form reported in Equation~\eqref{SO_noise_model}. The four parameters ($N_{\rm white}$, $N_{\rm red}$, $\ell_{\rm knee}$ and $\alpha_{\rm knee}$) were retrieved for each \planck frequency by fitting the model to the $EE$ and $BB$ angular power spectra of the publicly available FFP10 noise simulated maps\footnote{The noise simulated maps are available on the Planck Legacy Archive: \url{http://pla.esac.esa.int/pla}}, which also include the contribution of instrumental systematic effects. Our analysis is therefore similar to that of \citet{Planck_2018_XI}, who employed data splits to model the noise power spectra for the Planck Low Frequency Instrument (LFI) and WMAP; we applied the same procedure to Planck E2E simulations for both LFI and the High Frequency Instrument (HFI). All fits were performed on full-sky data, but we found no qualitative differences in the noise power spectra when restricting to the LAT or SAT footprints.

The noise model for WMAP is constructed by first computing the $EE$ and $BB$ noise power spectra of the K and Ka band maps in the LAT and SAT observing regions after masking the Galactic plane (Galactic latitudes $|b| < 10^\circ$). We then fit the same four parameter model as for \planck to these noise power spectra.

The C-band All-Sky Survey (C-BASS) is an on-going, full sky polarimetric survey at 5\,GHz \citep{Jones2018}. When simulating C-BASS observations, we assume a uniform noise rms of 4.5\,mK-arcmin and a resolution of 45$^\prime$ following \citet{Jones2018}.

The S-band Polarization All-Sky Survey \citep[S-PASS;][]{Carretti19} is a 2.3\,GHz survey of the Southern Sky (Dec. $< -1^\circ$) in polarization. As the survey was conducted with the 64\,m Parkes radio telescope, these maps have a resolution of 8.9$^\prime$ (FWHM). When simulating S-PASS observations, we assume a uniform noise rms of 8\,mK-arcmin following \citet{Krachmalnicoff2018} \citep[see also][]{Carretti19}.

\subsection{Sky Coverage}
\label{subsect:masks}
\begin{figure}
\centering
 \includegraphics[width=\columnwidth]{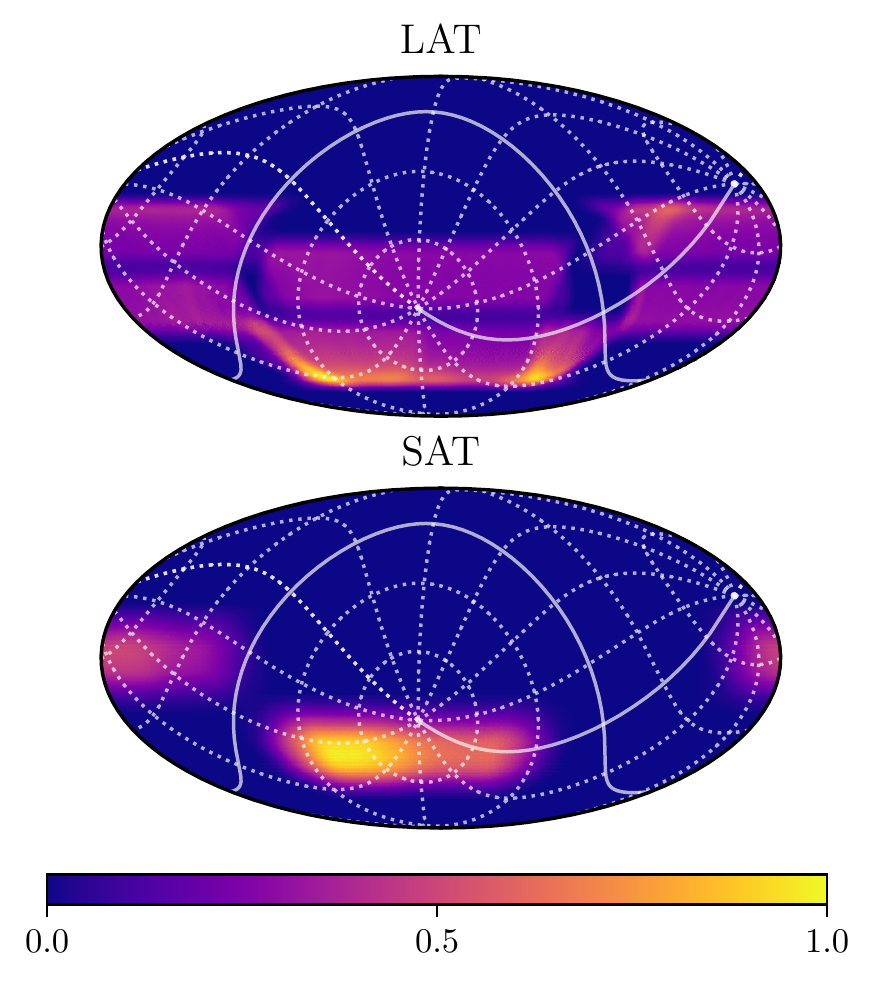}
\caption{Counts-weighted and apodized masks for the LAT (top) and SAT (bottom) Surveys used in the forecasts presented here. Both masks are shown in Equatorial projection with gridlines corresponding to Galactic coordinates.} \label{fig:masks}
\end{figure}

In all forecasts we employ the masks corresponding to the nominal SO survey regions for the LAT and SAT surveys presented in \citet{Stevens2018}. The nominal total sky fractions $f_{\rm sky}$ for the LAT and SAT are 57.5 and 34.4\%, respectively. These masks, pixellated to a HEALPix\footnote{\url{http://healpix.sourceforge.net}} grid \citep{Gorski2005} having $N_{\rm side} = 512$, are presented in Figure~\ref{fig:masks}.

\section{Power Spectrum Analysis of Multi-frequency Galactic Emission}
\label{sec:galactic_emission}

The sensitive, large-area observations of diffuse Galactic emission by SO will provide detailed tests of the physical models of these emission mechanisms that have not been possible with intensity-only observations. After introducing our power spectrum-based forecasting framework (Section~\ref{subsect:ps_forecast}), we forecast the ability of SO to measure the component SEDs in polarization and highlight what can be learned about the underlying emission physics. This includes properties of Galactic cosmic ray electrons (Section~\ref{subsec:sync_sed}), tests of single versus multi-component dust models (Section~\ref{subsec:dust_sed}), and the difference in ISM phases probed by polarized dust and synchrotron emission (Section~\ref{subsec:dust_sync_cor}).

\subsection{Power Spectrum Forecasting Framework} \label{subsect:ps_forecast}

\subsubsection{Galactic emission model} \label{subsect:pysmmodels}
We begin our forecasting with map-domain simulations of Galactic emission constructed with the Python Sky Model \citep[\pysm;][]{Thorne2017,pysm3}. We focus exclusively on the polarization data, where the emission is dominated by the CMB, dust emission, and synchrotron emission. Other components that contribute to the total intensity signal---such the cosmic infrared background, free-free emission, AME, and CO emission---are largely unpolarized \cite[e.g.,][and references therein]{Planck_2018_IV}. Indeed, searches for polarized CO emission and AME are the subjects of Sections~\ref{subsec:co_pol} and \ref{subsec:ame_pol}, respectively.

As Galactic emission has the most power on large scales \citep[e.g.,][]{Dunkley2013,Planck_2018_XI}, we focus our analyses on $\ell < 1000$. Therefore, we generate and analyze maps with $N_{\rm side} = 512$, corresponding to a resolution of 6.9$^\prime$.

Dust emission is simulated with the \pysm ``d0'' model, based on the \texttt{Commander} dust parameter maps \citep{Planck_2015_X}. The dust SED in each pixel is described by a modified blackbody having an amplitude parameter $A_d$ in each of Stokes $Q$ and $U$, a dust temperature $T_d$, and a spectral index $\beta_d$, i.e.,

\begin{equation} \label{eq:pysm_dust}
    S_{\nu,{\rm d}}^{[QU]} = A_d^{[QU]} \left(\frac{\nu}{\nu_{0,d}}\right)^{\beta_d-2} \frac{B_\nu\left(T_d
\right)}{B_{\nu_{0,d}}\left(T_d\right)}
~~~,
\end{equation}
where $S_{\nu,d}$ is one of Stokes $Q$ or $U$ in brightness temperature units (e.g., $\mu$K$_{\rm RJ}$), $\nu_{0,d}$ is an arbitrary reference frequency taken to be 353\,GHz, and $B_\nu\left(T\right)$ is the Planck function. The emission templates are smoothed to a resolution of 2.6$^\circ$ to which small-scale Gaussian fluctuations are added as described in \citet{Thorne2017}. In the adopted model, $\beta_d = 1.54$ and $T_d = 20$\,K for all pixels, i.e., the dust spectrum is not spatially variable.

Synchrotron emission is simulated with the \pysm ``s0'' model, based on the WMAP 9-year $Q$ and $U$ maps \citep{Bennett_2013}. The synchrotron SED in each pixel is described by amplitude parameters $A_s$ in $Q$ and $U$ based on these maps, and by a spectral index $\beta_s$ taken to be $-3$ over the full sky. Thus,

\begin{equation} \label{eq:pysm_sync}
    S_{\nu,s}^{[QU]} = A_s^{[QU]} \left(\frac{\nu}{\nu_{0,s}}\right)^{\beta_s}
    ~~~,
\end{equation}
where, in analogy with Equation~\eqref{eq:pysm_dust}, $S_{\nu, s}$ is one of Stokes $Q$ or $U$ in brightness temperature units and $\nu_{0,s}$ is an arbitrary reference frequency. We adopt $\nu_{0,s} = 23$\,GHz. The synchrotron polarization templates are smoothed to a scale of 5$^\circ$ and then, as with dust, smaller scales are added assuming Gaussian fluctuations as described in \citet{Thorne2017}.

To the Galactic emission we add a realization of the CMB signal using the \pysm ``c1'' model. This model draws a Gaussian CMB realization from a  primordial unlensed CMB power spectrum computed with \texttt{CAMB}\footnote{\url{https://github.com/cmbant/CAMB}} \cite{Lewis:1999bs} and then applies lensing in pixel space with the \texttt{Taylens} code\footnote{\url{https://github.com/amaurea/taylens}} \citep[][see \citet{Thorne2017} for a more detailed description of the ``c1'' model]{taylens}. 

\subsubsection{Simulated Power Spectra}
\label{SPS}
Following the framework employed for cosmological analyses both in other experiments \citep{Choi2015,BICEP2018,Planck_2018_XI,CMBS42020} and in \soforecast, we constrain the frequency dependence of each emission mechanism using the combination of all auto- and cross-power spectra that can be constructed from the set of observed frequencies. In this formulation, the cross-spectrum $C_\ell^{XX}\left(\nu_1\times\nu_2\right)$ has the parametric form

\begin{align}
    & C_\ell^{XX}\left(\nu_1\times\nu_2\right) = C_{\ell,{\rm CMB}}^{XX}\left(\nu_1\times\nu_2\right) +\nonumber \nonumber \\ &A_d^{XX} S_{\nu_1,d} S_{\nu_2,d} \left(\frac{\ell}{\ell_0}\right)^{\alpha_d} 
    + A_s^{XX} S_{\nu_1,s} S_{\nu_2,s} \left(\frac{\ell}{\ell_0}\right)^{\alpha_s} +\nonumber \nonumber \\
    &\rho^{XX}\sqrt{A_d A_s} \left[S_{\nu_1,d} S_{\nu_2,s} + S_{\nu_1,s} S_{\nu_2,d}\right] \left(\frac{\ell}{\ell_0}\right)^{\left(\alpha_d+\alpha_s\right)/2} \label{eq:cross_spectra}
\end{align}
where $X$ is one of $E$ or $B$, $C_{\ell,{\rm CMB}}^{XX}$ is the CMB power spectrum, $A_d^{XX}$ and $A_s^{XX}$ are the amplitudes of the $XX$ dust and synchrotron auto-spectra at 353 and 23\,GHz, respectively, and $\rho^{XX}$ is the correlation coefficient between dust and synchrotron emission, taken here to be independent of $\ell$. We normalize the amplitude parameters at $\ell_0 = 84$.

In this formulation, we are implicitly assuming perfect correlation across frequencies of both dust and synchrotron emission. While ``frequency decorrelation'' has yet to be observed in the dust $BB$ spectrum \citep{Planck_2018_XI}, variations in dust spectral parameters are well-attested \citep{Planck_2018_IV,Pelgrims:2021}, and even small levels of frequency decorrelation can influence constraints on the tensor-to-scalar ratio $r$ \citep{BICEP2018,CMBS42020}. For forecasting purposes, we do not include frequency decorrelation in both our simulated maps and our parametric fits. Nevertheless, searching for frequency decorrelation in dust and synchrotron emission is a potential Galactic science objective using SO data.

Using the sky simulations presented in Section~\ref{subsect:pysmmodels}, $C_\ell^{XX}\left(\nu_1\times\nu_2\right)$ for all combinations of $\nu_1$ and $\nu_2$ are computed using the \texttt{NaMaster} software\footnote{\url{https://github.com/LSSTDESC/NaMaster}} \citep{Alonso2019}. We employ a constant bandpower binning width $\Delta\ell = 15$ and use the masks described in Section~\ref{subsect:masks} apodized with the ``C1'' method and an apodization scale of 3$^\circ$. $E$- and $B$-mode purification is used when computing $EE$ and $BB$ spectra, respectively.

Finally, we use the noise models presented in Section~\ref{subsect:noise} to estimate the noise $\sigma\left(C_\ell^{XX}\right)$, which we add to the computed spectra. Explicitly \citep[e.g.,][]{Knox1995},

\begin{equation}
    \sigma\left(C_\ell^{XX}\right) = \sqrt{\frac{2}{\left(2\ell+1\right)f_{\rm sky}\Delta\ell}\left[\left(C_\ell^{XX}\right)^2 + \mathcal{N}_\ell^2\right]}
    ~~~,
\end{equation}
where the noise power spectra $N_\ell$ from Section~\ref{subsect:noise} are combined following

\begin{align}
    \mathcal{N}_\ell^2 &= N_\ell\left(\nu_1\right)^2, &\nu_1 = \nu_2 \\
    \mathcal{N}_\ell^2 &= \frac{1}{2}N_\ell\left(\nu_1\right)N_\ell\left(\nu_2\right), &\nu_1 \neq \nu_2
\end{align}
for auto and cross-spectra, respectively.

\subsubsection{Model Fitting}
As described in Section~\ref{subsect:pysmmodels}, the dust SED in each pixel of our simulated sky maps is a modified blackbody and the synchrotron emission in each pixel is a power law. Thus, it is most natural to model the $S_{\nu,d}$ and $S_{\nu,s}$ terms in Equation~\eqref{eq:cross_spectra} using the parametric forms corresponding to modified blackbody and power law emission (Equations~\eqref{eq:pysm_dust} and \eqref{eq:pysm_sync}). As we have adopted input simulations that have spatially uniform frequency spectra for dust and synchrotron, we expect our fits to the cross-spectra to reproduce these values. However, the $\ell$-dependence of the simulated emission at large angular scales is based on observational data and thus does not conform precisely to the power laws in Equation~\eqref{eq:cross_spectra}. Therefore, Equation~\eqref{eq:cross_spectra} is not an exact description of our input. Nevertheless, we find this parameterization adequate for all of the forecasting presented here and sufficient for assessing the constraining power of the SO observations.

We make two additional approximations to simplify the model fitting. First, given the lack of constraining power on the dust temperature at low frequencies where dust emission is in the Rayleigh-Jeans limit, we fix $T_d$ to its input value of 20\,K in all analysis. Second, since determination of the CMB spectrum is a principal aim for cosmological analyses in SO, we assume for our purposes that it is perfectly known and thus do not include it as a free parameter in the fit. For future analysis on SO data, we anticipate combining the framework presented here with that detailed in \soforecast\ to measure jointly both cosmological and astrophysical parameters. 

With these assumptions, the most general parametric fit to the ensemble of auto- and cross-power spectra in $EE$ or $BB$ involves seven parameters: $A_s$, $\alpha_s$, $\beta_s$, $A_d$, $\alpha_d$, $\beta_d$, and $\rho$. The simulated power spectra are fit using the \texttt{PyMC} software \citep{PyMC}.

We are interested both in how well the parameters of this model can be constrained with SO data and how well extensions to this model can be constrained. In particular, Section~\ref{subsec:sync_sed} explores sensitivity to curvature in the synchrotron SED, Section~\ref{subsec:dust_sed} constraints on the dust SED relative to existing data, and Section~\ref{subsec:dust_sync_cor} the spatial correlation between dust and synchrotron emission. More detailed study of spatial variability is possible both by separately analyzing different sub-regions of the sky or by map-level modeling of the SEDs, but these more sophisticated approaches are beyond the scope of the present study.

\subsection{The Galactic Synchrotron SED}
\label{subsec:sync_sed}

\subsubsection{Motivation}
The Galactic synchrotron radiation that dominates the radio sky at $\nu \lesssim 70$\,GHz arises primarily from cosmic ray electrons accelerated by the Galactic magnetic field. A power law energy distribution of the cosmic ray electrons yields a synchrotron SED that is also a power law. This functional form has proven effective at modeling synchrotron emission in CMB analyses, even when utilizing data with as low frequency as the 408\,MHz Haslam map \citep{Planck_2015_X}. Synchrotron emission is intrinsically linearly polarized with a microwave polarization fraction of a few percent in the Galactic plane and typically $\lesssim 15$\% at intermediate and high Galactic latitudes \citep{Page2007,Planck_2015_XXV}.

Radio observations of Galactic synchrotron emission have long provided evidence for a spatially variable spectral index, with a tendency for regions in the Galactic plane to have a shallower spectrum that those at higher latitudes \citep[e.g.,][]{Lawson1987}. However, analyses of total intensity data at GHz frequencies is complicated by the presence of other emission mechanisms, making interpretation difficult. Recently, owing to the availability of ground-based surveys of synchrotron polarization like the Q, U, I Joint Experiment in Tenerife \citep[QUIJOTE,][]{2021MNRAS.503.2927C}, C-BASS \citep{Jones2018}, and S-PASS \citep{Carretti19} in addition to all-sky measurements from WMAP and \planck, constraints on synchrotron spectral parameters have been obtained in polarized intensity as well. Such analyses have suggested that the power law index of polarized synchrotron emission is fairly uniform over much of the sky \citep{Dunkley2009,Svalheim2020}, though some level of variation has been observed \citep{Planck_2015_XXV,Krachmalnicoff2018,Fuskeland2021}. For example, \citet{Krachmalnicoff2018} reports a mean value of synchrotron spectral index $\beta_s\sim -3.2$ with spatial variation of the order of few percent in the frequency range 2.3--33\,GHz, by combining S-PASS with WMAP and \planck data.

The idealization of the synchrotron SED as a power law is expected to break down in detail. The cosmic ray electron energy distribution is likely to have a high energy cutoff, resulting in a exponential fall-off in the synchrotron spectrum at sufficiently high frequency. As the electrons lose energy via radiation, the spectrum steepens, thus making the synchrotron spectral index a probe of the time since injection \citep[e.g.,][]{Lisenfeld2000}. An additional complication is that multiple synchrotron emitting regions along the line of sight may have different slopes of the energy distribution. While the SED of each emitting region may itself be a power law, the integrated emission will not be. These effects motivate a search for curvature in the synchrotron spectrum.

Suggestions of curvature in the synchrotron spectrum have been reported in analyses of WMAP combined with radio data in total intensity \citep{Dickinson2009,Kogut2012}. In addition to probing the energetics of cosmic ray electrons, curvature in the synchrotron SED complicates removal of polarized synchrotron emission as a CMB foreground. Indeed, analysis of WMAP and Planck data has indicated that there is neither a region of the sky nor a frequency below 100\,GHz in which synchrotron emission is sub-dominant to CMB $B$-modes at angular scales of $\sim 1^\circ$ \citep{Krachmalnicoff2016}.

We therefore quantify the power of SO data to improve upon existing and forthcoming constraints on synchrotron emission from S-PASS, C-BASS, WMAP, and \planck. We find that the additional sensitivity and frequency coverage provided by the lowest frequency SO bands in combination with the other data sets provides a stringent test of a simple power law model of synchrotron polarization and breaks the degeneracy between the synchrotron amplitude and spectral index.

\subsubsection{Forecasting Framework}
As synchrotron emission dominates the low-frequency sky, we focus our analysis in this section on the low frequency data only. This consists of S-PASS (2.3\,GHz), C-BASS (5\,GHz), WMAP (K and Ka bands at 23 and 33\,GHz, respectively), and \planck (30\,GHz) in addition to the SO 27 and 39\,GHz bands. Although our simulated maps contain emission from dust, it is sufficiently subdominant at these frequencies that it can be neglected in the parametric fitting. We assume that CMB spectrum is perfectly known and so do not include it as a free parameter in the fit.

We focus our synchrotron forecast on the SAT survey. Although the LAT covers a greater sky area, including synchrotron-bright regions near the Galactic plane, Faraday rotation complicates analysis of the low frequency ancillary data, particularly S-PASS. The requisite masking negates much of the LAT's advantage over the more sensitive SAT. Even the SAT survey footprint overlaps with some potentially problematic regions, and so we augment our SAT mask (Section~\ref{subsect:masks}) with an additional Galactic latitude cut of $|b| < 30^\circ$, which, after apodization, reduces our mask-weighted $f_{\rm sky}$ to 6.8\%. We note that masking the low Galactic latitudes does not significantly impact the derived parameter constraints, as verified by analyses with less aggressive masking of the Galactic plane. Given the limited sky area and 91$^\prime$ and 63$^\prime$ resolutions of the 27 and 39\,GHz channels on the SAT, respectively, we restrict our analysis to $70 < \ell < 300$. 

To assess sensitivity to curvature in the synchrotron SED, we add the curvature parameter $s_{\rm run}$ to our parametric model of synchrotron emission in Equation~\eqref{eq:pysm_sync}:

\begin{equation} \label{eq:s_run}
    S_{\nu,s} = A_{s}\left( \frac{\nu}{\nu_{0,s}}\right)^{\beta_s + s_{\rm run} \log (\nu/\nu_{0,s})}
    ~~~,
\end{equation}
where $\nu_{0,s} = 23\,$GHz. The simulated maps have $s_{\rm run} = 0$, as described in Section~\ref{subsect:pysmmodels}. 

As we are neglecting dust and fixing the CMB, the full model of Equation~\eqref{eq:cross_spectra} for a given set of $EE$ or $BB$ spectra has only four free parameters to be fit: $A_s$, $\alpha_s$, $\beta_s$, and $s_{\rm run}$. The $Q$ and $U$ maps in the full set of seven frequency channels yield 28 auto- and cross-spectra for each of $EE$ and $BB$, while the reduced set of five frequencies without the two SO channels yields 15 auto- and cross-spectra for each of $EE$ and $BB$.

\subsubsection{Results}
\begin{figure}
\centering
 \includegraphics[width=\columnwidth]{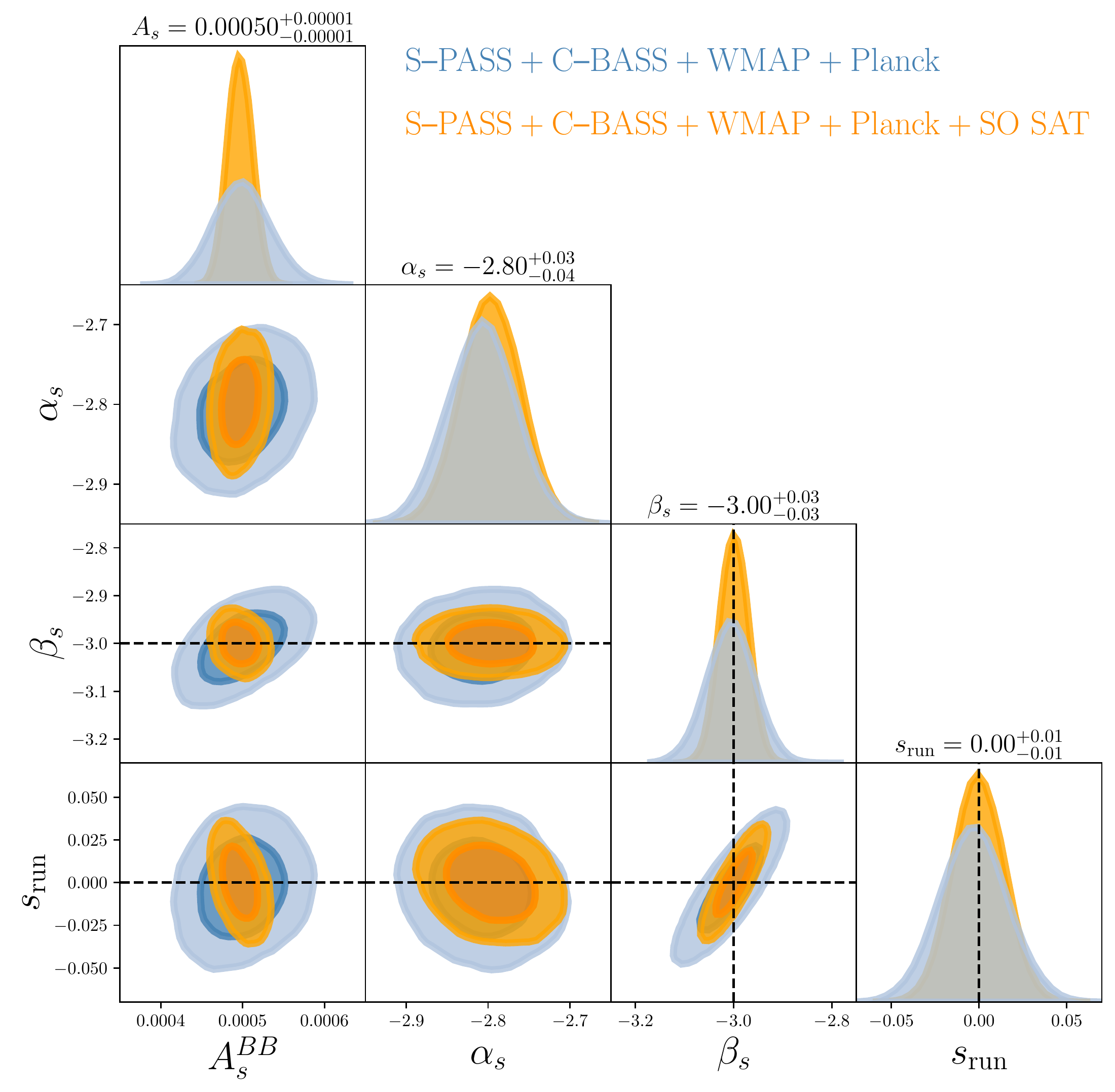}
 \caption{Posterior distributions on the synchrotron amplitude ($A^{BB}_s$, in $\mu$K$_{\rm RJ}$), spectral index in $\ell$ ($\alpha_s$), spectral index in frequency ($\beta_s$), and curvature of the spectral index ($s_{\rm run}$, see Equation~\eqref{eq:s_run}), obtained from fits to the $BB$ cross-spectra between $70 < \ell < 300$. The addition of SO SAT data to S-PASS, C-BASS, WMAP, and \planck data improves parameter constraints on $A_s$, $\alpha_s$, $\beta_s$ and $s_{\rm run}$ by factors of 2.3, 1.2, 1.7, and 1.3, respectively, as illustrated by the orange (including SO) versus blue (without SO) contours.} \label{fig:triangle_sync}
\end{figure}

We focus primarily on fits to the $BB$ spectrum, both for the importance of accurate foreground modeling for $B$-mode science as well as the fact that a $BB$ analysis is less sensitive to treatment of the CMB component itself. The results of the full fit to the simulated $BB$ spectra with and without the SO frequency bands are presented in Figure~\ref{fig:triangle_sync}. The input parameters are recovered without bias in all cases. The posteriors on all model parameters tighten with the addition of SO data. In particular, the constraints on $A_s$ improve by a factor 1.6 for the $EE$ spectra, and by a factor 2.3 for the $BB$ spectra; the constraints on $\beta_s$ improve by a factor 1.5 for the $EE$ spectra, and by a factor 1.7 for the $BB$ spectra. The constraints on $s_{\rm run}$ tighten by a factor 1.3 for both the $EE$ and $BB$ spectra.

Upcoming data from both SO and C-BASS will provide significant improvement on current constraints on the Galactic synchrotron SED that employ S-PASS, WMAP, and \planck data alone \citep{Krachmalnicoff2018}. Figure~\ref{fig:triangle_sync} highlights, for instance, how the sensitivity of the SO data at comparatively high radio frequencies can break the degeneracy between $A_s$ and $\beta_s$, sharpening constraints on the level of synchrotron emission. In addition to furnishing new constraints on the synchrotron SED, this helps enable searches for other polarized emission mechanisms at these frequencies, notably AME (see Section~\ref{subsec:ame_pol}).

\subsection{The Composition of Interstellar Dust}
\label{subsec:dust_sed}

\subsubsection{Motivation}
Recent analyses of polarized dust emission have found that its frequency dependence at millimeter wavelengths is well-fit by a modified blackbody having temperature $T_d$ and an opacity law scaling as $\nu^{\beta_d}$ with $\beta_d \simeq 1.5$ \citep{Planck_2015_X,Planck_2018_XI}. This simple parameterization provides a good description at both the map level and the power spectrum level at current sensitivities. The same values of $T_d$ and $\beta_d$ are found for both temperature and polarization to within measurement uncertainties \citep{Planck_2018_XI}. Balloon-borne observations from BLASTPol extending to sub-millimeter wavelengths likewise find consistency between the dust SED in total intensity and polarization, with deviations not exceeding $\sim 10\%$ \citep{Ashton_2018}.

Historically, most physical dust models have posited separate populations of silicate and carbonaceous grains \citep[e.g.,][]{Mathis1977,Draine1984,Zubko2004,Siebenmorgen2014,Jones2017,Guillet2018}. Being made of different materials, the grains have distinct opacity laws (i.e., different $\beta_d$) and come to different temperatures even when exposed to the same radiation field. Pre-\planck models anticipated pronounced differences in the dust SED in total intensity versus polarization \citep{Draine_2009}, which have not been observed \citep{Ashton_2018,Planck_2018_XI}.

Only recently have dust models consistent with Planck and BLASTPol observations been put forward. \citet{Guillet2018} presented a suite of four models based on separate populations of highly elongated (3:1) silicate and carbonaceous grains. These models are consistent with the observed frequency independence of the dust polarization fraction at the $\sim$10\% level, but with distinct variations at the few percent level. In contrast, \citet{Draine2021} introduced a single component ``astrodust'' model that posits that the submillimeter emission and polarization arises from a single homogeneous grain type. This model predicts a nearly constant polarization fraction across submillimeter and microwave frequencies, departing from this behavior only at THz frequencies.

The models of \citet{Guillet2018} and \citet{Draine2021}, as well as two versus one component models more broadly, can therefore be tested through differences in the dust frequency spectrum in total intensity vis-a-vis polarization. \citet{Planck_2018_XI} found $\beta_P = 1.53\pm0.02$ in polarized intensity and $\beta_I = 1.48$ in total intensity with negligible relative uncertainty, and so single component models with $\beta_I \simeq \beta_P$ remain viable and perhaps favored. The additional frequency coverage and polarization sensitivity of SO will result in tighter constraints particularly on $\beta_P$, and thus on the nature of interstellar dust, as we quantify below.

\subsubsection{Forecasting Framework}
We focus our analysis in this section on data from 23 to 353\,GHz. This consists of WMAP (K and Ka bands at 23 and 33\,GHz, respectively), and \planck (30, 44.1, 70.4, 100, 143, 217 and 353\,GHz) in addition to all SO bands: 27, 39, 93, 145, 225 and 280\,GHz. Our simulated maps contain CMB, synchrotron, and dust emission. Noise is added at the power spectrum level--we quote results for a single noise realization.

The SO LAT survey covers a greater sky area than the SO SAT survey and is thus better suited for this analysis. Unlike the forecast presented in Section~\ref{subsec:sync_sed}, we do not employ ancillary low frequency radio data and so are not concerned about Faraday rotation on sightlines near the Galactic plane. The high angular resolution of the LAT, ranging from 7.4$^\prime$ at 27\,GHz to 0.9$^\prime$ at 280\,GHz, permits signal-dominated forecasts on Galactic emission up to high $\ell$ values. Here we analyze $70 < \ell < 600$. 

The full SED model has seven free parameters (see Section~\ref{subsect:pysmmodels}): $A_s$, $\alpha_s$, $\beta_s$, $A_d$, $\alpha_d$, $\beta_d$, and $\rho$ to be fit using the ensemble of 120 auto- and cross-spectra constructed from maps in fifteen frequency channels for each of $EE$ and $BB$. The reduced set of nine frequencies without the six SO channels yields 45 auto- and cross-spectra for each of $EE$ and $BB$.

\subsubsection{Results}
\begin{figure*}
\centering
 \includegraphics[width=\textwidth]{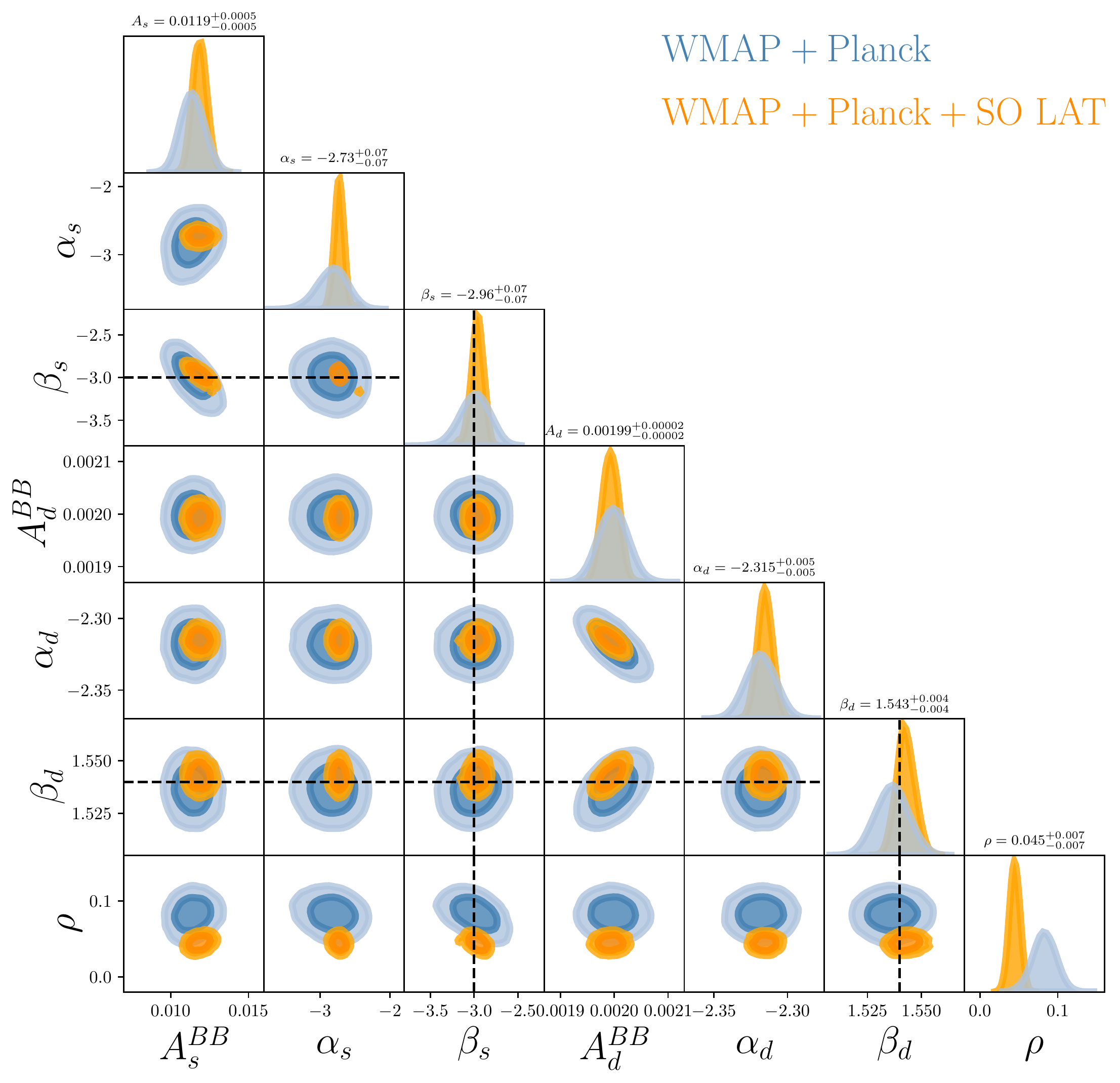}
 \caption{Posterior distribution of the parameters obtained by full fit of $BB$ cross-spectra model in Equation~\eqref{eq:cross_spectra}, of WMAP (23 and 33\,GHz), \planck (30, 44.1, 70.4, 100, 143, 217 and 353\,GHz) and all SO-LAT bands (27, 39, 93, 145, 225 and 280\,GHz), from $\ell = 70$ to $\ell = 600$. The addition of SO LAT data to existing WMAP and Planck data improve parameter constraints on $A_s^{BB}$ (in $\mu$K$_{\rm RJ}$), $\alpha_s$, $\beta_s$, $A_d^{BB}$ (in $\mu$K$_{\rm RJ}$), $\alpha_d$, $\beta_d$, and $\rho$ by factors of 1.7, 3.1, 2.4, 1.7, 2.0, 1.8 and 2.2, respectively as illustrated by the relative sizes of the blue and orange contours. Values quoted atop the 1D histograms are the 1$\sigma$ constraints on each parameter when including SO LAT data.} \label{fig:triangle_dust}
\end{figure*}

The results of the full fit to the simulated $BB$ spectra with and without the SO frequency bands are presented in Figure~\ref{fig:triangle_dust}, illustrating significant improvement on all parameter constraints with the inclusion of SO observations. We find that constraints on the synchrotron and dust amplitudes ($A_s$ and $A_d$, respectively), the synchrotron and dust spectral indices ($\beta_s$ and $\beta_d$), the scale dependence of the dust emission ($\alpha_d$), and the correlation between synchrotron and dust emission ($\rho$) all tighten at the factor of two level. The constraint on the scale dependence of the synchrotron emission ($\alpha_s$) improves by more than a factor of three due to the coverage and sensitivity of the SO data at low frequencies.

In absolute terms, the uncertainty on $\beta_d$ of $\Delta\beta_d = 0.01$ derived here with only WMAP and Planck data is only slightly more optimistic than the $\Delta\beta_d = 0.02$ derived from analysis of $BB$ spectra from a much narrower $\ell$ range ($40 < \ell < 59$) by \citet{Planck_2018_XI}, lending credence to this framework. The $\Delta\beta_d = 0.004$ achievable with SO as forecasted here is more than sufficient to discern whether the mean $\beta_d = 1.48$ measured in total intensity is indeed discrepant with the mean $\beta_d = 1.53$ measured in polarization \citep{Planck_2018_XI}, and thus whether the interstellar dust responsible for the FIR emission and polarization has indeed a largely homogeneous composition.

Even in the simplified sky in our simulations, the parametric model of Equation~\eqref{eq:cross_spectra} is an imperfect description. In particular, since the simulations are based on the observed sky at large angular scales, the $\ell$-dependence of the Galactic emission is not a perfect law, nor is the correlation between dust and synchrotron emission scale-independent. These limitations of the model are a possible source of the very slight bias ($<1\sigma$) in the recovered model parameters $\beta_s$ and $\beta_d$. More strikingly, the parameter degeneracies inherent in this model likely underlie the different, but not necessarily conflicting, posteriors on $\rho$ with and without the inclusion of SO. This underscores the important role of additional sensitive observations in both sharpening parameter constraints as well as testing the validity of the underlying model. In the case of the simulations employed here, we find no need to resort to more sophisticated models to accommodate the additional data, finding instead that our input parameters are recovered with even greater fidelity. This may not be the case for the real sky, where SO data will allow us to assess the need to elaborate our models beyond what has sufficed for the lower sensitivity observations of WMAP and \planck.

\subsubsection{Synergies with Other Experiments}

Higher frequency measurements beyond SO are well motivated to further constrain dust models and probe the relationship between $\beta_P$ and $\beta_I$ as well as other aspects of the Galactic polarization spectrum through the dust emission peak at THz frequencies. A number of CMB satellites have been proposed with polarization sensitivity at frequencies above 300 GHz such as PIXIE \citep{PIXIE2016} and PICO \citep{PICO2018}. Currently, the only funded satellite is the LiteBIRD CMB mission which covers frequencies up to 448\,GHz 
\citep{LiteBIRD2020}, leaving wide sky-area high-frequency measurements at higher resolutions to ground-based and balloon-borne observatories over the next decade. The Prime-Cam receiver on the Fred Young Submillimeter Telescope (FYST) has five frequency bands spanning from 220 to 850\,GHz \citep{Choi2020}. With similar sky coverage and ability to measure at higher frequencies, it will provide highly complementary data for many of the SO Galactic science goals \citep{CCATp_2021}. 

Balloon-borne experiments have the potential to make significant contributions. For instance, PIPER has frequency coverage up to 600\;GHz \citep{PIPER2016}, OLIMPO observes up to 460\,GHz \citep{OLIMPO2020}, and PILOT extends to 1.2\,THz \citep{Bernard2016}. Submillimeter experiments such as the proposed Balloon-Borne Large Aperture Submillimeter Telescope (BLAST) Observatory \citep{BLASTObs2020} would have the capability to survey hundreds of square degrees at frequencies between 850\;GHz and 1.7\;THz, providing a strong lever arm to distinguish between proposed dust models. Further exploration of the synergies between SO and other experiments is left for future investigations.

\subsection{The Correlation Between Synchrotron and Dust Emission}
\label{subsec:dust_sync_cor}

The Galactic magnetic field is fundamental to the polarization properties of both synchrotron radiation and thermal dust emission. The direction of linear polarization for both emission mechanisms is set by the orientation of the local magnetic field, while synchrotron emission is also sensitive to the field strength. Therefore, we expect the polarized synchrotron and dust emission to be correlated to some extent, as has been observed at the $\sim20\%$ level at large angular scales \citep{Choi2015,Krachmalnicoff2018}.

It remains unclear, however, to what extent the synchrotron and dust polarization signals probe different phases of the ISM and different regions of the Galaxy. The synchrotron emission depends upon the distribution of cosmic ray electrons, which may extend to large Galactic scale heights. In contrast, the distribution of dust grains is correlated with the atomic and molecular gas in the ISM. The polarized dust emission arises largely from the Galactic disk and, at high latitudes, from gas within a few hundred parsecs of the Solar neighborhood \citep[e.g.,][]{Alves2018,Skalidis2019}. Therefore, the imperfect correlation between polarized synchrotron and dust emission is not unexpected, and may change qualitatively depending upon region of the sky and angular scale probed. Large sky area, high angular resolution, and high sensitivity polarimetry of the microwave sky provides means of disentangling these correlations and clarifying the interrelationships between interstellar cosmic rays, dust, and magnetic fields.

The data model presented in Section~\ref{subsec:dust_sed} includes an explicit parameter $\rho$ governing the synchrotron-dust correlation. Constraints on $\rho$ from the $BB$ spectra are presented in Figure~\ref{fig:triangle_dust}, where we find that the inclusion of SO data can improve existing constraints on $\rho$ at the factor of two level. Further, we find that the posteriors on $\rho$ are quite sensitive to the inclusion of additional data, shifting from a larger degree of correlation ($\rho = 0.08\pm0.02$) to less ($\rho = 0.04\pm0.01$) when the simulated SO data are added to the analysis.

Which value of $\rho$ is correct? As discussed in Section~\ref{subsec:dust_sed}, the scale dependence of the simulated dust and synchrotron emission at large angular scales is set by observations of the Galaxy, not an analytic formula, and so the parametric fit of Equation~\eqref{eq:cross_spectra} can only approximate the input sky. Thus, a key role for the SO data is not simply tightening constraints on $\rho$, but testing whether the correlation between the two emission mechanisms can be adequately modeled as scale-independent. In the sky simulated here, we find this to be an excellent approximation, with the analysis successfully recovering the input parameters $\beta_s$ and $\beta_d$. Whether a scale-independent model is sufficient for the real sky, and what the implications are for where the observed dust and synchrotron emission originate in the Galaxy, require new observational data to answer.

\section{Probing Galactic Emission from Disks to Clouds to the Diffuse ISM}
\label{sec:multiscaleISM}
The deep sensitivity, high angular resolution, and large sky coverage of the SO surveys enable Galactic science cases spanning a wide range of physical scales and interstellar environments. In this section, we first quantify the expected signal to noise on measurements of Galactic dust emission across the sky (Section~\ref{subsec:snr_maps}), then present quantitative forecasts on the detectability of exo-Oort clouds (Section~\ref{sec:exooort}), the ability to map magnetic fields in a statistical sample of molecular clouds (Section~\ref{subsect:mol_clouds}), the prospects for detecting polarized CO emission (Section~\ref{subsec:co_pol}) and polarized AME (Section~\ref{subsec:ame_pol}), large-scale correlation analyses between SO dust polarization and upcoming stellar polarization surveys (Section~\ref{sec:starlightpol}), and finally the constraining power on the properties of MHD turbulence on small angular scales (Section~\ref{sec:turbulence}).

\subsection{Mapping Galactic Dust Emission with SO} \label{subsec:snr_maps}

\begin{figure}
\centering
 \includegraphics[width=\columnwidth]{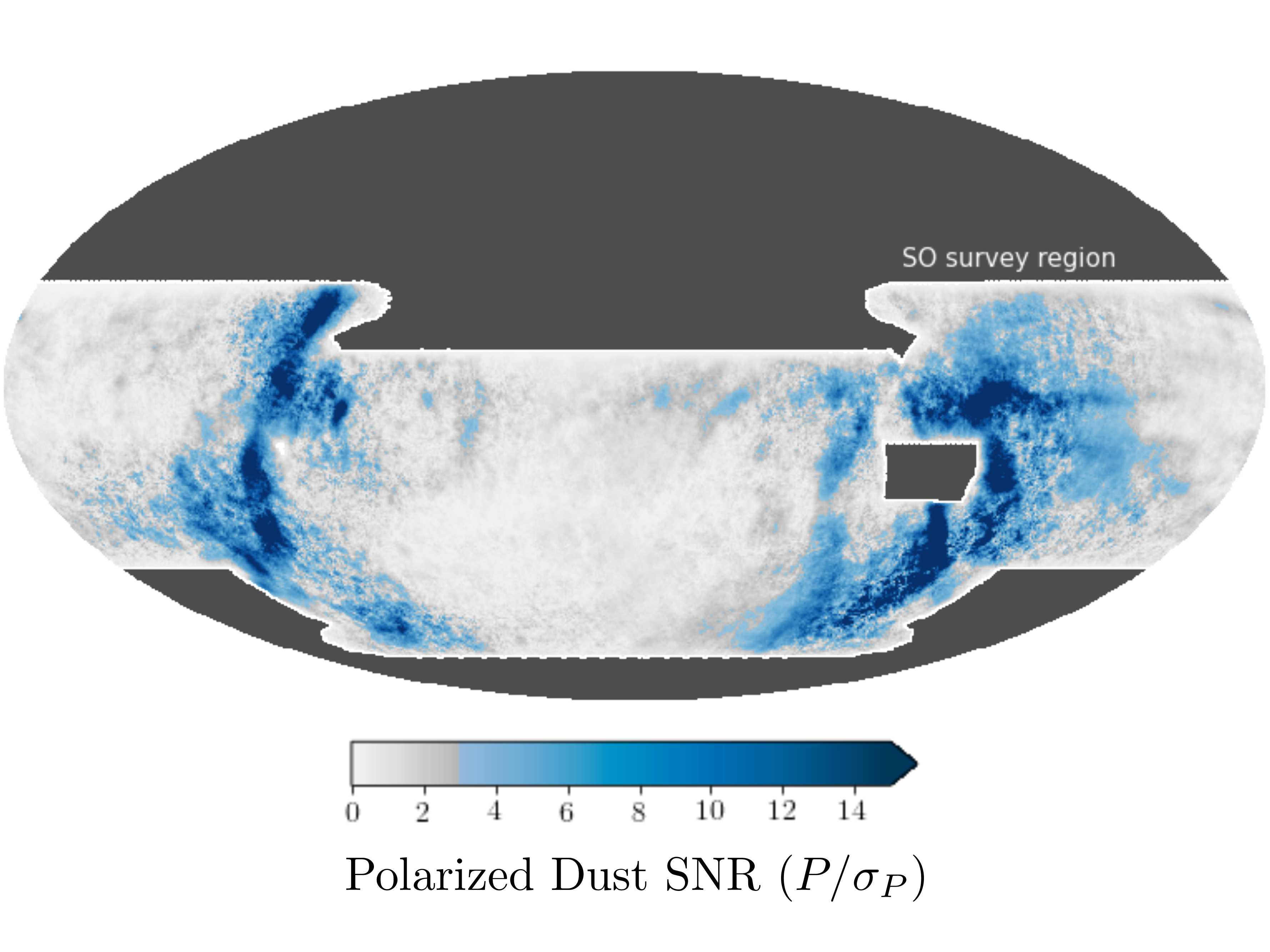}
 \caption{Map of the forecasted SNR on the polarized dust intensity at 280 GHz, for simulated data at a uniform $3.4'$ angular resolution. Regions where the polarized SNR is $<3$ are mapped in grayscale; blue indicates SNR $\geq3$, and saturates at SNR $=15$. Regions outside the nominal SO survey region are masked. The SNR calculation includes a model for LAT white noise and the sky-variable hit rate shown in Figure~\ref{fig:masks}.} \label{fig:SNR_280}
\end{figure}

We forecast the expected signal-to-noise ratio (SNR) for SO LAT measurements of polarized dust at 280\,GHz, simulated as described in Section~\ref{subsect:pysmmodels}. Figure~\ref{fig:SNR_280} shows the dust SNR, calculated as the ratio of the simulated 280\,GHz polarized intensity to the rms noise in each pixel from the noise model in Section~\ref{subsubsect:SOnoise}. The polarized dust SNR can be increased by degrading the resolution of the data, so analyses that involve maps of the spatial structure of dust polarization can optimize this inherent trade-off between sensitivity and map resolution. 

As Figure~\ref{fig:SNR_280} shows, at $3.4'$ resolution we expect the baseline SO survey to make $>3\sigma$ detections of polarized dust (blue regions) for an appreciable fraction of the low-Galactic latitude sky: about $12\%$ of the celestial sphere. For many lines of sight the SNR at this resolution will be much higher than three, or alternatively, for many regions SO will make high SNR dust polarization maps at higher ($>3.4'$) angular resolution.

\subsection{Exo-Oort clouds and Debris Disks}\label{sec:exooort}

Several processes associated with planet formation are thought to produce large quantities of dust in orbit around stars.  Dust produced in collisions between planetesimals, for example, can lead to the formation of debris disks with sizes of tens to hundreds of au \citep{Hughes:2018}.  Similarly, in our own solar system, planetesimals ejected from the inner solar system via interactions with the giant planets are believed to have formed the Oort cloud, which is also expected to have significant quantities of dust, and likely extends to tens of thousands of au \citep{Oort:1950}.   Dust at large distances ($\gtrsim 100\,{\rm au}$) from a central star is radiatively heated to temperatures of few to tens of Kelvin, resulting in an  emission spectrum that is fairly well matched to the frequency bands of SO. 

SO has the potential to detect thermal emission from dust in orbit at large distances around nearby stars.    We focus on the possibility of using SO to detect such emission from Oort clouds around distant stars, but will later comment on the ability of SO to probe debris disks.  While thermal emission from our own Oort cloud could potentially be detectable, the signal is expected to be roughly isotropic, and therefore difficult to distinguish from backgrounds (see \citealt{2009NewA...14..166B} for discussion of the signal from an anisotropic Oort cloud).  The possibility of detecting thermal emission from Oort clouds around other stars (exo-Oort clouds) has been explored previously by \citet{Stern:1991} and \citet{Baxter:2018}.  In \citet{Baxter:2018}, limits on the properties of exo-Oort clouds were set using \planck observations near \gaia-detected stars. Still, little is known about our own Oort cloud, let alone exo-Oort clouds.  A detection of such emission by SO would therefore represent an important advancement in planetary science.

To forecast the ability of SO to detect thermal emission from exo-Oort clouds, we generate and analyze simulated sky maps.  We begin with a full-sky galactic dust emission map generated at 280~GHz using \pysm.  Note that this simulated dust emission map does not include contributions from exo-Oort clouds. Additionally, we include a mock cosmic infrared background map from the Websky Extragalactic CMB Simulation \citep{2019MNRAS.483.2236S,2020JCAP...10..012S} and a realization of instrumental noise (see \S\ref{subsect:noise} for a discussion of the SO noise models). 

Simulated Oort cloud emission profiles generated using the model developed in \citet{Baxter:2018} are then inserted into the maps with distances sampled from \gaia detected main sequence stars. For scale, at $50~\rm{pc}$ a spherical Oort cloud with a typical radius of $50000~\rm{au}$ has an angular size $\sim 30^\prime$.  We note that emission from stars themselves is expected to be negligible in SO maps, except perhaps for some extreme giant stars.  Here we consider only the exo-Oort signal around main sequence stars, so we can safely ignore all stellar emission in our analysis.  We adopt a fiducial Oort cloud mass of $M_{\rm Oort} = 100~M_{\rm Earth}$ and a minimum grain size of $5~\mu\rm{m}$.   This mass is consistent with early estimates of the mass of our own Oort cloud \citep[e.g.,][]{1988Sci...242..547M}, but larger than more recent constraints \citep{2005ApJ...635.1348F}.  We emphasize that the mass of our own Oort cloud, let alone the typical mass of exo-Oort clouds, is poorly constrained; our fiducial model therefore provides a useful basis of comparison.  Adopting a model for the grain size distribution based on \citet{2005Icar..173..342P}, we can then calculate the total Oort cloud signal as described in \citet{Baxter:2018}.  We add Oort cloud signals around the roughly 4000 stars within $70~\rm{pc}$ that are detected by \gaia in the SO observation footprint.

In order to reduce the impact of diffuse galactic emission, we limit our analysis to regions of the sky (within the SO footprint) with low levels of emission from galactic cirrus, using a neutral hydrogen column density map from \citet{2016A&A...594A.116H}. In our fiducial analysis, we leave unmasked any pixels within the SO footprint whose column densities are in the bottom 25th percentile of this map. The remaining area is $\sim 5500$ square degrees.

\begin{figure}
    \centering
    \includegraphics[width=\columnwidth]{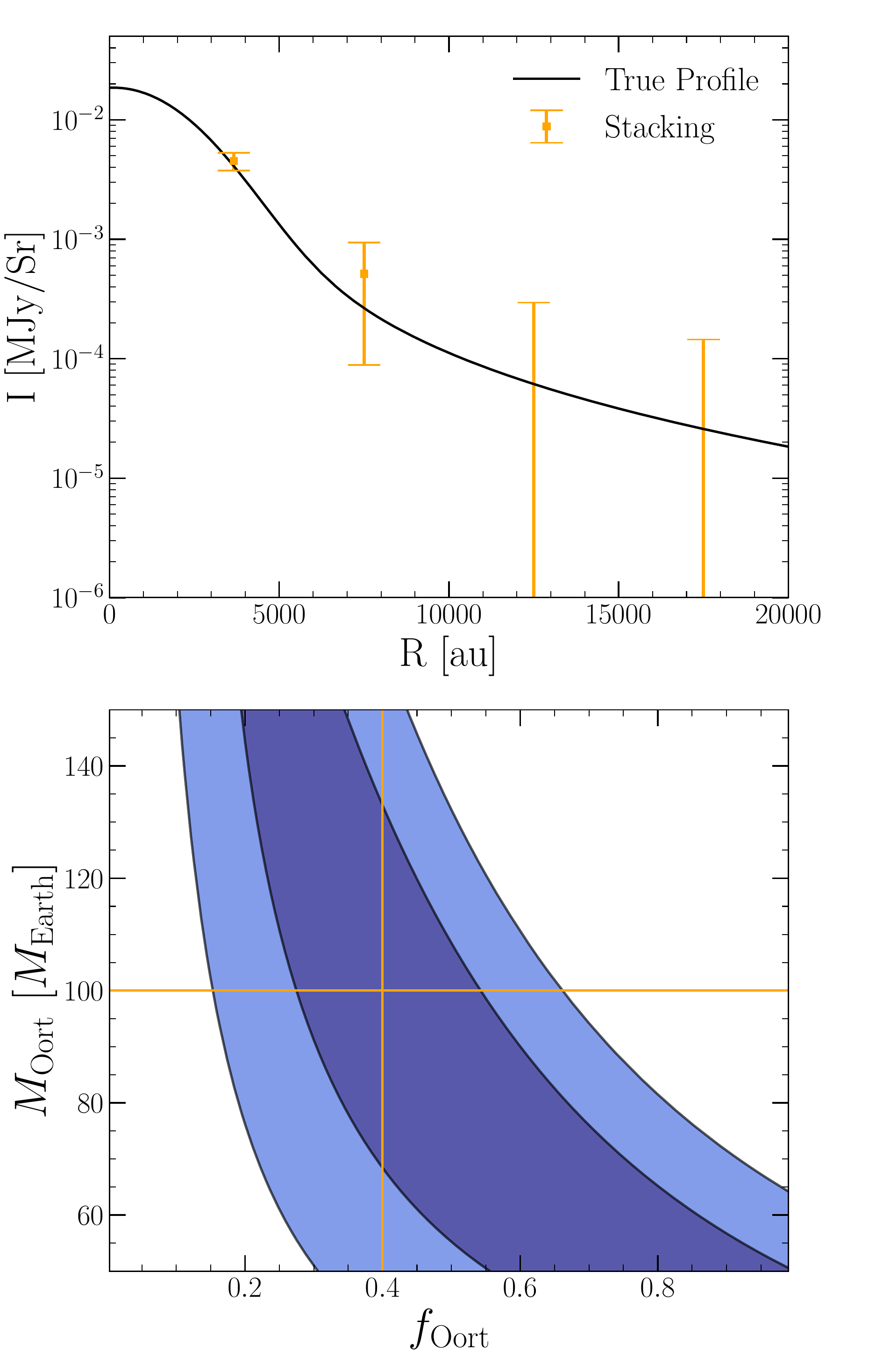}
    \caption{{\it Top}: model Oort cloud emission intensity profile as a function of projected radius (black curve), and the recovered profile from our analysis of simulated data (orange points with error bars). The orange points are obtained by averaging the measurements across the entire population of stars. Here we assume that every star hosts an Oort cloud.
    {\it Bottom}: results of a simulated analysis that assumes only a fraction $f_{\rm Oort} = 0.4$ of stars host Oort clouds.  Here, we treat $f_{\rm Oort}$ and the Oort cloud mass, $M_{\rm Oort}$, as free parameters.  Orange lines depict the true input parameters used to generate the mock data. Dark and light blue regions correspond to 68 and 95\% confidence regions, respectively.}
    \label{fig:SO_oort_forecast}
\end{figure}

Detection of an individual exo-Oort cloud is unlikely, unless it is very massive, contains very small grains, or is very nearby.  However, by averaging the emission profile around many distant stars, a detection can potentially be obtained. We refer to this method as stacking. In Figure~\ref{fig:SO_oort_forecast}, we show expected constraints on the Oort cloud emission profile, averaged across $\sim 4000$ stars (top panel).  These constraints are computed by azimuthally averaging the simulated intensity maps around all stars with Oort clouds in four evenly spaced radial bins of projected distance from the parent star, with a maximum radial extent of 20000~au. Although the SNR is expected to peak near the central star, signal at these scales could be confused with possible debris disk emission.  We therefore remove the pixels immediately surrounding the central star in our analysis. Background emission is estimated on a star-by-star basis in the simulated maps by averaging over an annulus with outer radius of 10$^\prime$ and inner radius given by the star's outermost radial bin.     The background estimate is then subtracted from the measurements in each radial bin. We note that our ability to extract the exo-Oort cloud signal is limited by our ability to estimate small-scale fluctuations in the Galactic dust emission. Improvements on our simple annulus-based estimates of the Galactic backgrounds could in principle enable the analysis to be extended to more distant stars, for which the accuracy of the small-scale background modeling becomes more important.

Given the analysis choices described above, we forecast a [5.9,  1.2, 0.06, 0.48]$\sigma$ measurement of exo-Oort cloud emission in each radial bin, from smallest to largest scale.

The formation of our own Oort cloud is believed to be connected to the presence of the giant planets.  Consequently, it is not necessarily the case that all stars host Oort clouds, and the stacking methodology described previously could result in a dilution of the exo-Oort cloud signal.  To account for this possibility, we re-generate the simulated data assuming only a fraction $f_{\rm Oort} = 0.4$ of stars host exo-Oort clouds, somewhat larger than the occurrence rate of giant planets \citep{2019ApJ...874...81F,2020MNRAS.492..377W}. 

Rather than averaging measurements across all stars, we now take measurements around each star individually, and fit them using a two-parameter mixture model similar to that developed in \citet{Nibauer:2020} and \citet{2021ApJ...907..116N} (in the context of debris disks and solar analog stars, respectively).  Constraints on the two parameters ---  $f_{\rm Oort}$ and $M_{\rm Oort}$ --- are shown in the lower panel of Figure~\ref{fig:SO_oort_forecast}.  A likelihood ratio test shows that the best fit parameters are within $1\sigma$ of the true input parameters. We have tested that for larger amplitude signals, the input parameters are still recovered to within the errorbars.  There is significant degeneracy between $f_{\rm Oort}$ and $\rm M_{Oort}$ in our constraints. A likelihood ratio test shows that $f_{\rm Oort} = 0$ is excluded at roughly 2.9$\sigma$.  In addition to providing constraints on  Oort cloud parameters, the same techniques could identify the most probable exo-Oort cloud candidates, which could then be provided to the community for follow up.

SO also has the potential to place constraints on the ensemble statistical properties of debris disks around nearby stars, complementing existing submillimeter measurements of individual disks from surveys like those made by ALMA \citep[e.g.,][]{2017ApJ...842....8M,2021ApJ...917....5N}. Submillimeter observations using CMB surveys such as SO are well suited to characterizing disks around faint stars, such as M dwarfs, or disks at large distances from their host stars. 

\cite{Nibauer:2020} used \planck observations to place constraints on the fraction of nearby stars hosting debris disks, the majority of which were M dwarfs.  The higher resolution of SO offers the potential of  improved constraints: with roughly five times better angular resolution than \planck, we expect to be able to probe roughly an order of magnitude more debris disks (assuming that the measurements are confusion-limited). Given our currently limited knowledge of the debris disk population around M dwarfs, such constraints would provide valuable insight into the evolution of planetary systems. SO is also likely to detect many individual debris disks (indeed, individual disks can even be detected in the lower resolution and sensitivity \planck data). Debris disk candidates could be provided to the community for higher resolution follow-up. 

\subsection{Molecular Cloud Magnetic Fields}
\label{subsect:mol_clouds}

Star formation takes place in molecular clouds, via gravitational collapse mediated by turbulence, feedback, and magnetic fields \citep[e.g.,][]{Shu1987, McKee2007, Federrath2015, Krause2020, Girichidis2020}. The relative importance of these processes as regulators of star formation has been a topic of much debate over the years. Theoretical models predict very different roles for magnetic fields. At one extreme are models where molecular clouds are magnetically supported, and star formation proceeds only when ambipolar diffusion has sufficiently decoupled the neutral material from the magnetic field to precipitate gravitational collapse \citep{Mouschovias1999, Hennebelle2019}. Other models hold that supersonic turbulence is the dominant regulator of star formation, and that cloud-scale magnetic fields are too weak to have much influence \citep{Padoan1999, MacLow2004}. Some other models find that turbulence and magnetic fields are both important \citep{Nakamura2005, Vazquez2011}. Other studies invoke feedback effects such as protostellar outflows and ionization due to the presence of nearby stars \citep{Cunningham2018,Krumholz2019}. 

Progress on a predictive theory of star formation requires detailed observations of magnetic fields in molecular clouds. Of particular interest are well-resolved maps of the magnetic field structure in clouds as measured by polarized dust emission, which probes the magnetic field orientation (but gives no direct measurement of the magnetic field strength). However, since dust polarization is sensitive only to the plane-of-sky component of the magnetic field, polarization measurements of molecular clouds are sensitive to the (unknown) angle between the line of sight and the local magnetic field orientation. Robust inferences about the role of magnetic fields in molecular clouds necessitate observations of enough molecular clouds to marginalize over this uncertainty. Current measurements do not provide a large sample.

To estimate the number of molecular clouds in the SO field, we scaled the number of clouds observed in \citet{Miville_2017} by the ratio of the Galactic plane coverage of the two surveys. Specifically, we define the SO molecular cloud survey area as the intersection between the SO coverage and a stripe centered at $0^\circ$ Galactic latitude and width corresponding to the highest latitude cloud found in the catalogue for both Galactic hemispheres. We then estimated the polarized dust emission at 225\,GHz and 280\,GHz from each cloud using the methods presented in Section~\ref{subsect:pysmmodels}. Finally, we selected the clouds that can be observed by SO with 1\,pc resolution or better with a signal to noise ratio greater than three. This value corresponds approximately to an error in polarization angle of 10$^\circ$ \citep{fissel2013}.

With these assumptions, we find that a SO survey will include more than 1300 molecular clouds that can be observed at 1\,pc resolution at $3\sigma$. For comparison, \planck only observed tens of molecular clouds with such resolution \citep{Planck_Int_XXXV}. 

Access to a large sample of molecular clouds is essential for understanding the magnetic fields in these objects. The observed polarized dust emission probes only the projection of the magnetic field on the plane of the sky, so the resulting polarized maps for clouds observed at different viewing angles will be significantly different. Moreover, it is possible that the influence of the magnetic field in a cloud is a function of age and mass \citep{Sullivan:2021}. Of the clouds with $>3\sigma$ detections, 850 will have measurements of at least 50 independent polarization vectors.

\begin{figure}
    \centering
    \includegraphics[width=\columnwidth]{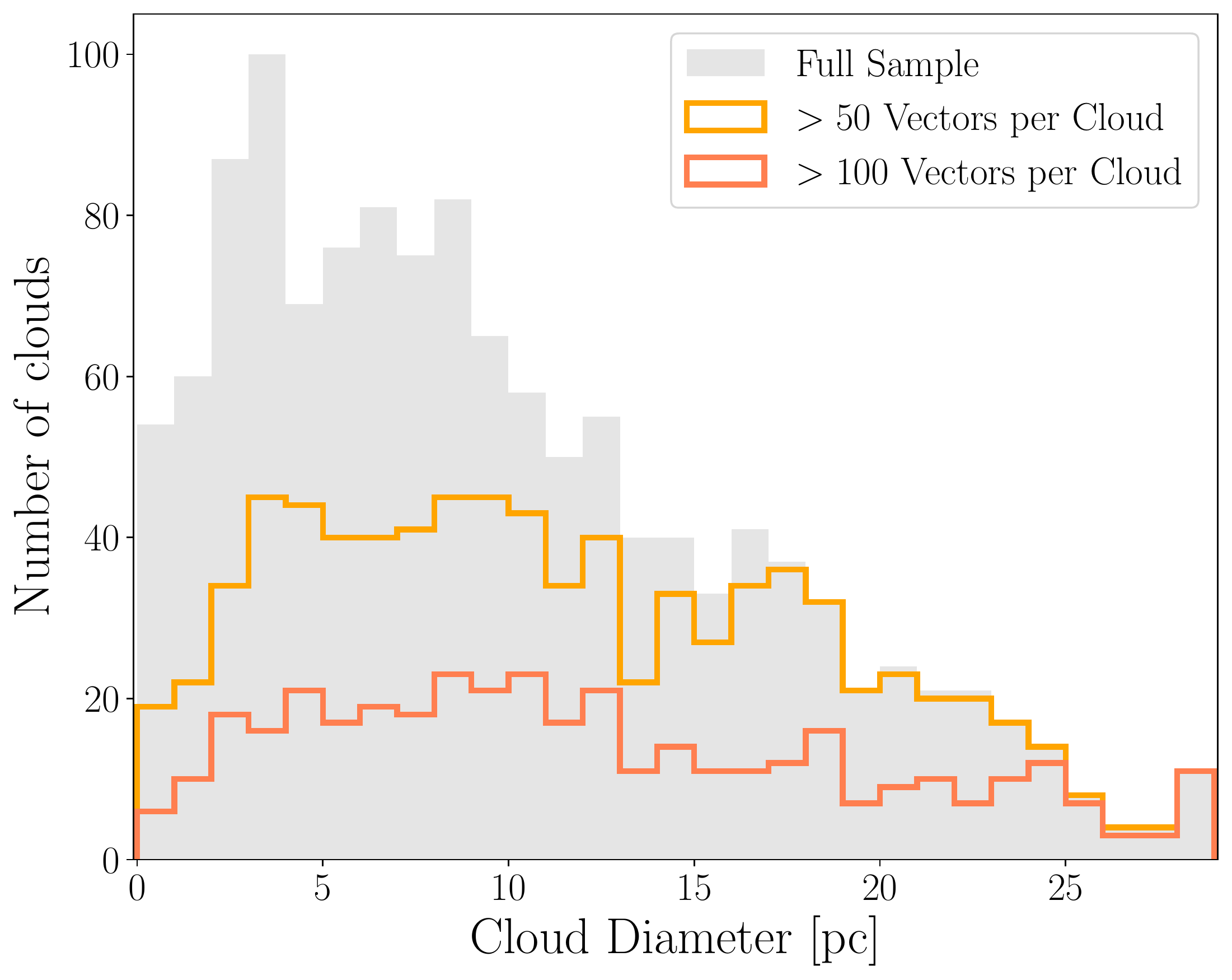}
    \caption{Histogram of the number of clouds with resolved polarization vectors (i.e., SNR higher than 3) for a nominal scanning strategy at 280\,GHz. The gray color represents the full sample of clouds with at least one pixel over the threshold, while the orange and red represent the subset of clouds with at least 50 and 100 pixels over the threshold, respectively. SO will map 850 clouds with $1$\,pc resolution and at least 50 high-resolution polarization measurements per cloud (orange line). }
    \label{fig:nominal_MC}
\end{figure}

Results for the nominal sky coverage for the 280\,GHz frequency band are presented in Figure~\ref{fig:nominal_MC} for the baseline noise scenario. The distribution for the 220\,GHz band is nearly identical and is not shown here. The only difference between the two bands is in the number of molecular clouds with more than 100 polarization vectors. In particular, we have 10\% more clouds at 280\,GHz versus the 220\,GHz band.

\subsection{CO line emission and polarization}
\label{subsec:co_pol}

\subsubsection{Motivation}
The cold molecular component of the ISM forms the reservoir of gas for star formation. The most abundant interstellar molecule, H$_2$, has no emission lines readily observable from the ground. In contrast, the microwave rotational lines of carbon monoxide (CO) are an excellent and accessible tracer of the molecular ISM. Under typical interstellar conditions, the brightest CO rotational transition lines are the $J = 1\rightarrow0, 2\rightarrow1$, and $3\rightarrow2$ transitions at 115.3, 230.6, and 345.8\,GHz, respectively.

Large-area CO line emission surveys have mainly observed a strip of the Galactic plane ($|b| \lesssim 15^\circ$) at moderate resolution \citep[e.g.,][]{dame2001}. \planck demonstrated that broad-band, multi-frequency CMB observations can be used to map CO in total intensity. By applying component separation algorithms to these data, \planck provided the first all-sky CO maps of the first three rotational lines, with an angular resolution ranging from 5 to 15$^\prime$ \citep[][hereafter \citetalias{Planck_2013_XIII}]{Planck_2013_XIII}. The maps include regions at $|b| \gtrsim 30^\circ$, where direct measurement of CO lines is challenging. While maps from this study contain a mixture of CO emission from different isotopologues, a subsequent analysis disentangled the $^{13}$CO and $^{12}$CO emission in the CO$(1\hbox{--}0)$ line over the whole sky \citep{hurier2019}.

CO line emission can be linearly polarized via the Goldreich-Kylafis effect \citep{Goldreich1981, Crutcher2012}. The presence of a magnetic field causes Zeeman splitting of the CO rotational levels $J$ into magnetic sublevels $M$. Unequal population of these sublevels gives rise to net linear polarization of the CO line. The levels may be differentially populated due to an anisotropic radiation field or the presence of a velocity gradient that causes the line optical depth to be anisotropic. The net effect is that the CO line can be polarized either parallel or perpendicular to the local magnetic field. This ambiguity may be resolved in practice if there is other information available on the system anisotropy \citep[e.g.,][]{Greaves:2002}. Despite this limitation, the Goldreich-Kylafis effect is an independent probe of magnetic field strength and orientation within molecular clouds that is, alone or in conjunction with other magnetic field tracers, a powerful tool for 3D magnetic field mapping in molecular and star-forming regions \citep{Greaves1999, 2006ApJ...653.1358K, 2008ApJ...676..464C}. 

The first detection of CO polarization in molecular clouds was obtained by \citet{Greaves1999} in complexes near the Galactic center, with a polarization fraction ranging from 0.5\% to 2.5\%. More recently, \citet{2006ApJ...653.1358K} detected linear polarization from \cotwoone in the multiple protostar system L1448 IRS3, \citet{2008ApJ...676..464C} mapped both dust and CO$(1\hbox{--}0)$ linearly polarized emission in the proximity of star-forming region G34.4+0.23~MM, and \citet{Teague2021} reported a detection of polarized emission from both $^{12}$\cothreetwo $^{13}$\cothreetwo and in the protoplanetary disk TW~Hya.

Wide CO polarization surveys are hard to undertake in practice given the intrinsically small degree of polarization and the long integration time required to achieve a significant detection of the signal in presence of atmospheric emission correlated in time. In principle, the same component separation approach could be applied to CO polarization as has been used in total intensity. However, the limited sensitivity of \planck data have so far prevented the extraction of any polarized CO emission. \citet{Puglisi2016a} presented a model to simulate the polarized emission of CO lines in molecular clouds, taking into account the 3D spatial distribution of CO in the Galaxy. The model was able to successfully reproduce the angular power spectrum of the observed \planck CO intensity maps \citepalias{Planck_2013_XIII}. However, in the absence of solid observational constraints, the model had to assume a strong correlation between the CO polarization and the polarized galactic dust emission to forecast the amplitude of CO polarized emission.

The SO frequency channels are designed to avoid the CO$(1\hbox{--}0)$ rotational line (the transmission at the line frequency is $10^{-4}$ and $10^{-3}$ for 90 and 150\,GHz LAT frequency channels, respectively). This is also the case for the CO$(3\hbox{--}2)$ line, with transmission around $10^{-3}$ in the 280\,GHz band. In contrast, the SO 220\,GHz channel has a transmission of 0.8 at $\sim 230.6$\,GHz, and is thus sensitive to the \cotwoone line. 

In this section we demonstrate that SO, with its low noise level and multi-frequency coverage, will constrain polarized CO emission at the level of polarization fractions $\sim 0.1\%$ in the brightest molecular clouds. The combination of LAT and SAT data will deliver observations of CO at unprecedented resolution and sensitivity across a large sky fraction, improving measurements of CO in the most diffuse regions by an order of magnitude or more on sub-degree scales.

\subsubsection{Forecasting approach and results}\label{sec:co-method}
\citetalias{Planck_2013_XIII} delivered maps of the intensity of the \cotwoone line with a resolution raging from 5 to 15$^\prime$, while \citet{Planck_2015_X} delivered updated CO maps at resolutions of $1^\circ$ and $7.5^\prime$ consistent with the results obtained with the latest reprocessing of the Planck data \citep{NPIPE2020}. These maps were extracted using targeted component separation methods and are subject to various degrees of foreground contamination. Given that the polarization of the CO line is largely uncharted territory, no wide survey exists that can be used as a template to assess any detection significance. For the purpose of forecasting SO performance, we use a template of the \cotwoone emission constructed from the CO$(1\hbox{--}0)$ map of \citet{dame2001}\footnote{The map is available in HEALPix pixelization at \url{https://lambda.gsfc.nasa.gov/product/foreground/fg_WCO_get.cfm}.}. This survey covers the Galactic plane and has an angular resolution comparable to the best \planck observations and, most importantly, was assembled from spectroscopic surveys. As such, it is less affected by contamination from other foreground emissions relative to the \planck broad-band measurements and has a slightly higher overall SNR.

In order to convert the CO$(1\hbox{--}0)$ \citet{dame2001} map into a \cotwoone template, we apply a constant multiplicative conversion factor of $0.595$ corresponding to the mean line ratio observed by \planck in bright CO clouds. The spatial variation of this mean factor introduces an overall error in the amplitude of our \cotwoone template $\lesssim 35\%$ along the Galactic mid-plane and lower elsewhere (see Section~4.2.3 and 6.2.2 of \citetalias{Planck_2013_XIII}). We refer to this \cotwoone template as $I_{\rm CO}$.

We assess the performance of SO in terms of the smallest polarization fraction of the CO emission that permits a 3$\sigma$ or greater detection, i.e.,
\begin{equation}
    \pco= \frac{ 3 \sigma_{\rm CO}}{I_{\rm CO}},
    \label{eq:PiCO}
\end{equation}
where $\sigma_{\rm CO}$ is the standard deviation of the noise expected in each pixel after a component separation step. Outside the Galactic plane where detections of CO polarization cannot be made on single lines of sight, we characterize the SO performance terms of amplitude of $\sigma_{\rm CO}$ only. Component separation is necessary to disentangle the CO from other sky emissions. For this purpose, we adopted a maximum likelihood parametric approach as implemented in the \texttt{fgbuster} package\footnote{\url{https://github.com/fgbuster/fgbuster}} \citep{stompor2009}. We assume that we have for each sky pixel a measurement from each of the SO frequencies $\nu_i$ with a Gaussian noise level in the pixel $p$ $\sigma^{\nu_i}_p$ given by the instantaneous reference noise values of the survey, modulated by the number of observations in each sky pixel within the footprint (hits map) as presented in Section~\ref{subsect:masks}. Our data model therefore reads

\begin{equation}
\mathbf{d}_p = \mathbf{A}_p\mathbf{s}_p + \mathbf{n}_p,
\end{equation}
where $\mathbf{d}_p$ is a data vector containing the measured signal for all the SO frequencies and Stokes parameters, $\mathbf{s}_p$ is a vector of the underlying sky component to be estimated from the data, and $\mathbf{A}_p\equiv \mathbf{A}_p(\boldsymbol\beta)$ is the component mixing matrix. 

A set of unknown parameters $\{\boldsymbol\beta_i\}$ describe the emission laws of each component analogous to the one described in previous sections (e.g., Section~\ref{subsect:ps_forecast}). The elements of the mixing matrix express the amplitude of each sky component at a given frequency, with each column representing a sky component and each row an observation frequency. We assume the noise variance to be uncorrelated between pixels and different frequencies, i.e., $\langle\mathbf{n}^{\nu_i}_p\mathbf{n}^{T,\nu_j}_{p^\prime}\rangle = \delta_{pp^\prime}\delta_{\nu_i\nu_j}\sigma^{2,\nu_i}_p$. 

For our baseline setup we assume that we can extract three signals from the measurements of the sky in the four highest SO frequencies. These components and their SEDs are: CO emission, assumed to be proportional to a delta function at the central frequency; dust emission, assumed to be a modified blackbody spectrum with fixed $\beta = 1.54$ and $T_d = 20$\,K, consistent with the model employed in Section~\ref{subsect:pysmmodels}; and finally the CMB.
As in \cite{Planck_2013_XIII}, the proportionality factor for the CO SED converts from K\,km\,s$^{-1}$ to K$_{\rm CMB}$ and is computed as the ratio of the CO line emission and CMB SED integrated across a SO reference bandpass \citep{2021JCAP...05..032A}.
In Section~\ref{subsubsec:compsep} we comment on how the analysis can be further improved by leveraging on the differences in the bandpass of the detectors.
Finally, note that this baseline model neglects low frequency emission mechanisms, which we explore further in Section~\ref{subsubsec:systematics}.  

Assuming all spectral parameters are fixed, the statistical noise in the CO map extracted through a minimum-variance, generalized least square estimation is
\begin{equation}
    \sigma^2_{\rm CO} = \left[ \mathbf{A}^t\ {\rm diag}({\{\sigma_{\nu_i}^{-2}\}})\ \mathbf{A} \right]^{-1}_{\rm CO,CO},
   \label{eq:costd}
\end{equation}
where the dependence on the pixel index has been omitted for simplicity. We note that, since $\mathbf{A}$ has dimensions of number of frequency channels times number of sky components, the two subscripts on the right hand side indicate the sky-component indices of the inverse matrix. This uncertainty level is based only on the spectral information of the signals and does not exploit prior expectations on the amplitude of the components nor their morphological properties. In this respect, it can be regarded as a conservative estimate, relatively robust to the expected significant CO-dust correlation. We elaborate on this aspect together with other possible shortcomings of the component separation assumptions in Section~\ref{subsubsec:compsep}.

\begin{figure*}
\centering
 \includegraphics[width=.9\textwidth]{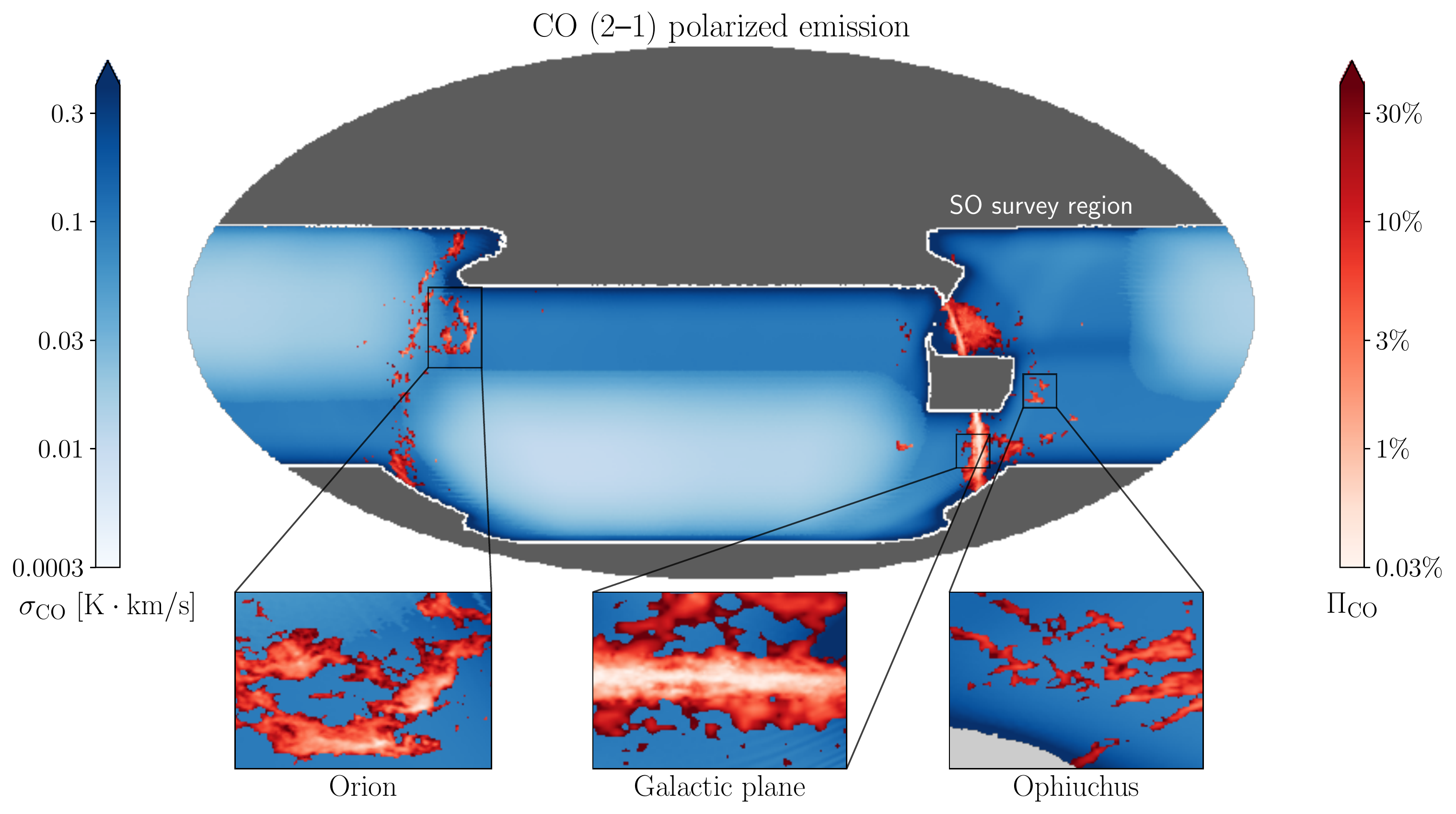}\\
 \caption{Sensitivity of the SO \cotwoone polarized emission map in equatorial coordinates. The gray mask identifies the sky regions not observed by SO LAT or SAT. The detectable $\pco$ with a SNR $\geq3$ is shown in red for regions of the sky covered by our \cotwoone intensity template $I_{\rm CO}$ (see Section~\ref{sec:co-method} for more details). Regions within the SO footprint display the noise rms in blue. The light blue areas correspond to the sky area where both SAT and LAT observations are available and the sensitivity is dominated by the SAT measurements. The zoom-in regions of Orion, Ophiuchus and the Galactic plane are centered on $(l,b)=(210^\circ,-14^\circ), (l,b)=(0^\circ,17^\circ), (l,b)=(-14^\circ,0^\circ)$ respectively and cover roughly 225\,deg$^2$.} 
 \label{fig:CO_detections}
\end{figure*}

We performed this analysis separately for the LAT and SAT surveys and computed the combined sensitivity of SO as the inverse variance combination of both surveys in the commonly observed sky area. Since the SAT is designed to target primordial B-modes, its survey avoids the Galactic plane but provides the deepest observations at intermediate resolution. The LAT conversely provides shallower high-resolution observations but observes the Galactic plane. In Figure~\ref{fig:CO_detections}, we show the expected noise level of the combined CO survey as well as the polarization fraction computed with Equation~\eqref{eq:PiCO} in different molecular clouds. The low noise level of the SO survey permits detection of the CO polarization at the sub-percent level and as low as $\sim 0.3\%$ in the most CO-bright regions in the Galactic plane. Along with the measured dust polarization, detection of CO polarization will allow us to study the interplay between magnetic fields and molecular gas as well as to assess the potential of the Goldreich-Kylafis effect as a means of mapping molecular-phase magnetic fields.

\begin{figure*}
\includegraphics[width=\textwidth]{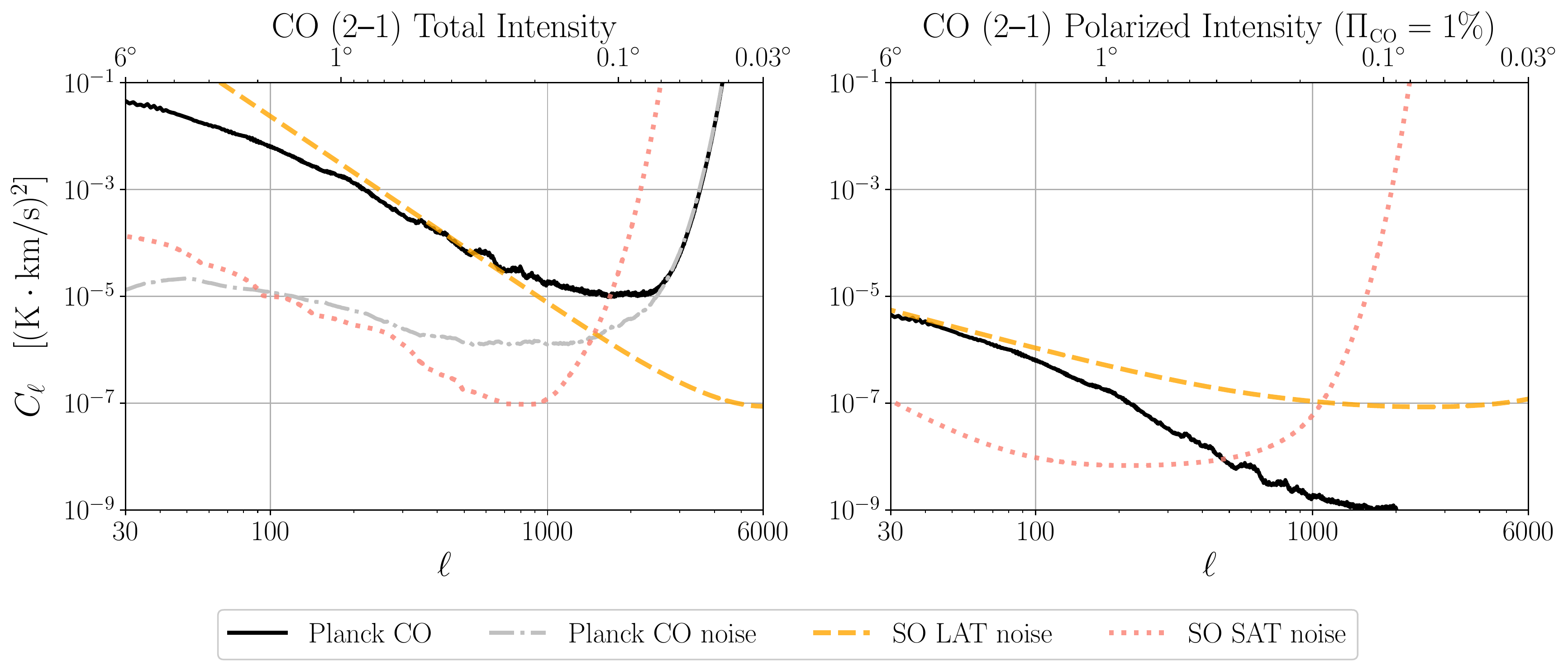}
\caption{Forecasted angular power spectra for \cotwoone intensity (left) and polarization (right) compared to noise power spectra. The solid black line is the power spectrum of the \planck \texttt{Commander} 2015 CO map \citep[][left]{Planck_2015_X} and the forecasted polarization power spectrum if the CO is polarized at a uniform $1\%$ level (right). These curves are compared to the noise power spectra of the SO LAT (orange dashed), SO SAT (pink dotted), and the noise power spectrum of the \planck CO map in total intensity (black dot-dashed). }
\label{fig:co-cls}
\end{figure*}

In order to understand at which angular scales SO will improve the most over current observations, it is useful to evaluate the overall noise of the CO map as a function of angular scale. We obtain the noise power spectrum of the SO CO map by evaluating Equation~\eqref{eq:costd} for each multipole $\ell$ with $\sigma^2_{\nu_i}$ replaced by the power spectrum $N_\ell^{(\nu_i)}$ of each SO frequency $\nu_i$. Figure~\ref{fig:co-cls} compares the CO intensity and polarization noise of SO to that of different \planck CO data products\footnote{We estimate the noise power spectrum of the Planck products from the publicly available CO null maps associated to each component separation approach adopted for the CO analysis.}. In intensity, at large angular scales the correlated noise induced by the atmosphere degrades the sensitivity of our data and Planck data dominate the sensitivity. However, our baseline LAT survey will have an almost three times lower noise than the most sensitive Planck data at scales $\sim 5\arcmin$ and improvements will reach two orders of magnitudes at scales $\sim 3\arcmin$. A combination of Planck and LAT observations in the Galactic plane could therefore deliver signal dominated measurements of CO clouds from degree to arcminute scales. 

In polarization, we see less degradation of the sensitivity at large angular scales compared to the intensity case as the atmosphere is largely unpolarized. The forecasted noise levels should allow, on average, a detection of CO line emission with a polarization fraction of $\pco\lesssim 1\%$ from $1^\circ$ to $10\arcmin$ scales. A hypothetical mean polarization fraction of CO emission of $\pco=1\%$ (shown in Figure~\ref{fig:co-cls}) could be measured at about $5\sigma$ significance even if more complex component separation approaches compared to the baseline case have to be used (see discussion below) and if foreground residuals are sufficiently low.  

The SAT CO survey improves significantly on \planck observations on scales as large as one degree, where the signal of typical CO clouds peaks \citep{Puglisi2016a}, and reaches a factor of four lower noise at $\sim 10\arcmin$. This survey will allow us to extend the search for low brightness clouds far from the Galactic plane and to set upper limits on their polarization properties. The improved sensitivity at large angular scales compared to the LAT is due to its larger field of view and to its half wave plate, which render the instrument less sensitive to correlated atmospheric noise. 
 
\subsubsection{Limitations due to intensity-to-polarization leakage}\label{subsubsec:systematics}
Given the SO sensitivity to the \cotwoone line, the main factors limiting the survey quality might be the systematic effects proper of a CMB instrument having a broad frequency response and residual Galactic emission after component separation. 

The most important systematic for this analysis is the temperature to polarization leakage in the 220\,GHz channel. The SO LAT telescope, contrary to the SAT, will not have any polarization modulator that will ease the separation of the Stokes parameter signal from single detector measurements. Since the details of the mapmaking approach for SO are not yet fixed, we assume the analysis of the LAT data will employ detector pair-differencing techniques to produce maps of the polarized sky, as commonly done for ground-based experiments. Despite being very effective in minimizing the dominant unpolarized emission in the data (mainly due to the atmosphere), this approach is sensitive to differences between the properties of the detectors that measure the incoming radiation across two orthogonal directions in a single focal plane pixel.  A mismatch in the bandpasses of the detectors translates in a direct leakage of unpolarized emission into the $Q$ and $U$ Stokes parameter maps ($I\rightarrow P$ leakage). Each focal plane pixel has in principle different bandpass mismatch properties and therefore the effect is expected to average out in the final map. However, due to the strength of the Galactic emission, this averaging effect might not be sufficient to prevent such leakage to be unimportant and, as such, the amplitude of the effect has to be carefully quantified. 

Differences in the beam shapes of the detectors can similarly cause $I\rightarrow P$ leakage. This effect is easier to account for in the analysis steps \citep[e.g.,][]{2021MNRAS.501..802M,2015ApJ...814..110B}, and preliminary studies indicate that the such contamination is expected to be minimal for SO \citep{2018SPIE10708E..3ZC,2021PhRvD.103l3540M}. 

\citet{matsuda2019} performed an extensive characterization of the bandpasses of bolometric instruments in the field in Atacama using a dedicated Fourier Transform Spectrometer coupled to the POLARBEAR telescope. The results showed that bandpass mismatch is fairly low for modern detectors and has a high level of stochasticity across the focal plane. Thus, we assumed a fixed bandpass leakage of $0.4\%$, consistent with the achieved upper limit on array-averaged bandpass measurements quoted in that work. This might be insufficient for regions close to the Galactic plane where the dominant unpolarized dust emission is very intense. We used the \texttt{Commander} \planck 2015 dust intensity map \citep{Planck_2015_X} as a template and compared its amplitude multiplied by the bandpass leakage with the $3\sigma_{\rm CO}$ detection threshold. In regions not covered by the $I_{\textrm{CO}}$ template we found that the median leakage corresponds to $\sim 4\%$ of any detected polarization fraction, while in the Galactic plane the median leakage is potentially higher and close to $20\%$ of the detected polarization fraction. This estimate is conservative and assumes that no temperature to polarization leakage can be mitigated through data analysis techniques or by the cross-linking  properties of the scanning strategy. The estimated leakage levels do not affect significantly our science case in particular for a blind survey of low emission regions outside the Galactic plane where no CO cloud has been detected so far. However, it might require more careful analyses in the Galactic plane where this effect can become proportionally more important. We note however that the most severe degradation happens for regions where $\pco\lesssim 1\%$ and thus SO measurements would still remain extremely competitive.

Finally, we note that in the evaluation of the CO noise variance in Equation~\eqref{eq:costd} we neglect any extra noise variance induced by correlated atmospheric noise. We estimated the impact of correlated noise computing the expected noise variance in real space from the CO noise power spectrum obtained with and without including the correlated noise in SO frequency channel prior to in the component separation. For this purpose we included scales around the peak of the CO emission $30<\ell<150$ and found that for the forecast shown of Figure~\ref{fig:CO_detections} the correlated noise can degrade the achievable $\pco$ by a factor $\sim 2$. 

\subsubsection{Limitations due to component separation effects}\label{subsubsec:compsep}
Dust-to-CO leakage might also occur due to an oversimplified model for the dust SED. For example, the value of the dust spectral index might be slightly different from the reference one and it cannot be fit per pixel (at least without priors) because it is highly degenerate with the amplitude of the polarized CO. One can imagine fitting the spectral index on scales larger than those of interest for the CO emission. This would effectively result in fixing the dust spectral index for the CO estimation. However, the emitting regions of the thermal dust and CO line emission are potentially the same, resulting in a variation of the spectral properties of the thermal dust emission precisely at the location and with the morphology of the CO emission. 

We estimated the importance of such effect computing the distribution of the dust amplitude at 230\,GHz $A_d^{230}$ relative to its value at 280\,GHz after a rescaling with a modified blackbody SED having a spectral index $\beta_d$ randomly drawn from a Gaussian distribution having a mean $\bar{\beta}_d=1.565$ and standard deviation $\sigma_{\beta_d} = 0.04$. These values correspond to the mean and standard deviation of the pixel values of the \planck 2015 \texttt{Commander} dust polarization spectral index map across the SO footprint. The ratio between the standard deviation of these $A_d^{230}$ with respect to the dust amplitude obtained using the mean $\beta_d=1.565$ give us an estimate of the fraction of the dust emission $\epsilon_{d}$ that can be left in the map due to a mismodeling of the dust SED. We estimate the amplitude of the dust bias on the detected CO emission rescaling the \texttt{Commander} dust polarization map $P=\sqrt{Q^2+U^2}$ to 230\,GHz using the \texttt{Commander} $\beta_d$ map and multiplying it by $\epsilon_{d}$. The median value of this bias across the footprint is $\sim 1.5\%$ of the detected CO signal and about $\sim 0.8\%$ of the polarized flux measured at $3\sigma$ outside the Galactic plane. For the bright regions shown in Figure \ref{fig:CO_detections}, the median value of the dust bias on the detected signal can go up to $\sim 3.5\%$. 

The reason why the number is relatively modest is the fact that our dust anchor is at 280\,GHz, thus close to the 230\,GHz of the CO emission. This makes the extrapolation of the dust amplitude only mildly incorrect when assuming an imperfect spectral index. The same effect is also present in the dust intensity, which is then converted into polarization by bandpass mismatch. We verified that this is negligible for the expected level of bandpass mismatch leakage ($\lesssim 0.2$\% in the Galactic plane and lower elsewhere prior to any mitigation induced by the cross-linking).

The complexity of Galactic emission may require us to adopt a more complex approach compared to the one implemented in our baseline analysis in order to produce CO maps with minimal contamination. We therefore investigated two alternative setups where we fit (1) $\beta_d$ and (2) $\beta_d$ plus the amplitude of a low frequency foreground. For the latter case we considered a synchrotron component with power law SED with $\beta_s=-3$ in Rayleigh-Jeans units for the polarization or free-free emission with spectral index -2 for the intensity. For both these two new cases we performed component separation by adding extra columns to the mixing matrix $\mathbf{A}$, the SED of the low frequency component and the derivative of the thermal dust modified black body with respect to $\beta_d$. When fitting for a low frequency component we include all the SO frequency bands, even the lowest two that we exclude from the baseline analysis.

Increasing the complexity of the foreground model reduces potential foreground biases but comes with the penalty of increased noise in the final CO map. An excess in the 220\,GHz channel can be interpreted either as a smaller $\beta_d$ or as CO emission. Therefore, the two parameters are highly correlated and fitting for $\beta_d$ in the component separation degrades the noise level of the CO map significantly: a factor of $\sim$2--3. Moreover, in order to fit for the CO amplitude and the dust amplitude and spectral index, we also have to leverage the intermediate frequencies. As their noise starts to rise exponentially due to their lower resolution, the amplitude of the CO noise also diverges. In other words, fitting for $\beta_d$ not only increases the noise of the CO map, but it also reduces its resolution.

This effect is even more apparent when fitting for low-frequency foregrounds. They are expected to be negligible at the \cotwoone frequency, but we now ignore this prior information and fit for their amplitude. As long as the low frequency channels have sufficient resolution, fitting the CO and the low frequency foregrounds, are decoupled problems: no increase in the CO noise is observed. However, as the noise in the low-frequency channels grows exponentially and we require these data alone to constrain all of the parameters, the uncertainty on the recovered CO emission increases dramatically. Summarizing, fitting for $\beta_d$ degrades the CO resolution by a factor of $\sim 3$, which becomes a factor of $\sim10$ if low-frequency foregrounds are also included. 

Finally, we note that a component separation approach similar to the one we employed has also been used by \planck to produce CO maps. Our approach also makes minimal assumption about the knowledge of the instrument bandpasses, requiring a characterization only of the mean response. This approach has been shown to provide more sensitive measurements optimized for low emission regions but it might be subject to a higher residual contamination due to Galactic emission in the brightest regions of the Galactic plane. Where the foreground contamination is high, alternative approaches based on blind methods exploiting the difference in the bandpasses of individual detectors in a given frequency band \citep[e.g., MILCA;][]{milca} have been shown to be potentially more robust (see Section~4.2.1 of \citetalias{Planck_2013_XIII}). We did not explore these methods because they require the knowledge of the bandpasses of each detector (a task challenging for an instrument like SO) and the details of the bandpass characterization of SO during its observational campaign have not yet been fixed. The Planck CO analysis also showed that they lead to higher noise levels compared to our approach (consistent with a white noise power spectrum $C_\ell \approx 2\times10^{-5}\,(\rm{K}\,\rm{km\,s}^{-1})^2$ up to $\ell\sim 1000$). We also note that the knowledge of the bandpass of the individual detectors would also improve the optimality of our CO estimate compared to our baseline approach that assumes the average response of the array.

\subsection{Is Anomalous Microwave Emission Polarized?}
\label{subsec:ame_pol}
\begin{figure*}
\centering
 \includegraphics[width=\textwidth]{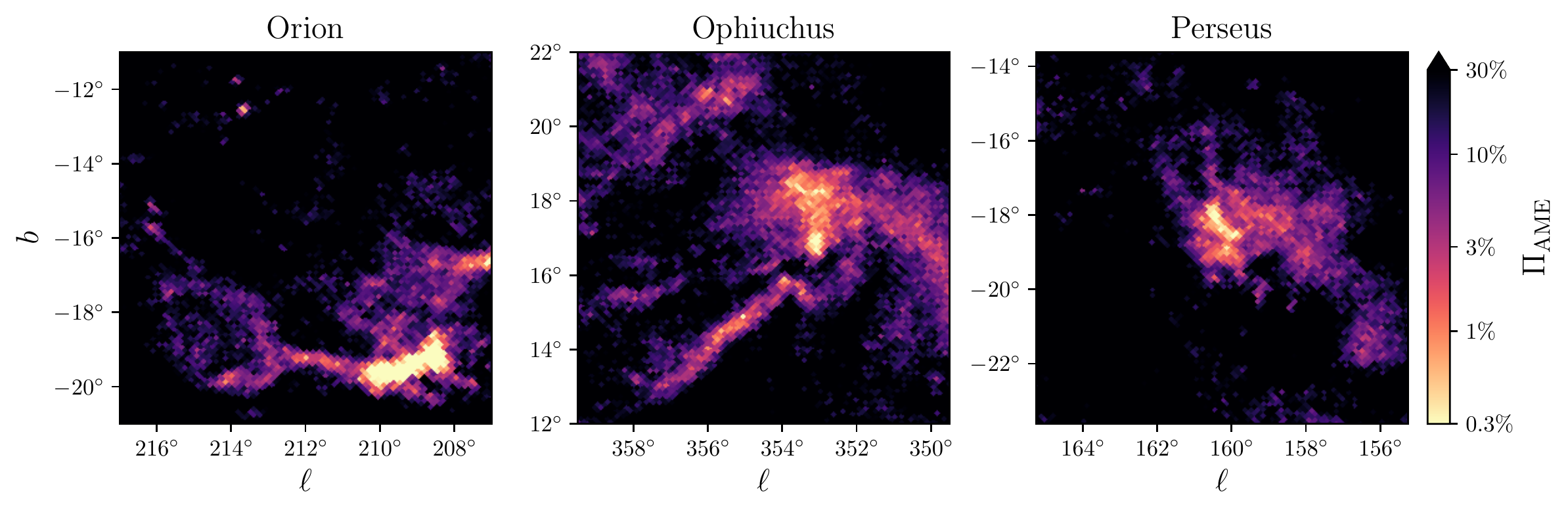}\\
 \caption{ $10\times10^{\circ}$ cutouts of Galactic regions that are promising targets for detection of AME polarization. The color corresponds to the minimum AME polarization fraction that could be detected at $3\sigma$ with SO LAT observations following Equation~\eqref{eq:ame_forecasts} evaluated at 27\,GHz. Constraints at the $p_{\rm AME} \simeq 0.1$\% level can be achieved in particularly bright regions.} \label{fig:AME_detections}
\end{figure*}

\subsubsection{Motivation}
The anomalous microwave emission (AME) was discovered as dust-correlated excess emission near 30\,GHz that exceeded model fits to dust emission at higher frequencies and free-free emission at lower frequencies \citep{Kogut_1996,deOliveiraCosta_1997}. It has since been established that AME is ubiquitous in the ISM, found wherever far-infrared dust emission is observed \citep{Planck_2015_X,Dickinson2018}.

Following the discovery of AME, it was quickly realized that electric dipole emission from rapidly-spinning dust grains could explain both its observed strength and frequency dependence \citep{Draine_1998a}. For grains to spin at frequencies of $\sim$30\,GHz, they must be $\lesssim 1$\,nm in radius and therefore the smallest of the interstellar grains \citep{Draine_1998b,AliHaimoud_2009}. 

It remains unknown whether AME is polarized \citep[see][for a recent review]{Dickinson2018}, though searches are ongoing \citep[e.g.,][]{Abitbol2018}. Current observational upper limits of both individual clouds \citep[e.g.,][]{GenovaSantos_2017} and the large-scale diffuse emission \citep{Macellari_2011,Planck_2015_XXV} suggest it must have a polarization fraction of no more than a few percent. This is consistent with indirect evidence from interstellar extinction. From the lack of polarization in dust extinction at ultraviolet wavelengths, it is known empirically that grain alignment is much less effective for small grains \citep{Kim+Martin_1995}. Thus, if sub-nm grains are powering the AME, the polarization must be similarly low.

Nevertheless, a small amount of residual polarization could result from enhanced magnetic relaxation in small grains heated to high temperatures \citep{Hoang2014}. In this scenario, the AME can achieve a maximum polarization fraction $p$ of nearly 1\%, with the polarized SED peaking at systematically lower frequencies than the total emission \citep{Hoang2013}. There is tentative observational evidence of alignment in small grains, including polarization in the 2175\,\AA\ extinction feature \citep{Clayton_1992,Wolff_1997} and in the 11.2\,$\mu$m emission feature attributed to polycyclic aromatic hydrocarbons \citep{Zhang_2017}. On the other hand, \citet{Draine_2016} argued that quantization of the vibrational energy levels in ultrasmall grains greatly suppresses alignment, resulting in negligible polarization ($p \ll 0.01$\%).

While spinning dust emission remains the favored explanation of the AME, other explanations have not been definitively excluded, particularly if they are present as subdominant emission components at these frequencies. These include magnetic dipole emission from ferromagnetic grains \citep{Draine_1999,Draine_2013,Hensley_2017} and thermal emission associated with structural transitions in amorphous grains \citep{Jones_2009,Nashimoto_2020}. Such emission mechanisms can be constrained by looking for departures from a single modified-blackbody fit to the polarized dust SED, as employed in Section~\ref{subsec:dust_sed}, e.g., sharp steepening or flattening of the dust emission spectrum at low frequencies.

Clarity on the degree of polarization of the AME, and thus the alignment physics of ultrasmall grains, is possible with the sensitive polarimetry of SO which extends to 27\,GHz. We describe in the following section how analysis of SO data will improve on existing upper limits, or perhaps make the first definitive detection, of AME polarization in select regions of interest.

\subsubsection{Forecasting AME fractional polarization }
Given the instrumental specifications of the low frequency SO LAT channels, we can estimate the SNR on the AME polarization fraction measured in any given pixel as the ratio of the AME polarization signal to the other polarized emission at that frequency:

\begin{equation}
{\rm SNR} = \sqrt{ \frac{  \left(p_{\rm AME} I_{\rm AME}\right)^2 } {  {\rm Var} \left[  P_{\rm noise} +P_{\rm cmb} \right]_{\rm MC}     + P^2 _{s}+P^2_{d}}  }~.
\label{eq:ame_forecasts}
\end{equation}
Here $I_{\rm AME}$ is the total intensity of the AME as given by the \texttt{Commander} AME map \citep{Planck_2015_X}, $p_{\rm AME}$ is the AME polarization fraction, $P_{s}$ and $P_d$ are the polarized synchrotron  and dust emission, respectively, evaluated at 27\,GHz. $P_{\rm noise}$ and $P_{\rm CMB}$ are the contributions to the measured polarization from instrumental noise and the CMB, respectively. The variance on noise and CMB maps is estimated from 20 Monte-Carlo (MC) realizations. Once the SNR map is built following Equation~\ref{eq:ame_forecasts}, we can then forecast the fractional polarization level $p_{\rm AME}$ required for SO to detect a polarized signal from AME at a given confidence level (cf. a similar approach for CO in Equation~\eqref{eq:PiCO}). We define $\Pi_{\rm AME}$ as the value of $p_{\rm AME}$ that would permit detection of AME polarization in a given pixel with SNR $= 3$.

To validate this forecasting methodology, we consider the upper limits on the AME polarization fraction set by \citet{Planck_2015_XXV} in the Pegasus Plume and Perseus. First we apply our methods to the \planck nominal noise of the LFI 30\,GHz channel assuming uniform coverage with 210\,$\mu$K-arcmin sensitivity in these regions. We evaluate the $2\sigma$ upper limit on $p_{\rm AME}$ from Equation~\ref{eq:ame_forecasts} by averaging the input quantities in a $\sim 30^\prime$ beam (the LFI beam at 30\,GHz) centered on the regions of interest. Our resulting values for Pegasus of $< 10.1\%$ agrees well with the reported $<12.8$ in \citet{Planck_2015_XXV}. The agreement is less good in Perseus, where our $<6.2\%$ is more conservative than the reported $<1.6\%$. This mismatch is likely due to the fact that the emission in Perseus is less localized than in the Pegasus Plume. We therefore anticipate that the simple approach employed here yields realistic to somewhat conservative forecasts when compared to more detailed analyses on real data.

Figure~\ref{fig:AME_detections} presents three regions where AME polarization can be detected above $3\sigma$ from SO LAT observations. In particular, the Ophiuchus and Orion regions are promising targets given the brightness of their dust emission. As demonstrated in Section~\ref{subsec:co_pol}, these are also excellent regions to search for CO polarization. We also show the Perseus region as it has been identified by \planck as a potential target for detecting AME polarization \citep{Planck_2015_XXV}.

The $\Pi_{\rm AME}$ computed with Equation~\eqref{eq:ame_forecasts} is conservative since the SNR can potentially be increased by averaging over larger regions. Considering the full LAT footprint outside the Galactic plane, we find that SO could make a $3\sigma$ detection of polarized AME in the mean Galactic SED if $p_{\rm AME}\gtrsim 0.1 \%$.

\subsection{Probing dust physics with SO and starlight polarization surveys}\label{sec:starlightpol}

In addition to producing polarized emission, aligned populations of asymmetric dust grains preferentially absorb and scatter optical light of different polarizations. Thus initially unpolarized starlight becomes polarized when attenuated by intervening dust between star and observer. The polarized intensity of microwave dust emission has been shown to correlate closely with optical polarization in both orientation and magnitude \citep{Planck_XXI}.

The characteristic ratio of microwave polarized intensity $P$ per unit optical polarization $p$, denoted $R_{P/p}$, depends on the size, shape, and composition of interstellar grains. Using all-sky 353\,GHz polarimetry from \planck and a sample of 1505 stars curated from various stellar polarization catalogues in the V band, \citet{Planck_2018_XII} found $R_{P/p} = 5.42\pm 0.05\,\mathrm{MJy}\,\mathrm{sr}^{-1}$ with no significant dependence on either dust column density or Galactic latitude at precision of the data. The observed value of this ratio is in sharp conflict with predictions from pre-\planck dust models \citep{Draine_2009}. \citet{Guillet2018} developed a new suite of dust models capable of reproducing the observed value of $R_{P/p}$, while \citet{Draine_2021} used it to derive constraints on the axial ratios and porosities of interstellar grains, finding a preference for oblate, relatively compact grains.

However, recent data have complicated this picture. Measuring polarization of 22 stars in a region of high 353\,GHz polarization fraction and comparing to the Planck polarized intensity, \citet{Panopoulou_2019} found $R_{P/p} = 4.1\pm0.1$\,MJy\,sr$^{-1}$, significantly lower than the all-sky value of \citet{Planck_2018_XII}. It is unclear whether this discrepancy arises from variations at spatial scales smaller than the \planck beam, genuine spatial variations in $R_{P/p}$ throughout the Galaxy, or unmodeled systematic errors. Polarimetry of more stars and higher sensitivity, higher angular resolution microwave polarization data are both needed to constrain the value of this quantity and its potential variation with interstellar environment.

Fortunately, the upcoming stellar polarization survey PASIPHAE \citep{PASIPHAE} will measure the optical polarization of millions of stars at high Galactic latitudes. We forecast here how combining these data with large area SO polarimetry will enable mapping of $R_{P/p}$ and thus the physical properties of Galactic dust. To do so, we consider an idealized 1\,deg$^2$ sky patch at some column density $N_{\rm H}$. The expected uncertainty on $R_{P/p}$ is

\begin{equation}
    \sigma_{R_{P/p}} = R_{P/p} \sqrt{\left(\frac{\sigma_P}{P}\right)^2 + \left(\frac{\sigma_p}{p}\right)^2}
    ~~~,
\end{equation}
where $\sigma_P$ and $\sigma_p$ are uncertainties on the measurements of $P$ and $p$, respectively. We adopt the baseline 280\,GHz LAT sensitivity $\sigma_P = 76\,\mu$K-arcmin (see Table~\ref{table:so_noise_model}).

The maximum monochromatic 353\,GHz polarized dust spectral energy density per atom H has been measured to be $2.5\times10^{-28}$\,erg\,s$^{-1}$\,sr$^{-1}$\,H$^{-1}$ \citep{Planck_2018_XI,Hensley2021}, which we scale to 280\,GHz assuming a modified blackbody emission law with $\beta = 1.54$ and $T_d = 20$\,K (see Section~\ref{subsect:pysmmodels}). To convert from a maximum polarized intensity to a mean value, we divide by three to average over line of sight inclination angles, finding $P/N_{\rm H} = 1.2\times10^{-23}\,$MJy\,sr$^{-1}$\,cm$^{2}$\,H$^{-1}$ at 280\,GHz. The corresponding $p/N_{\rm H} = 4.9\times10^{-24}$\,cm$^2$\,H$^{-1}$ is computed from the adopted 353\,GHz mean polarized intensity assuming the fiducial $R_{P/p} = 5.42/1.11$\,MJy\,sr$^{-1}$ at 353\,GHz, where the factor of 1.11 is the color correction from the Planck 353\,GHz band to a monochromatic value \citep[see discussion in][]{Hensley2021}. With these adopted values, $R_{P/p} = 2.38$\,MJy\,sr$^{-1}$ at 280\,GHz.

Finally, we estimate $\sigma_p$ from the forecasted performance of the PASIPHAE survey. The polarization fraction of each star is expected to be measured with an accuracy of $\sigma_p^\star = 0.2\%$. The most diffuse, high latitude sightlines have roughly 30 stars per square degree that lie behind the bulk of the dust column and that are bright enough for a high signal-to-noise detection \citep{PASIPHAE}. We assume this value corresponds to a column density of $1\times10^{20}\,$cm$^{-2}$ and that the stellar density increases linearly with column density. By limiting the stellar sample to sufficiently distant stars, we avoid complications from the stellar polarization not tracing all of the grains responsible for the polarized emission. The final $\sigma_p$ is then determined by dividing $\sigma_p^\star$ by the square root of the number of usable stars in the 1\,deg$^2$ patch. 

With these assumptions $\sigma_{R_{P/p}}$ is an analytic function of column density:

\begin{align} \label{eq:RPp_NH}
    \sigma_{R_{P/p}}^{\rm LAT} &= 2.38\,{\rm MJy}\,{\rm sr}^{-1}\times \nonumber \\ &\sqrt{0.112\left(\frac{10^{20}\,{\rm cm}^{-2}}{N_{\rm H}}\right)^2 + 0.56\left(\frac{10^{20}\,{\rm cm}^{-2}}{N_{\rm H}}\right)^3} 
    ~~~.
\end{align}
Thus, $R_{P/p}$ can be measured at 3$\sigma$ in a 1\,deg$^2$ patch for all column densities greater than $2\times10^{20}$\,cm$^{-2}$. Even the most diffuse regions \citep[$N_{\rm H} \simeq 5\times10^{19}$\,cm$^{-2}$;][]{2017ApJ...846...38L} can be accessed by modest increase of the patch size.

Equation~\eqref{eq:RPp_NH} makes apparent that measurements on diffuse sightlines ($N_{\rm H} < 5\times10^{20}$\,cm$^{-2}$) are most limited by the stellar polarization data, both the number of accessible stars and the low level of signal, rather than microwave emission at the LAT sensitivity. However, this analysis necessarily compares microwave emission integrated in a finite beam to stellar polarization measured in a pencil beam. Inhomogeneities of the dust properties and magnetic field geometry on scales smaller than the beam induce scatter in the relation that is not modeled by Equation~\eqref{eq:RPp_NH}. Thus we expect the SO observations not only to extend this analysis to lower frequencies than what can be done with the Planck data, but also to produce higher fidelity correlations by making more sensitive measurements at higher angular resolution.

A signal-to-noise greater than three on a single 1\,deg$^2$ patch over most of the sky illustrates the viability of {\it mapping} $R_{P/p}$. As $R_{P/p}$ is a probe of the shape, porosity, and composition of interstellar grains, correlations might be expected with other observables, such as the spectral index $\beta$ or the shape of the optical extinction law. The novel analyses enabled by these sensitive, large area datasets promise new insights into the properties of interstellar grains and the processes that shape them.

\subsection{ISM Turbulence}\label{sec:turbulence}

\begin{figure*}
\centering
 \includegraphics[width=\textwidth]{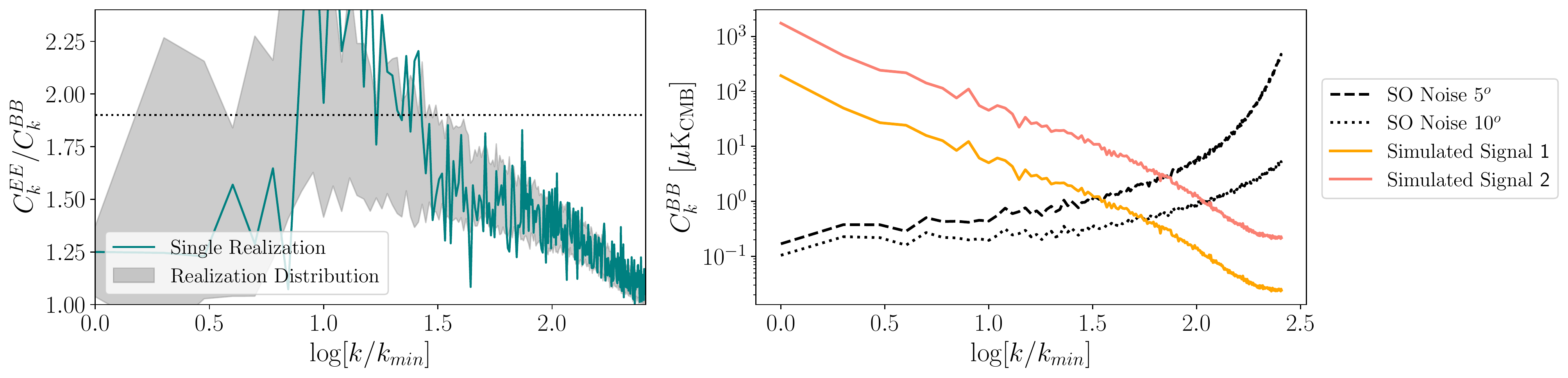}
\caption{Left: The $EE/BB$ ratio for synthetic polarization maps constructed from MHD simulations described in \citet{Kritsuk_2017}. A single timestamp of the MHD simulation is shown in teal, and the $1\sigma$ distribution over many timestamps is shown in gray. Right: $C_\ell^{BB}$ power spectrum for one (noiseless) realization of the MHD simulation, scaled by an arbitrary factor such that the median polarized intensity is 94 (orange) or 282 $\mu\mathrm{K_{CMB}}$ (pink). These theoretical power spectra are compared to the SO LAT polarization noise power spectra mapped onto the simulation domain, assuming that the simulation geometry is $5^\circ\times5^\circ$ (dashed line) or $10^\circ\times10^\circ$ (dotted line).   } \label{fig:simPS}
\end{figure*}

Turbulence, a physical phenomenon in which a fluid flow devolves into a cascade of swirls and eddies, is a ubiquitous state of  astrophysical fluids \citep{GS95,Elmegreen:2004,krumreview2014}. Interstellar structure is shaped by a supersonic turbulent cascade of energy, from sites of energy injection down to the dissipation scale \citep[e.g.,][]{Krumholz_2016}. Turbulence in the ISM can be driven by a wide variety of energetic events acting on different scales. On large ($\sim$0.1--1\,kpc) scales this may include supernovae explosions, the magnetorotational instability, and gravitational disk instabilities. On intermediate and small scales ($\lesssim 10$\,pc), stellar winds and jets from young stars are sources of turbulent energy injection in the ISM.

Turbulent motions are self-similar, with eddies forming smaller eddies, and this gives rise to a power-law behavior in the power spectra of ISM density or velocity tracers until dissipation mechanisms, such as ambipolar diffusion, damp the cascade \citep[e.g.,][]{BurL15}. Because turbulence correlates density, magnetic field, and velocity structures across scales, many observational diagnostics of turbulence involve correlations between observable quantities as a function of spatial scale \citep[e.g.,][]{Burkhart09a,Kritsuk_2017}. Thus, one key to observational diagnostics of turbulence is large spatial dynamic range, or high-fidelity measurements over a large range of scales. 

High-resolution maps of polarized Galactic emission offer a new window into the statistical properties of interstellar turbulence. \planck data at frequencies dominated by dust polarization show an asymmetry in the amplitude of the $EE$ and $BB$ autocorrelation spectra, $EE/BB \sim 2$, and a positive $TE$ correlation on large angular scales \citep{Planck_2018_XI}. Theoretical work suggests that the properties of the dust polarization power spectra may be related to parameters of MHD turbulence such as the sonic and Alfv\'enic Mach numbers, or other physics of the turbulent ISM \citep{Caldwell:2017, Kritsuk_2017, Kim:2019}. SO will extend \planck constraints on the polarization power spectra to higher multipoles, and SO's polarization sensitivity on small angular scales will enable analysis of these statistical quantities as a function of Galactic environment. 

To explore the theoretical variability of the polarized dust power spectra in turbulent environments, we compute power spectra of individual realizations of the simulations described in \citep{Kritsuk_2017}. As illustrated by the lefthand panel of Figure~\ref{fig:simPS}, the power spectra of individual simulation snapshots can deviate substantially from the mean over many snapshots of the simulation.

The real sky is likely to include regions described by very different MHD turbulence parameters, and these simulations suggest that we should expect substantial sky variability. Thus, measurements of the polarization power spectrum over small regions of sky are necessary. To forecast SO's ability to measure these power spectra, we need to translate the simulation domain (a line-of-sight-integrated cube $200~\mathrm{pc}$ on a side) to an angular scale. In other words, we need to place the simulated domain at some distance. In Figure~\ref{fig:simPS} we show the SO polarization noise power spectrum if we assume that the simulated dust polarization corresponds to a domain $5^\circ$ on a side or $10^\circ$ on a side. Here we face the limitations of state-of-the-art MHD simulations. Much of the high Galactic latitude dust lies within $\sim500$\,pc of the Sun \citep[e.g.,][]{Lallement:2019}. At a distance of 100\,pc, the $0.9'$ resolution of the SO 280\,GHz channel corresponds to a physical scale of $\sim0.03$\,pc, an order of magnitude higher than the physical resolution of the \citet{Kritsuk_2017} simulations. We show the measured $C_k^{BB}$ for one simulation realization, with the arbitrary simulation amplitude rescaled, for comparison to the noise power spectra. For bright regions of the polarized sky at 280\,GHz, SO will make high-fidelity measurements of the polarized power spectra over a wavenumber range where the \citet{Kritsuk_2017} simulations see substantial variation.   

One link between measurements of polarized cross-power spectra and properties of interstellar turbulence seems to lie in the filamentary structure of the magnetic ISM. The ISM is highly anisotropic, and high angular resolution observations of interstellar dust and gas reveal a network of filamentary structures on many scales \citep[e.g.,][]{Andre:2014}. In the low-column density ISM, density structures tend to be aligned with the sky-projected magnetic field orientation, whether traced by neutral hydrogen \citep{Clark:2014, Clark:2015, Martin:2015, Kalberla:2016} or dust emission \citep{Planck_Int_XXXII, Panopoulou:2016}. Toward denser sightlines, e.g., molecular cloud filaments, there is evidence that the relative orientation between density structures and the magnetic field becomes preferentially perpendicular \citep{Planck_Int_XXXV, Malinen:2016, Cox:2016, Fissel:2019}. The relative orientation between ISM gas and dust filaments and the magnetic field traced by polarized dust emission is thus a powerful point of comparison between theory and observations, and there is great theoretical interest in explaining the data and linking these insights to the physics of star formation \citep[e.g.,][]{Soler:2017, Seifried:2020, Barreto-Mota:2021}. SO dust polarization maps can be used to investigate correlations between density structures and the magnetic field, including within star-forming molecular clouds.

Recent work has shown that the preferential alignment between dust density structures and the ambient magnetic field drives the $EE/BB > 1$ and $TE > 0$ correlations measured in the diffuse ISM \citep{Clark:2015, Planck_Int_XXXVIII, ClarkHensley:2019, Huffenberger:2020, HerviasCaimapo2021}. This suggests that, for instance, the $TE$ correlation may be a sensitive probe of the physics of structure formation on different angular scales. SO data can be used to determine, for example, whether a negative $TE$ correlation is observed toward molecular clouds, where the sky-projected density structure is more perpendicular to the measured magnetic field orientation. Then $TE$ can be used in conjunction with estimators like the Histogram of Relative Orientations \citep{Soler:2013} or the Projected Rayleigh Statistic \citep{Jow:2018} to make detailed comparisons between theory and observations. Such measurements can be made from component-separated SO maps, or by cross-correlating SO polarization maps with total intensity maps derived from frequencies that are not contaminated by the CMB.  

\planck also measured a nonzero $TB$ correlation in the polarized dust emission \citep{Planck_2018_XI, Weiland:2020}. Recent work demonstrates that this parity-odd signal is driven by misalignment between dust filaments and the sky-projected magnetic field, with the sign of the $TB$ signal driven by the handedness of magnetic misalignment over a given region of sky \citep{Huffenberger:2020,Clark:2021}. This model predicts a nonzero $EB$ correlation in the dust polarization as well, which is a possible target for SO. Small misalignments between density structures and the magnetic field are ubiquitous in a turbulent ISM, and it remains an open question whether the positive $TB$ signal over large areas of sky is simply a statistical fluctuation away from $TB = 0$, or whether some ISM physics sets a preferred misalignment handedness. This question can be addressed with SO maps of the polarized dust emission at 280\,GHz. 

In addition to power spectra and other statistics mentioned above, SO synchrotron polarization data can be analyzed via polarization gradients \citep{Gaensler:2011}, which are sensitive to turbulent Mach numbers \citep{2012ApJ...749..145B,2018ApJ...855...29H,2019MNRAS.486.4813Z,2020ApJ...905..130C}. Understanding the physical properties of interstellar turbulence is a fundamentally multi-scale problem, and wavelet-based or otherwise hierarchical analyses of SO synchrotron and dust polarization data will be particularly valuable \citep[e.g.,][]{Robitaille:2017, Regaldo-SaintBlancard:2020}. Statistics of dust polarization also probe the 3D distribution of density and magnetic fields in interstellar environments; a critical consideration for disentangling inferences of turbulent properties from projection effects \citep{Fissel:2016, Ghosh:2017, Clark:2018, Sullivan:2021, Pelgrims:2021}. 

\section{Conclusions}\label{sec:conclusions}
SO will make sensitive measurements of Galactic emission that can be used to probe a diverse set of science questions. We briefly summarize the main forecasts considered in this paper below.

\begin{enumerate}
    \item The frequency-space coverage of SO will enable detailed constraints on the physics of Galactic emission. Constraints on the frequency spectra of dust and synchrotron emission, including the value of $\beta_d$ in polarization and curvature of the synchrotron spectrum, will be improved at a factor of two level relative to current constraints. Achieving the forecasted $\Delta\beta_d \lesssim 0.01$ would establish definitively whether current best determinations of $\beta_d = 1.48$ and 1.53 in total intensity and polarization, respectively \citep{Planck_2018_XI}, are a real difference, thereby testing one-component dust models which predict nearly identical $\beta_d$ for both.
    \item SO's multi-frequency view of Galactic magnetic fields probing both dust and synchrotron emission will help unravel the field structure both in different ISM phases and in different regions of the Galaxy. SO, in combination with other microwave and radio polarimetry, can measure the correlation coefficient between polarized dust and synchrotron emission with a factor of two greater precision than is possible with current data.
    \item SO will map the Galactic polarized dust emission at 280\,GHz over a nominal survey region that covers 40\% of the sky, with a sensitivity that enables $> 3\sigma$ measurements of dust polarization at $5'$ over about 12\% of the celestial sphere. The SO sensitivity and $1'$ native resolution will bridge a critical gap between \planck measurements of dust polarization on large angular scales and the sub-core scales probed by ALMA.
    \item Dynamical processes involved in planet formation generate debris disks and clouds of dust and rocky bodies out to hundreds or thousands of au around stars. SO will constrain the population of exo-Oort clouds around nearby \gaia stars. Oort cloud formation may be related to the presence of giant planets. Joint SO constraints on the abundance fraction and masses of exo-Oort clouds will exclude the non-existence of exo-Oort clouds at roughly $2.9\sigma$ if the true fraction is similar to the detection rate of giant planets. 
    \item Understanding the role of magnetic fields in star formation requires high significance measurements of polarization in a statistical sample of molecular clouds. SO will map more than 850 molecular clouds with at least 50 independent polarization measurements at $1$\,pc resolution.
    \item In select regions including Ophiuchus, Orion, and Perseus, SO polarimetry can constrain the presence of polarized CO emission and AME at sub-percent levels. The SO 220\,GHz band permits sensitive searches for polarized CO(2--1) line emission in other dense clouds with potential to expand significantly the sample of known sources. AME polarization can be detected in the mean Galactic polarization spectrum in the LAT footprint if its polarization fraction is $\gtrsim 0.1$\%.
    \item The sensitivity and large sky coverage of SO will enable joint analyses of polarized thermal dust emission and current and forthcoming optical starlight polarization surveys. The combination of SO and the PASIPHAE starlight polarization measurements can make resolved maps of the emission-to-starlight correlation coefficient at $1^\circ$ resolution for all sightlines with $N_{\rm H} \gtrsim 2\times10^{20}$\,cm$^{-2}$. This will furnish new constraints on the shape and porosity of interstellar grains.
\end{enumerate}

The SO maps will be used to study a number of topics and objects of interest in addition to those enumerated here, many of which build upon the foundations laid by current ground-based experiments. For instance, ACT has observed supernova remnants and pulsar wind nebulae in the Galactic center \citep{Guan:2021}. SO is poised not only to detect many more of these over the full survey area, but to characterize their frequency spectrum over a full order of magnitude in both total and polarized intensity. These environments are convenient laboratories for studying the physics of particle acceleration and of the supernova itself, as well as the shocked ambient ISM.

The Galactic center itself is a noteworthy target itself, being an extreme region of the Galactic disk ideal for testing star formation theories. As ACT maps demonstrate the utility of microwave intensity and polarization maps for studying the Galactic center region \citep{Guan:2021}, SO is poised to provide robust component separation with its broader frequency coverage and higher sensitivity.

Other regions of interest in the SO footprint include the Magellanic system, enabling observations of the dust emission in the Large and Small Magellanic Clouds. Total intensity maps at similar frequencies were created using SPT and \planck data \citep{Crawford:2016}. The Magellanic Clouds offer a nearby view of star formation and magnetic fields in galaxies with very different properties from the Milky Way \citep[e.g.,][]{McClure-Griffiths:2018}. Additionally, they are known to harbor dust with microwave emission in excess of predictions based on Galactic dust models \citep{Bot2010,Israel2010,Planck_Early_XVII}, which may point to more exotic emission mechanisms such as from amorphous \citep{Paradis2011} or ferromagnetic \citep{Draine_2012} grains.

A recent study employing ACT data has demonstrated the utility of CMB telescopes for studying the Solar System \citep{naess2021planet}. Based on the clustering of Kuiper Belt objects, it has been proposed that a 5--10\,$M_\Earth$ Planet 9 may be orbiting the Sun at 400--800\,au \citep{Batygin_2016,Batygin_2019}. While faint at optical wavelengths, the putative Planet 9 could be detected from its thermal emission in the far-infrared and microwave. Although \citet{naess2021planet} reported no significant detection, they were able to shrink the allowed parameter space by 17 and 9\% for 5 and 10\,$M_\Earth$ planets, respectively. With increased sensitivity and longer time baselines, SO data can complement and extend this analysis. Additionally, microwave observations can constrain models of Solar System planet temperatures \citep[e.g.,][]{Weiland_2011,Hasselfield_2013,Planck_Int_LII}.

Finally, both ACT and SPT have revealed dramatic time variability in the microwave sky and thus a vast discovery space for next-generation CMB experiments \citep{Guns2021,Naess2021transient}. Microwave transient science is an active area of development within SO, but as it lies outside the scope of the science cases and forecasts presented here, a detailed exploration is left for future work. 

This paper makes specific forecasts for Galactic Science with SO. Other next-generation CMB experiments can also provide rich information on the physics of the magnetized ISM of the Galaxy. Satellite experiments like LiteBIRD \citep{LiteBIRD2020} and PICO \citep{Hanany2019} can extend some of these science cases to all-sky data. Many of the sciences cases detailed here are directly applicable to the future ground-based experiment CMB-S4. The landscape of Galactic science with CMB experiments is also enriched by balloon-borne observatories like SPIDER \citep{Crill_2008} and the proposed BLAST Observatory \citep{BLASTObs2020}. We anticipate that the analysis presented in this work will serve as a roadmap for Galactic science with other microwave polarization experiments as well.

\acknowledgments 
This work was supported in part by a grant from
the Simons Foundation (Award \#457687, BK). We thank Alexei Kritsuk for sharing the numerical simulation data used in this work. 

S.E.C. acknowledges support by the National Science Foundation under Grant No. 2106607. NK, DP, GP, CB acknowledge support from the \href{cosmosnet.it}{COSMOS Network from the Italian Space Agency}. NK, DP, and CB also acknowledge support by the \href{web.infn.it/CSN4/IS/Linea5/InDark}{INDARK INFN Initiative}. GF acknowledges the support of the European Research Council under the Marie Sk\l{}odowska Curie actions through the Individual Global Fellowship No.~892401 PiCOGAMBAS. GC is supported by the European Research Council under the Marie Sk\l{}odowska Curie actions through the Individual European Fellowship No. 892174 PROTOCALC. SKC acknowledges support from NSF award AST-2001866. 

PCA was supported by the World Premier International Research Center Initiative (WPI), MEXT, Japan. CB acknowledges support from the RADIOFOREGROUNDS grant of the European Unions Horizon 2020 research and innovation programme (COMPET-05-2015, grant agreement number 687312) as well as by the LiteBIRD network of the Italian Space Agency (cosmosnet.it). B.B.~is grateful for funded support by the Simons Foundation, Sloan Foundation, and the Packard Foundation. EC acknowledges support from the STFC Ernest Rutherford Fellowship ST/M004856/2, STFC Consolidated Grant ST/S00033X/1 and the European Research Council (ERC) under the European Union’s Horizon 2020 research and innovation programme (Grant agreement No. 849169). JC was supported by the ERC Consolidator Grant {\it CMBSPEC} (No.~725456) and the Royal Society as a Royal Society University Research Fellow (URF/R/191023). KMH acknowledges support from NSF awards 1815887 and 2009870 and NASA award NNX17AF87G. ZX is supported by the Gordon and Betty Moore Foundation through grant GBMF5215 to the Massachusetts Institute of Technology.

Some of the results in this paper have been derived using the healpy and HEALPix package.

\software{healpy \citep{Gorski2005,Zonca2019}, Matplotlib \citep{Matplotlib}, NaMaster \citep{Alonso2019}, NumPy \citep{NumPy}, PySM3 \citep{Thorne2017,pysm3}, SciPy \citep{SciPy}}

\bibliography{gs.bib}

\begin{thebibliography}{}
\expandafter\ifx\csname natexlab\endcsname\relax\def\natexlab#1{#1}\fi
\providecommand{\url}[1]{\href{#1}{#1}}
\providecommand{\dodoi}[1]{doi:~\href{http://doi.org/#1}{\nolinkurl{#1}}}
\providecommand{\doeprint}[1]{\href{http://ascl.net/#1}{\nolinkurl{http://ascl.net/#1}}}
\providecommand{\doarXiv}[1]{\href{https://arxiv.org/abs/#1}{\nolinkurl{https://arxiv.org/abs/#1}}}

\bibitem[{{Abazajian} {et~al.}(2019){Abazajian}, {Addison}, {Adshead}, {Ahmed},
  {Allen}, {Alonso}, {Alvarez}, {Anderson}, {Arnold}, {Baccigalupi}, {Bailey},
  {Barkats}, {Barron}, {Barry}, {Bartlett}, {Basu Thakur}, {Battaglia},
  {Baxter}, {Bean}, {Bebek}, {Bender}, {Benson}, {Berger}, {Bhimani},
  {Bischoff}, {Bleem}, {Bocquet}, {Boddy}, {Bonato}, {Bond}, {Borrill},
  {Bouchet}, {Brown}, {Bryan}, {Burkhart}, {Buza}, {Byrum}, {Calabrese},
  {Calafut}, {Caldwell}, {Carlstrom}, {Carron}, {Cecil}, {Challinor}, {Chang},
  {Chinone}, {Cho}, {Cooray}, {Crawford}, {Crites}, {Cukierman}, {Cyr-Racine},
  {de Haan}, {de Zotti}, {Delabrouille}, {Demarteau}, {Devlin}, {Di Valentino},
  {Dobbs}, {Duff}, {Duivenvoorden}, {Dvorkin}, {Edwards}, {Eimer}, {Errard},
  {Essinger-Hileman}, {Fabbian}, {Feng}, {Ferraro}, {Filippini}, {Flauger},
  {Flaugher}, {Fraisse}, {Frolov}, {Galitzki}, {Galli}, {Ganga}, {Gerbino},
  {Gilchriese}, {Gluscevic}, {Green}, {Grin}, {Grohs}, {Gualtieri}, {Guarino},
  {Gudmundsson}, {Habib}, {Haller}, {Halpern}, {Halverson}, {Hanany},
  {Harrington}, {Hasegawa}, {Hasselfield}, {Hazumi}, {Heitmann}, {Henderson},
  {Henning}, {Hill}, {Hlozek}, {Holder}, {Holzapfel}, {Hubmayr},
  {Huffenberger}, {Huffer}, {Hui}, {Irwin}, {Johnson}, {Johnstone}, {Jones},
  {Karkare}, {Katayama}, {Kerby}, {Kernovsky}, {Keskitalo}, {Kisner}, {Knox},
  {Kosowsky}, {Kovac}, {Kovetz}, {Kuhlmann}, {Kuo}, {Kurita}, {Kusaka},
  {Lahteenmaki}, {Lawrence}, {Lee}, {Lewis}, {Li}, {Linder}, {Loverde},
  {Lowitz}, {Madhavacheril}, {Mantz}, {Matsuda}, {Mauskopf}, {McMahon},
  {McQuinn}, {Meerburg}, {Melin}, {Meyers}, {Millea}, {Mohr}, {Moncelsi},
  {Mroczkowski}, {Mukherjee}, {M{\"u}nchmeyer}, {Nagai}, {Nagy}, {Namikawa},
  {Nati}, {Natoli}, {Negrello}, {Newburgh}, {Niemack}, {Nishino}, {Nordby},
  {Novosad}, {O'Connor}, {Obied}, {Padin}, {Pandey}, {Partridge}, {Pierpaoli},
  {Pogosian}, {Pryke}, {Puglisi}, {Racine}, {Raghunathan}, {Rahlin},
  {Rajagopalan}, {Raveri}, {Reichanadter}, {Reichardt}, {Remazeilles}, {Rocha},
  {Roe}, {Roy}, {Ruhl}, {Salatino}, {Saliwanchik}, {Schaan}, {Schillaci},
  {Schmittfull}, {Scott}, {Sehgal}, {Shandera}, {Sheehy}, {Sherwin},
  {Shirokoff}, {Simon}, {Slosar}, {Somerville}, {Spergel}, {Staggs}, {Stark},
  {Stompor}, {Story}, {Stoughton}, {Suzuki}, {Tajima}, {Teply}, {Thompson},
  {Timbie}, {Tomasi}, {Treu}, {Tristram}, {Tucker}, {Umilt{\`a}}, {van
  Engelen}, {Vieira}, {Vieregg}, {Vogelsberger}, {Wang}, {Watson}, {White},
  {Whitehorn}, {Wollack}, {Kimmy Wu}, {Xu}, {Yasini}, {Yeck}, {Yoon}, {Young},
  \& {Zonca}}]{CMBS42019}
{Abazajian}, K., {Addison}, G., {Adshead}, P., {et~al.} 2019, arXiv e-prints,
  arXiv:1907.04473.
\newblock \doarXiv{1907.04473}

\bibitem[{{Abitbol} {et~al.}(2018){Abitbol}, {Johnson}, {Jones}, {Dickinson},
  \& {Harper}}]{Abitbol2018}
{Abitbol}, M.~H., {Johnson}, B.~R., {Jones}, G., {Dickinson}, C., \& {Harper},
  S. 2018, \apj, 864, 97, \dodoi{10.3847/1538-4357/aad548}

\bibitem[{{Abitbol} {et~al.}(2021){Abitbol}, {Alonso}, {Simon}, {Lashner},
  {Crowley}, {Ali}, {Azzoni}, {Baccigalupi}, {Barron}, {Brown}, {Calabrese},
  {Carron}, {Chinone}, {Chluba}, {Coppi}, {Crowley}, {Devlin}, {Dunkley},
  {Errard}, {Fanfani}, {Galitzki}, {Gerbino}, {Hill}, {Johnson}, {Jost},
  {Keating}, {Krachmalnicoff}, {Kusaka}, {Lee}, {Louis}, {Madhavacheril},
  {McCarrick}, {McMahon}, {Meerburg}, {Nati}, {Nishino}, {Page}, {Poletti},
  {Puglisi}, {Randall}, {Rotti}, {Spisak}, {Suzuki}, {Teply}, {Verg{\`e}s},
  {Wollack}, {Xu}, \& {Zannoni}}]{2021JCAP...05..032A}
{Abitbol}, M.~H., {Alonso}, D., {Simon}, S.~M., {et~al.} 2021, \jcap, 2021,
  032, \dodoi{10.1088/1475-7516/2021/05/032}

\bibitem[{{Adachi} {et~al.}(2020){Adachi}, {Aguilar Fa{\'u}ndez}, {Arnold},
  {Baccigalupi}, {Barron}, {Beck}, {Bianchini}, {Chapman}, {Cheung}, {Chinone},
  {Crowley}, {Dobbs}, {El Bouhargani}, {Elleflot}, {Errard}, {Fabbian}, {Feng},
  {Fujino}, {Galitzki}, {Goeckner-Wald}, {Groh}, {Hall}, {Hasegawa}, {Hazumi},
  {Hirose}, {Jaffe}, {Jeong}, {Kaneko}, {Katayama}, {Keating}, {Kikuchi},
  {Kisner}, {Kusaka}, {Lee}, {Leon}, {Linder}, {Lowry}, {Matsuda}, {Matsumura},
  {Minami}, {Navaroli}, {Nishino}, {Pham}, {Poletti}, {Reichardt}, {Segawa},
  {Siritanasak}, {Tajima}, {Takakura}, {Takatori}, {Tanabe}, {Teply}, {Tsai},
  {Verg{\`e}s}, {Westbrook}, {Zhou}, \& {Polarbear Collaboration}}]{PB2020}
{Adachi}, S., {Aguilar Fa{\'u}ndez}, M.~A.~O., {Arnold}, K., {et~al.} 2020,
  \apj, 904, 65, \dodoi{10.3847/1538-4357/abbacd}

\bibitem[{{Aiola} {et~al.}(2020){Aiola}, {Calabrese}, {Maurin}, {Naess},
  {Schmitt}, {Abitbol}, {Addison}, {Ade}, {Alonso}, {Amiri}, {Amodeo},
  {Angile}, {Austermann}, {Baildon}, {Battaglia}, {Beall}, {Bean}, {Becker},
  {Bond}, {Bruno}, {Calafut}, {Campusano}, {Carrero}, {Chesmore}, {Cho},
  {Choi}, {Clark}, {Cothard}, {Crichton}, {Crowley}, {Darwish}, {Datta},
  {Denison}, {Devlin}, {Duell}, {Duff}, {Duivenvoorden}, {Dunkley},
  {D{\"u}nner}, {Essinger-Hileman}, {Fankhanel}, {Ferraro}, {Fox}, {Fuzia},
  {Gallardo}, {Gluscevic}, {Golec}, {Grace}, {Gralla}, {Guan}, {Hall},
  {Halpern}, {Han}, {Hargrave}, {Hasselfield}, {Helton}, {Henderson},
  {Hensley}, {Hill}, {Hilton}, {Hilton}, {Hincks}, {Hlo{\v{z}}ek}, {Ho},
  {Hubmayr}, {Huffenberger}, {Hughes}, {Infante}, {Irwin}, {Jackson}, {Klein},
  {Knowles}, {Koopman}, {Kosowsky}, {Lakey}, {Li}, {Li}, {Li}, {Lokken},
  {Louis}, {Lungu}, {MacInnis}, {Madhavacheril}, {Maldonado}, {Mallaby-Kay},
  {Marsden}, {McMahon}, {Menanteau}, {Moodley}, {Morton}, {Namikawa}, {Nati},
  {Newburgh}, {Nibarger}, {Nicola}, {Niemack}, {Nolta}, {Orlowski-Sherer},
  {Page}, {Pappas}, {Partridge}, {Phakathi}, {Pisano}, {Prince}, {Puddu}, {Qu},
  {Rivera}, {Robertson}, {Rojas}, {Salatino}, {Schaan}, {Schillaci}, {Sehgal},
  {Sherwin}, {Sierra}, {Sievers}, {Sifon}, {Sikhosana}, {Simon}, {Spergel},
  {Staggs}, {Stevens}, {Storer}, {Sunder}, {Switzer}, {Thorne}, {Thornton},
  {Trac}, {Treu}, {Tucker}, {Vale}, {Van Engelen}, {Van Lanen}, {Vavagiakis},
  {Wagoner}, {Wang}, {Ward}, {Wollack}, {Xu}, {Zago}, \& {Zhu}}]{Aiola_2020}
{Aiola}, S., {Calabrese}, E., {Maurin}, L., {et~al.} 2020, \jcap, 2020, 047,
  \dodoi{10.1088/1475-7516/2020/12/047}

\bibitem[{{Ali} {et~al.}(2020){Ali}, {Adachi}, {Arnold}, {Ashton}, {Bazarko},
  {Chinone}, {Coppi}, {Corbett}, {Crowley}, {Crowley}, {Devlin}, {Dicker},
  {Duff}, {Ellis}, {Galitzki}, {Goeckner-Wald}, {Harrington}, {Healy}, {Hill},
  {Ho}, {Hubmayr}, {Keating}, {Kiuchi}, {Kusaka}, {Lee}, {Ludlam}, {Mangu},
  {Matsuda}, {McCarrick}, {Nati}, {Niemack}, {Nishino}, {Orlowski-Scherer},
  {Sathyanarayana Rao}, {Raum}, {Sakurai}, {Salatino}, {Sasse}, {Seibert},
  {Sierra}, {Silva-Feaver}, {Spisak}, {Simon}, {Staggs}, {Tajima}, {Teply},
  {Tsan}, {Wollack}, {Westbrook}, {Xu}, {Zannoni}, \& {Zhu}}]{SO_SAT}
{Ali}, A.~M., {Adachi}, S., {Arnold}, K., {et~al.} 2020, Journal of Low
  Temperature Physics, 200, 461, \dodoi{10.1007/s10909-020-02430-5}

\bibitem[{{Ali-Ha{\"\i}moud} {et~al.}(2009){Ali-Ha{\"\i}moud}, {Hirata}, \&
  {Dickinson}}]{AliHaimoud_2009}
{Ali-Ha{\"\i}moud}, Y., {Hirata}, C.~M., \& {Dickinson}, C. 2009, \mnras, 395,
  1055, \dodoi{10.1111/j.1365-2966.2009.14599.x}

\bibitem[{{Alonso} {et~al.}(2019){Alonso}, {Sanchez}, {Slosar}, \& {LSST Dark
  Energy Science Collaboration}}]{Alonso2019}
{Alonso}, D., {Sanchez}, J., {Slosar}, A., \& {LSST Dark Energy Science
  Collaboration}. 2019, \mnras, 484, 4127, \dodoi{10.1093/mnras/stz093}

\bibitem[{{Alves} {et~al.}(2018){Alves}, {Boulanger}, {Ferri{\`e}re}, \&
  {Montier}}]{Alves2018}
{Alves}, M.~I.~R., {Boulanger}, F., {Ferri{\`e}re}, K., \& {Montier}, L. 2018,
  \aap, 611, L5, \dodoi{10.1051/0004-6361/201832637}

\bibitem[{{Andr{\'e}} {et~al.}(2014){Andr{\'e}}, {Di Francesco},
  {Ward-Thompson}, {Inutsuka}, {Pudritz}, \& {Pineda}}]{Andre:2014}
{Andr{\'e}}, P., {Di Francesco}, J., {Ward-Thompson}, D., {et~al.} 2014, in
  Protostars and Planets VI, ed. H.~{Beuther}, R.~S. {Klessen}, C.~P.
  {Dullemond}, \& T.~{Henning}, 27,
  \dodoi{10.2458/azu\_uapress\_9780816531240-ch002}

\bibitem[{{Ashton} {et~al.}(2018){Ashton}, {Ade}, {Angil{\`e}}, {Benton},
  {Devlin}, {Dober}, {Fissel}, {Fukui}, {Galitzki}, {Gandilo}, {Klein},
  {Korotkov}, {Li}, {Martin}, {Matthews}, {Moncelsi}, {Nakamura},
  {Netterfield}, {Novak}, {Pascale}, {Poidevin}, {Santos}, {Savini}, {Scott},
  {Shariff}, {Soler}, {Thomas}, {Tucker}, {Tucker}, \&
  {Ward-Thompson}}]{Ashton_2018}
{Ashton}, P.~C., {Ade}, P. A.~R., {Angil{\`e}}, F.~E., {et~al.} 2018, \apj,
  857, 10, \dodoi{10.3847/1538-4357/aab3ca}

\bibitem[{{Babich} \& {Loeb}(2009)}]{2009NewA...14..166B}
{Babich}, D., \& {Loeb}, A. 2009, \na, 14, 166,
  \dodoi{10.1016/j.newast.2008.07.005}

\bibitem[{{Balkenhol} {et~al.}(2021){Balkenhol}, {Dutcher}, {Ade}, {Ahmed},
  {Anderes}, {Anderson}, {Archipley}, {Avva}, {Aylor}, {Barry}, {Basu Thakur},
  {Benabed}, {Bender}, {Benson}, {Bianchini}, {Bleem}, {Bouchet}, {Bryant},
  {Byrum}, {Carlstrom}, {Carter}, {Cecil}, {Chang}, {Chaubal}, {Chen}, {Cho},
  {Chou}, {Cliche}, {Crawford}, {Cukierman}, {Daley}, {de Haan}, {Denison},
  {Dibert}, {Ding}, {Dobbs}, {Everett}, {Feng}, {Ferguson}, {Foster}, {Fu},
  {Galli}, {Gambrel}, {Gardner}, {Goeckner-Wald}, {Gualtieri}, {Guns}, {Gupta},
  {Guyser}, {Halverson}, {Harke-Hosemann}, {Harrington}, {Henning}, {Hilton},
  {Hivon}, {Holder}, {Holzapfel}, {Hood}, {Howe}, {Huang}, {Irwin}, {Jeong},
  {Jonas}, {Jones}, {Khaire}, {Knox}, {Kofman}, {Korman}, {Kubik}, {Kuhlmann},
  {Kuo}, {Lee}, {Leitch}, {Lowitz}, {Lu}, {Meyer}, {Michalik}, {Millea},
  {Montgomery}, {Nadolski}, {Natoli}, {Nguyen}, {Noble}, {Novosad}, {Omori},
  {Padin}, {Pan}, {Paschos}, {Pearson}, {Posada}, {Prabhu}, {Quan}, {Rahlin},
  {Reichardt}, {Riebel}, {Riedel}, {Rouble}, {Ruhl}, {Sayre}, {Schiappucci},
  {Shirokoff}, {Smecher}, {Sobrin}, {Stark}, {Stephen}, {Story}, {Suzuki},
  {Thompson}, {Thorne}, {Tucker}, {Umilta}, {Vale}, {Vanderlinde}, {Vieira},
  {Wang}, {Whitehorn}, {Wu}, {Yefremenko}, {Yoon}, {Young}, \& {SPT-3G
  Collaboration}}]{SPT3g-ext2021}
{Balkenhol}, L., {Dutcher}, D., {Ade}, P.~A.~R., {et~al.} 2021, \prd, 104,
  083509, \dodoi{10.1103/PhysRevD.104.083509}

\bibitem[{{Barreto-Mota} {et~al.}(2021){Barreto-Mota}, {de Gouveia Dal Pino},
  {Burkhart}, {Melioli}, {Santos-Lima}, \& {Kadowaki}}]{Barreto-Mota:2021}
{Barreto-Mota}, L., {de Gouveia Dal Pino}, E.~M., {Burkhart}, B., {et~al.}
  2021, \mnras, 503, 5425, \dodoi{10.1093/mnras/stab798}

\bibitem[{Batygin {et~al.}(2019)Batygin, Adams, Brown, \&
  Becker}]{Batygin_2019}
Batygin, K., Adams, F.~C., Brown, M.~E., \& Becker, J.~C. 2019, Physics
  Reports, 805, 1–53, \dodoi{10.1016/j.physrep.2019.01.009}

\bibitem[{Batygin \& Brown(2016)}]{Batygin_2016}
Batygin, K., \& Brown, M.~E. 2016, The Astronomical Journal, 151, 22,
  \dodoi{10.3847/0004-6256/151/2/22}

\bibitem[{{Baxter} {et~al.}(2018){Baxter}, {Blake}, \& {Jain}}]{Baxter:2018}
{Baxter}, E.~J., {Blake}, C.~H., \& {Jain}, B. 2018, \aj, 156, 243,
  \dodoi{10.3847/1538-3881/aae64e}

\bibitem[{{Bennett} {et~al.}(2013){Bennett}, {Larson}, {Weiland}, {Jarosik},
  {Hinshaw}, {Odegard}, {Smith}, {Hill}, {Gold}, {Halpern}, {Komatsu}, {Nolta},
  {Page}, {Spergel}, {Wollack}, {Dunkley}, {Kogut}, {Limon}, {Meyer}, {Tucker},
  \& {Wright}}]{Bennett_2013}
{Bennett}, C.~L., {Larson}, D., {Weiland}, J.~L., {et~al.} 2013, \apjs, 208,
  20, \dodoi{10.1088/0067-0049/208/2/20}

\bibitem[{{Bernard} {et~al.}(2016){Bernard}, {Ade}, {Andr{\'e}}, {Aumont},
  {Bautista}, {Bray}, {Bernardis}, {Boulade}, {Bousquet}, {Bouzit}, {Buttice},
  {Caillat}, {Charra}, {Chaigneau}, {Crane}, {Crussaire}, {Douchin},
  {Doumayrou}, {Dubois}, {Engel}, {Etcheto}, {G{\'e}lot}, {Griffin}, {Foenard},
  {Grabarnik}, {Hargrave}, {Hughes}, {Laureijs}, {Lepennec}, {Leriche},
  {Longval}, {Maestre}, {Maffei}, {Martignac}, {Marty}, {Marty}, {Masi},
  {Mirc}, {Misawa}, {Montel}, {Montier}, {Mot}, {Narbonne}, {Nicot}, {Pajot},
  {Parot}, {P{\'e}rot}, {Pimentao}, {Pisano}, {Ponthieu}, {Ristorcelli},
  {Rodriguez}, {Roudil}, {Salatino}, {Savini}, {Simonella}, {Saccoccio},
  {Tapie}, {Tauber}, {Torre}, \& {Tucker}}]{Bernard2016}
{Bernard}, J.~P., {Ade}, P., {Andr{\'e}}, Y., {et~al.} 2016, Experimental
  Astronomy, 42, 199, \dodoi{10.1007/s10686-016-9506-1}

\bibitem[{{BICEP2 Collaboration} {et~al.}(2015){BICEP2 Collaboration}, {Ade},
  {Aikin}, {Barkats}, {Benton}, {Bischoff}, {Bock}, {Brevik}, {Buder},
  {Bullock}, {Dowell}, {Duband}, {Filippini}, {Fliescher}, {Golwala},
  {Halpern}, {Hasselfield}, {Hildebrandt}, {Hilton}, {Irwin}, {Karkare},
  {Kaufman}, {Keating}, {Kernasovskiy}, {Kovac}, {Kuo}, {Leitch}, {Lueker},
  {Netterfield}, {Nguyen}, {O'Brient}, {Ogburn}, {Orlando}, {Pryke}, {Richter},
  {Schwarz}, {Sheehy}, {Staniszewski}, {Sudiwala}, {Teply}, {Tolan}, {Turner},
  {Vieregg}, {Wong}, \& {Yoon}}]{2015ApJ...814..110B}
{BICEP2 Collaboration}, {Ade}, P.~A.~R., {Aikin}, R.~W., {et~al.} 2015, \apj,
  814, 110, \dodoi{10.1088/0004-637X/814/2/110}

\bibitem[{{BICEP2 Collaboration} {et~al.}(2018){BICEP2 Collaboration}, {Keck
  Array Collaboration}, {Ade}, {Ahmed}, {Aikin}, {Alexander}, {Barkats},
  {Benton}, {Bischoff}, {Bock}, {Bowens-Rubin}, {Brevik}, {Buder}, {Bullock},
  {Buza}, {Connors}, {Cornelison}, {Crill}, {Crumrine}, {Dierickx}, {Duband},
  {Dvorkin}, {Filippini}, {Fliescher}, {Grayson}, {Hall}, {Halpern},
  {Harrison}, {Hildebrandt}, {Hilton}, {Hui}, {Irwin}, {Kang}, {Karkare},
  {Karpel}, {Kaufman}, {Keating}, {Kefeli}, {Kernasovskiy}, {Kovac}, {Kuo},
  {Larsen}, {Lau}, {Leitch}, {Lueker}, {Megerian}, {Moncelsi}, {Namikawa},
  {Netterfield}, {Nguyen}, {O'Brient}, {Ogburn}, {Palladino}, {Pryke},
  {Racine}, {Richter}, {Schillaci}, {Schwarz}, {Sheehy}, {Soliman}, {St.
  Germaine}, {Staniszewski}, {Steinbach}, {Sudiwala}, {Teply}, {Thompson},
  {Tolan}, {Tucker}, {Turner}, {Umilt{\`a}}, {Vieregg}, {Wandui}, {Weber},
  {Wiebe}, {Willmert}, {Wong}, {Wu}, {Yang}, {Yoon}, \& {Zhang}}]{BICEP2018}
{BICEP2 Collaboration}, {Keck Array Collaboration}, {Ade}, P.~A.~R., {et~al.}
  2018, \prl, 121, 221301, \dodoi{10.1103/PhysRevLett.121.221301}

\bibitem[{{BICEP/Keck Collaboration} {et~al.}(2021){BICEP/Keck Collaboration},
  {SPTpol Collaboration}, {Ade}, {Ahmed}, {Amiri}, {Anderson}, {Austermann},
  {Avva}, {Barkats}, {Thakur}, {Beall}, {Bender}, {Benson}, {Bianchini},
  {Bischoff}, {Bleem}, {Bock}, {Boenish}, {Bullock}, {Buza}, {Carlstrom},
  {Chang}, {Cheshire}, {Chiang}, {Chou}, {Citron}, {Connors}, {Moran},
  {Cornelison}, {Crawford}, {Crites}, {Crumrine}, {Cukierman}, {de Haan},
  {Dierickx}, {Dobbs}, {Duband}, {Everett}, {Fatigoni}, {Filippini},
  {Fliescher}, {Gallicchio}, {George}, {Germaine}, {Goeckner-Wald},
  {Goldfinger}, {Grayson}, {Gupta}, {Hall}, {Halpern}, {Halverson}, {Harrison},
  {Henderson}, {Henning}, {Hildebrandt}, {Hilton}, {Holder}, {Holzapfel},
  {Hrubes}, {Huang}, {Hubmayr}, {Hui}, {Irwin}, {Kang}, {Karkare}, {Karpel},
  {Kefeli}, {Kernasovskiy}, {Knox}, {Kovac}, {Kuo}, {Lau}, {Lee}, {Leitch},
  {Li}, {Lowitz}, {Manzotti}, {McMahon}, {Megerian}, {Meyer}, {Millea},
  {Mocanu}, {Moncelsi}, {Montgomery}, {Nadolski}, {Namikawa}, {Natoli},
  {Netterfield}, {Nguyen}, {Nibarger}, {Noble}, {Novosad}, {O'Brient},
  {Ogburn}, {Omori}, {Padin}, {Palladino}, {Patil}, {Prouve}, {Pryke},
  {Racine}, {Reichardt}, {Reintsema}, {Richter}, {Ruhl}, {Saliwanchik},
  {Schaffer}, {Schillaci}, {Schmitt}, {Schwarz}, {Sheehy}, {Sievers},
  {Smecher}, {Soliman}, {Stark}, {Steinbach}, {Sudiwala}, {Teply}, {Thompson},
  {Tolan}, {Tucker}, {Turner}, {Umilt{\`a}}, {Veach}, {Vieira}, {Vieregg},
  {Wandui}, {Wang}, {Weber}, {Whitehorn}, {Wiebe}, {Willmert}, {Wong}, {Wu},
  {Yang}, {Yefremenko}, {Yoon}, {Young}, {Yu}, {Zeng}, {Zhang}, {Bicep/Keck},
  \& {Sptpol Collaborations}}]{BICEPSPT2021}
{BICEP/Keck Collaboration}, {SPTpol Collaboration}, {Ade}, P.~A.~R., {et~al.}
  2021, \prd, 103, 022004, \dodoi{10.1103/PhysRevD.103.022004}

\bibitem[{{Bot} {et~al.}(2010){Bot}, {Ysard}, {Paradis}, {Bernard}, {Lagache},
  {Israel}, \& {Wall}}]{Bot2010}
{Bot}, C., {Ysard}, N., {Paradis}, D., {et~al.} 2010, \aap, 523, A20,
  \dodoi{10.1051/0004-6361/201014986}

\bibitem[{{Burkhart} {et~al.}(2009){Burkhart}, {Falceta-Gon{\c c}alves},
  {Kowal}, \& {Lazarian}}]{Burkhart09a}
{Burkhart}, B., {Falceta-Gon{\c c}alves}, D., {Kowal}, G., \& {Lazarian}, A.
  2009, \apj, 693, 250, \dodoi{10.1088/0004-637X/693/1/250}

\bibitem[{{Burkhart} {et~al.}(2015){Burkhart}, {Lazarian}, {Balsara}, {Meyer},
  \& {Cho}}]{BurL15}
{Burkhart}, B., {Lazarian}, A., {Balsara}, D., {Meyer}, C., \& {Cho}, J. 2015,
  \apj, 805, 118, \dodoi{10.1088/0004-637X/805/2/118}

\bibitem[{{Burkhart} {et~al.}(2012){Burkhart}, {Lazarian}, \&
  {Gaensler}}]{2012ApJ...749..145B}
{Burkhart}, B., {Lazarian}, A., \& {Gaensler}, B.~M. 2012, \apj, 749, 145,
  \dodoi{10.1088/0004-637X/749/2/145}

\bibitem[{{Caldwell} {et~al.}(2017){Caldwell}, {Hirata}, \&
  {Kamionkowski}}]{Caldwell:2017}
{Caldwell}, R.~R., {Hirata}, C., \& {Kamionkowski}, M. 2017, \apj, 839, 91,
  \dodoi{10.3847/1538-4357/aa679c}

\bibitem[{{Carmo} {et~al.}(2020){Carmo}, {Gonz{\'a}lez-Casanova},
  {Falceta-Gon{\c{c}}alves}, {Lazarian}, {Jablonski}, {Zhang}, {Ferreira},
  {Castro}, \& {Yang}}]{2020ApJ...905..130C}
{Carmo}, L., {Gonz{\'a}lez-Casanova}, D.~F., {Falceta-Gon{\c{c}}alves}, D.,
  {et~al.} 2020, \apj, 905, 130, \dodoi{10.3847/1538-4357/abc331}

\bibitem[{{Carretti} {et~al.}(2019){Carretti}, {Haverkorn}, {Staveley-Smith},
  {Bernardi}, {Gaensler}, {Kesteven}, {Poppi}, {Brown}, {Crocker}, {Purcell},
  {Schnitzeler}, \& {Sun}}]{Carretti19}
{Carretti}, E., {Haverkorn}, M., {Staveley-Smith}, L., {et~al.} 2019, \mnras,
  489, 2330, \dodoi{10.1093/mnras/stz806}

\bibitem[{{CCAT-Prime collaboration} {et~al.}(2021){CCAT-Prime collaboration},
  {Aravena}, {Austermann}, {Basu}, {Battaglia}, {Beringue}, {Bertoldi},
  {Bigiel}, {Bond}, {Breysse}, {Broughton}, {Bustos}, {Chapman}, {Charmetant},
  {Choi}, {Chung}, {Clark}, {Cothard}, {Crites}, {Dev}, {Douglas}, {Duell},
  {Ebina}, {Erler}, {Fich}, {Fissel}, {Foreman}, {Gao}, {Garc{\'\i}a},
  {Giovanelli}, {Haynes}, {Hensley}, {Herter}, {Higgins}, {Huber}, {Hubmayr},
  {Johnstone}, {Karoumpis}, {Keating}, {Komatsu}, {Li}, {Magnelli}, {Matthews},
  {Meerburg}, {Meyers}, {Muralidhara}, {Murray}, {Niemack}, {Nikola}, {Okada},
  {Riechers}, {Rosolowsky}, {Roy}, {Sadavoy}, {Schaaf}, {Schilke}, {Scott},
  {Simon}, {Sinclair}, {Sivakoff}, {Stacey}, {Stutz}, {Stutzki}, {Tahani},
  {Thanjavur}, {Timmermann}, {Ullom}, {van Engelen}, {Vavagiakis}, {Vissers},
  {Wheeler}, {White}, {Zhu}, \& {Zou}}]{CCATp_2021}
{CCAT-Prime collaboration}, {Aravena}, M., {Austermann}, J.~E., {et~al.} 2021,
  arXiv e-prints, arXiv:2107.10364.
\newblock \doarXiv{2107.10364}

\bibitem[{{Cepeda-Arroita} {et~al.}(2021){Cepeda-Arroita}, {Harper},
  {Dickinson}, {Rubi{\~n}o-Mart{\'\i}n}, {G{\'e}nova-Santos}, {Taylor},
  {Pearson}, {Ashdown}, {Barr}, {Barreiro}, {Casaponsa}, {Casas}, {Chiang},
  {Fernandez-Cobos}, {Grumitt}, {Guidi}, {Heilgendorff}, {Herranz}, {Jew},
  {Jonas}, {Jones}, {Lasenby}, {Leech}, {Leahy}, {Mart{\'\i}nez-Gonz{\'a}lez},
  {Peel}, {Piccirillo}, {Poidevin}, {Readhead}, {Rebolo}, {Ruiz-Granados},
  {Sievers}, {Vansyngel}, {Vielva}, \& {Watson}}]{2021MNRAS.503.2927C}
{Cepeda-Arroita}, R., {Harper}, S.~E., {Dickinson}, C., {et~al.} 2021, \mnras,
  503, 2927, \dodoi{10.1093/mnras/stab583}

\bibitem[{{Choi} \& {Page}(2015)}]{Choi2015}
{Choi}, S.~K., \& {Page}, L.~A. 2015, \jcap, 2015, 020,
  \dodoi{10.1088/1475-7516/2015/12/020}

\bibitem[{{Choi} {et~al.}(2020){Choi}, {Austermann}, {Basu}, {Battaglia},
  {Bertoldi}, {Chung}, {Cothard}, {Duff}, {Duell}, {Gallardo}, {Gao}, {Herter},
  {Hubmayr}, {Niemack}, {Nikola}, {Riechers}, {Rossi}, {Stacey}, {Stevens},
  {Vavagiakis}, {Vissers}, \& {Walker}}]{Choi2020}
{Choi}, S.~K., {Austermann}, J., {Basu}, K., {et~al.} 2020, Journal of Low
  Temperature Physics, 199, 1089, \dodoi{10.1007/s10909-020-02428-z}

\bibitem[{{Clark}(2018)}]{Clark:2018}
{Clark}, S.~E. 2018, \apjl, 857, L10, \dodoi{10.3847/2041-8213/aabb54}

\bibitem[{{Clark} \& {Hensley}(2019)}]{ClarkHensley:2019}
{Clark}, S.~E., \& {Hensley}, B.~S. 2019, \apj, 887, 136,
  \dodoi{10.3847/1538-4357/ab5803}

\bibitem[{{Clark} {et~al.}(2015){Clark}, {Hill}, {Peek}, {Putman}, \&
  {Babler}}]{Clark:2015}
{Clark}, S.~E., {Hill}, J.~C., {Peek}, J.~E.~G., {Putman}, M.~E., \& {Babler},
  B.~L. 2015, \prl, 115, 241302, \dodoi{10.1103/PhysRevLett.115.241302}

\bibitem[{{Clark} {et~al.}(2021){Clark}, {Kim}, {Hill}, \&
  {Hensley}}]{Clark:2021}
{Clark}, S.~E., {Kim}, C.-G., {Hill}, J.~C., \& {Hensley}, B.~S. 2021, \apj,
  919, 53, \dodoi{10.3847/1538-4357/ac0e35}

\bibitem[{{Clark} {et~al.}(2014){Clark}, {Peek}, \& {Putman}}]{Clark:2014}
{Clark}, S.~E., {Peek}, J.~E.~G., \& {Putman}, M.~E. 2014, \apj, 789, 82,
  \dodoi{10.1088/0004-637X/789/1/82}

\bibitem[{{Clayton} {et~al.}(1992){Clayton}, {Anderson}, {Magalhaes}, {Code},
  {Nordsieck}, {Meade}, {Wolff}, {Babler}, {Bjorkman}, {Schulte-Ladbeck},
  {Taylor}, \& {Whitney}}]{Clayton_1992}
{Clayton}, G.~C., {Anderson}, C.~M., {Magalhaes}, A.~M., {et~al.} 1992, \apjl,
  385, L53, \dodoi{10.1086/186276}

\bibitem[{{Cortes} {et~al.}(2008){Cortes}, {Crutcher}, {Shepherd}, \&
  {Bronfman}}]{2008ApJ...676..464C}
{Cortes}, P.~C., {Crutcher}, R.~M., {Shepherd}, D.~S., \& {Bronfman}, L. 2008,
  \apj, 676, 464, \dodoi{10.1086/524355}

\bibitem[{{Cox} {et~al.}(2016){Cox}, {Arzoumanian}, {Andr{\'e}}, {Rygl},
  {Prusti}, {Men'shchikov}, {Royer}, {K{\'o}sp{\'a}l}, {Palmeirim}, {Ribas},
  {K{\"o}nyves}, {Bernard}, {Schneider}, {Bontemps}, {Merin}, {Vavrek}, {Alves
  de Oliveira}, {Didelon}, {Pilbratt}, \& {Waelkens}}]{Cox:2016}
{Cox}, N.~L.~J., {Arzoumanian}, D., {Andr{\'e}}, P., {et~al.} 2016, \aap, 590,
  A110, \dodoi{10.1051/0004-6361/201527068}

\bibitem[{{Crawford} {et~al.}(2016){Crawford}, {Chown}, {Holder}, {Aird},
  {Benson}, {Bleem}, {Carlstrom}, {Chang}, {Cho}, {Crites}, {de Haan}, {Dobbs},
  {George}, {Halverson}, {Harrington}, {Holzapfel}, {Hou}, {Hrubes}, {Keisler},
  {Knox}, {Lee}, {Leitch}, {Luong-Van}, {Marrone}, {McMahon}, {Meyer},
  {Mocanu}, {Mohr}, {Natoli}, {Padin}, {Pryke}, {Reichardt}, {Ruhl}, {Sayre},
  {Schaffer}, {Shirokoff}, {Staniszewski}, {Stark}, {Story}, {Vanderlinde},
  {Vieira}, \& {Williamson}}]{Crawford:2016}
{Crawford}, T.~M., {Chown}, R., {Holder}, G.~P., {et~al.} 2016, \apjs, 227, 23,
  \dodoi{10.3847/1538-4365/227/2/23}

\bibitem[{Crill {et~al.}(2008)Crill, Ade, Battistelli, Benton, Bihary, Bock,
  Bond, Brevik, Bryan, Contaldi, \& et~al.}]{Crill_2008}
Crill, B.~P., Ade, P. A.~R., Battistelli, E.~S., {et~al.} 2008, Space
  Telescopes and Instrumentation 2008: Optical, Infrared, and Millimeter,
  \dodoi{10.1117/12.787446}

\bibitem[{{Crowley} {et~al.}(2018){Crowley}, {Simon}, {Silva-Feaver},
  {Goeckner-Wald}, {Ali}, {Austermann}, {Brown}, {Chinone}, {Cukierman},
  {Dober}, {Duff}, {Dunkley}, {Errard}, {Fabbian}, {Gallardo}, {Ho}, {Hubmayr},
  {Keating}, {Kusaka}, {McCallum}, {McMahon}, {Nati}, {Niemack}, {Puglisi},
  {Sathyanarayana Rao}, {Reichardt}, {Salatino}, {Siritanasak}, {Staggs},
  {Suzuki}, {Teply}, {Thomas}, {Ullom}, {Verg{\`e}s}, {Vissers}, {Westbrook},
  {Wollack}, {Xu}, \& {Zhu}}]{2018SPIE10708E..3ZC}
{Crowley}, K.~T., {Simon}, S.~M., {Silva-Feaver}, M., {et~al.} 2018, in Society
  of Photo-Optical Instrumentation Engineers (SPIE) Conference Series, Vol.
  10708, Millimeter, Submillimeter, and Far-Infrared Detectors and
  Instrumentation for Astronomy IX, ed. J.~{Zmuidzinas} \& J.-R. {Gao},
  107083Z, \dodoi{10.1117/12.2313414}

\bibitem[{Crutcher(2012)}]{Crutcher2012}
Crutcher, R.~M. 2012, Annu. Rev. Astron. Astrophys, 50, 29,
  \dodoi{10.1146/annurev-astro-081811-125514}

\bibitem[{{Cunningham} {et~al.}(2018){Cunningham}, {Krumholz}, {McKee}, \&
  {Klein}}]{Cunningham2018}
{Cunningham}, A.~J., {Krumholz}, M.~R., {McKee}, C.~F., \& {Klein}, R.~I. 2018,
  \mnras, 476, 771, \dodoi{10.1093/mnras/sty154}

\bibitem[{{Dame} {et~al.}(2001){Dame}, {Hartmann}, \& {Thaddeus}}]{dame2001}
{Dame}, T.~M., {Hartmann}, D., \& {Thaddeus}, P. 2001, \apj, 547, 792,
  \dodoi{10.1086/318388}

\bibitem[{{de Oliveira-Costa} {et~al.}(1997){de Oliveira-Costa}, {Kogut},
  {Devlin}, {Netterfield}, {Page}, \& {Wollack}}]{deOliveiraCosta_1997}
{de Oliveira-Costa}, A., {Kogut}, A., {Devlin}, M.~J., {et~al.} 1997, \apjl,
  482, L17, \dodoi{10.1086/310684}

\bibitem[{{Dickinson} {et~al.}(2009){Dickinson}, {Eriksen}, {Banday}, {Jewell},
  {G{\'o}rski}, {Huey}, {Lawrence}, {O'Dwyer}, \& {Wandelt}}]{Dickinson2009}
{Dickinson}, C., {Eriksen}, H.~K., {Banday}, A.~J., {et~al.} 2009, \apj, 705,
  1607, \dodoi{10.1088/0004-637X/705/2/1607}

\bibitem[{{Dickinson} {et~al.}(2018){Dickinson}, {Ali-Ha{\"\i}moud}, {Barr},
  {Battistelli}, {Bell}, {Bernstein}, {Casassus}, {Cleary}, {Draine},
  {G{\'e}nova-Santos}, {Harper}, {Hensley}, {Hill-Valler}, {Hoang}, {Israel},
  {Jew}, {Lazarian}, {Leahy}, {Leech}, {L{\'o}pez-Caraballo}, {McDonald},
  {Murphy}, {Onaka}, {Paladini}, {Peel}, {Perrott}, {Poidevin}, {Readhead},
  {Rubi{\~n}o-Mart{\'\i}n}, {Taylor}, {Tibbs}, {Todorovi{\'c}}, \&
  {Vidal}}]{Dickinson2018}
{Dickinson}, C., {Ali-Ha{\"\i}moud}, Y., {Barr}, A., {et~al.} 2018, \nar, 80,
  1, \dodoi{10.1016/j.newar.2018.02.001}

\bibitem[{{Draine} \& {Fraisse}(2009)}]{Draine_2009}
{Draine}, B.~T., \& {Fraisse}, A.~A. 2009, \apj, 696, 1,
  \dodoi{10.1088/0004-637X/696/1/1}

\bibitem[{{Draine} \& {Hensley}(2012)}]{Draine_2012}
{Draine}, B.~T., \& {Hensley}, B. 2012, \apj, 757, 103,
  \dodoi{10.1088/0004-637X/757/1/103}

\bibitem[{{Draine} \& {Hensley}(2013)}]{Draine_2013}
---. 2013, \apj, 765, 159, \dodoi{10.1088/0004-637X/765/2/159}

\bibitem[{{Draine} \& {Hensley}(2016)}]{Draine_2016}
{Draine}, B.~T., \& {Hensley}, B.~S. 2016, \apj, 831, 59,
  \dodoi{10.3847/0004-637X/831/1/59}

\bibitem[{{Draine} \& {Hensley}(2021{\natexlab{a}})}]{Draine2021}
---. 2021{\natexlab{a}}, \apj, 909, 94, \dodoi{10.3847/1538-4357/abd6c6}

\bibitem[{{Draine} \& {Hensley}(2021{\natexlab{b}})}]{Draine_2021}
---. 2021{\natexlab{b}}, \apj, 919, 65, \dodoi{10.3847/1538-4357/ac0050}

\bibitem[{{Draine} \& {Lazarian}(1998{\natexlab{a}})}]{Draine_1998a}
{Draine}, B.~T., \& {Lazarian}, A. 1998{\natexlab{a}}, \apjl, 494, L19,
  \dodoi{10.1086/311167}

\bibitem[{{Draine} \& {Lazarian}(1998{\natexlab{b}})}]{Draine_1998b}
---. 1998{\natexlab{b}}, \apj, 508, 157, \dodoi{10.1086/306387}

\bibitem[{{Draine} \& {Lazarian}(1999)}]{Draine_1999}
---. 1999, \apj, 512, 740, \dodoi{10.1086/306809}

\bibitem[{{Draine} \& {Lee}(1984)}]{Draine1984}
{Draine}, B.~T., \& {Lee}, H.~M. 1984, \apj, 285, 89, \dodoi{10.1086/162480}

\bibitem[{{Dunkley} {et~al.}(2009){Dunkley}, {Spergel}, {Komatsu}, {Hinshaw},
  {Larson}, {Nolta}, {Odegard}, {Page}, {Bennett}, {Gold}, {Hill}, {Jarosik},
  {Weiland}, {Halpern}, {Kogut}, {Limon}, {Meyer}, {Tucker}, {Wollack}, \&
  {Wright}}]{Dunkley2009}
{Dunkley}, J., {Spergel}, D.~N., {Komatsu}, E., {et~al.} 2009, \apj, 701, 1804,
  \dodoi{10.1088/0004-637X/701/2/1804}

\bibitem[{{Dunkley} {et~al.}(2013){Dunkley}, {Calabrese}, {Sievers}, {Addison},
  {Battaglia}, {Battistelli}, {Bond}, {Das}, {Devlin}, {D{\"u}nner}, {Fowler},
  {Gralla}, {Hajian}, {Halpern}, {Hasselfield}, {Hincks}, {Hlozek}, {Hughes},
  {Irwin}, {Kosowsky}, {Louis}, {Marriage}, {Marsden}, {Menanteau}, {Moodley},
  {Niemack}, {Nolta}, {Page}, {Partridge}, {Sehgal}, {Spergel}, {Staggs},
  {Switzer}, {Trac}, \& {Wollack}}]{Dunkley2013}
{Dunkley}, J., {Calabrese}, E., {Sievers}, J., {et~al.} 2013, \jcap, 2013, 025,
  \dodoi{10.1088/1475-7516/2013/07/025}

\bibitem[{{Dutcher} {et~al.}(2021){Dutcher}, {Balkenhol}, {Ade}, {Ahmed},
  {Anderes}, {Anderson}, {Archipley}, {Avva}, {Aylor}, {Barry}, {Basu Thakur},
  {Benabed}, {Bender}, {Benson}, {Bianchini}, {Bleem}, {Bouchet}, {Bryant},
  {Byrum}, {Carlstrom}, {Carter}, {Cecil}, {Chang}, {Chaubal}, {Chen}, {Cho},
  {Chou}, {Cliche}, {Crawford}, {Cukierman}, {Daley}, {de Haan}, {Denison},
  {Dibert}, {Ding}, {Dobbs}, {Everett}, {Feng}, {Ferguson}, {Foster}, {Fu},
  {Galli}, {Gambrel}, {Gardner}, {Goeckner-Wald}, {Gualtieri}, {Guns}, {Gupta},
  {Guyser}, {Halverson}, {Harke-Hosemann}, {Harrington}, {Henning}, {Hilton},
  {Hivon}, {Holder}, {Holzapfel}, {Hood}, {Howe}, {Huang}, {Irwin}, {Jeong},
  {Jonas}, {Jones}, {Khaire}, {Knox}, {Kofman}, {Korman}, {Kubik}, {Kuhlmann},
  {Kuo}, {Lee}, {Leitch}, {Lowitz}, {Lu}, {Meyer}, {Michalik}, {Millea},
  {Montgomery}, {Nadolski}, {Natoli}, {Nguyen}, {Noble}, {Novosad}, {Omori},
  {Padin}, {Pan}, {Paschos}, {Pearson}, {Posada}, {Prabhu}, {Quan},
  {Raghunathan}, {Rahlin}, {Reichardt}, {Riebel}, {Riedel}, {Rouble}, {Ruhl},
  {Sayre}, {Schiappucci}, {Shirokoff}, {Smecher}, {Sobrin}, {Stark}, {Stephen},
  {Story}, {Suzuki}, {Thompson}, {Thorne}, {Tucker}, {Umilta}, {Vale},
  {Vanderlinde}, {Vieira}, {Wang}, {Whitehorn}, {Wu}, {Yefremenko}, {Yoon},
  {Young}, \& {SPT-3G Collaboration}}]{SPT3g2021}
{Dutcher}, D., {Balkenhol}, L., {Ade}, P.~A.~R., {et~al.} 2021, \prd, 104,
  022003, \dodoi{10.1103/PhysRevD.104.022003}

\bibitem[{{Dwek} {et~al.}(1997){Dwek}, {Arendt}, {Fixsen}, {Sodroski},
  {Odegard}, {Weiland}, {Reach}, {Hauser}, {Kelsall}, {Moseley}, {Silverberg},
  {Shafer}, {Ballester}, {Bazell}, \& {Isaacman}}]{Dwek:1997}
{Dwek}, E., {Arendt}, R.~G., {Fixsen}, D.~J., {et~al.} 1997, \apj, 475, 565,
  \dodoi{10.1086/303568}

\bibitem[{{Elmegreen} \& {Scalo}(2004)}]{Elmegreen:2004}
{Elmegreen}, B.~G., \& {Scalo}, J. 2004, \araa, 42, 211,
  \dodoi{10.1146/annurev.astro.41.011802.094859}

\bibitem[{{Essinger-Hileman} {et~al.}(2020){Essinger-Hileman}, {Ade},
  {Baildon}, {Bellis}, {Benford}, {Bennett}, {Chuss}, {Datta}, {Eimer},
  {Fixsen}, {Gandilo}, {Halpern}, {Hilton}, {Irwin}, {Jhabvala}, {Kimball},
  {Kogut}, {Lazear}, {Lowe}, {McMahon}, {Miller}, {Mirel}, {Moseley}, {Pawlyk},
  {Rodriguez}, {Sharp}, {Shirron}, {Staguhn}, {Sullivan}, {Switzer},
  {Taraschi}, {Tucker}, {Walts}, \& {Wollack}}]{PIPER2016}
{Essinger-Hileman}, T., {Ade}, P.~A., {Baildon}, T., {et~al.} 2020, in American
  Astronomical Society Meeting Abstracts, Vol. 236, American Astronomical
  Society Meeting Abstracts \#236, 244.01

\bibitem[{{Federrath}(2015)}]{Federrath2015}
{Federrath}, C. 2015, \mnras, 450, 4035, \dodoi{10.1093/mnras/stv941}

\bibitem[{{Fernandes} {et~al.}(2019){Fernandes}, {Mulders}, {Pascucci},
  {Mordasini}, \& {Emsenhuber}}]{2019ApJ...874...81F}
{Fernandes}, R.~B., {Mulders}, G.~D., {Pascucci}, I., {Mordasini}, C., \&
  {Emsenhuber}, A. 2019, \apj, 874, 81, \dodoi{10.3847/1538-4357/ab0300}

\bibitem[{{Finkbeiner} {et~al.}(1999){Finkbeiner}, {Davis}, \&
  {Schlegel}}]{Finkbeiner:1999}
{Finkbeiner}, D.~P., {Davis}, M., \& {Schlegel}, D.~J. 1999, \apj, 524, 867,
  \dodoi{10.1086/307852}

\bibitem[{{Fissel}(2013)}]{fissel2013}
{Fissel}, L.~M. 2013, PhD thesis, University of Toronto, Canada

\bibitem[{{Fissel} {et~al.}(2016){Fissel}, {Ade}, {Angil{\`e}}, {Ashton},
  {Benton}, {Devlin}, {Dober}, {Fukui}, {Galitzki}, {Gandilo}, {Klein},
  {Korotkov}, {Li}, {Martin}, {Matthews}, {Moncelsi}, {Nakamura},
  {Netterfield}, {Novak}, {Pascale}, {Poidevin}, {Santos}, {Savini}, {Scott},
  {Shariff}, {Diego Soler}, {Thomas}, {Tucker}, {Tucker}, \&
  {Ward-Thompson}}]{Fissel:2016}
{Fissel}, L.~M., {Ade}, P. A.~R., {Angil{\`e}}, F.~E., {et~al.} 2016, \apj,
  824, 134, \dodoi{10.3847/0004-637X/824/2/134}

\bibitem[{{Fissel} {et~al.}(2019){Fissel}, {Ade}, {Angil{\`e}}, {Ashton},
  {Benton}, {Chen}, {Cunningham}, {Devlin}, {Dober}, {Friesen}, {Fukui},
  {Galitzki}, {Gandilo}, {Goodman}, {Green}, {Jones}, {Klein}, {King},
  {Korotkov}, {Li}, {Lowe}, {Martin}, {Matthews}, {Moncelsi}, {Nakamura},
  {Netterfield}, {Newmark}, {Novak}, {Pascale}, {Poidevin}, {Santos}, {Savini},
  {Scott}, {Shariff}, {Soler}, {Thomas}, {Tucker}, {Tucker}, {Ward-Thompson},
  \& {Zucker}}]{Fissel:2019}
---. 2019, \apj, 878, 110, \dodoi{10.3847/1538-4357/ab1eb0}

\bibitem[{{Fixsen} {et~al.}(1999){Fixsen}, {Bennett}, \&
  {Mather}}]{Fixsen:1999}
{Fixsen}, D.~J., {Bennett}, C.~L., \& {Mather}, J.~C. 1999, \apj, 526, 207,
  \dodoi{10.1086/307962}

\bibitem[{{Fonnesbeck} {et~al.}(2015){Fonnesbeck}, {Patil}, {Huard}, \&
  {Salvatier}}]{PyMC}
{Fonnesbeck}, C., {Patil}, A., {Huard}, D., \& {Salvatier}, J. 2015, {PyMC:
  Bayesian Stochastic Modelling in Python}.
\newblock \doeprint{1506.005}

\bibitem[{{Francis}(2005)}]{2005ApJ...635.1348F}
{Francis}, P.~J. 2005, \apj, 635, 1348, \dodoi{10.1086/497684}

\bibitem[{{Fuskeland} {et~al.}(2021){Fuskeland}, {Andersen}, {Aurlien},
  {Banerji}, {Brilenkov}, {Eriksen}, {Galloway}, {Gjerl{\o}w}, {N{\ae}ss},
  {Svalheim}, \& {Wehus}}]{Fuskeland2021}
{Fuskeland}, U., {Andersen}, K.~J., {Aurlien}, R., {et~al.} 2021, \aap, 646,
  A69, \dodoi{10.1051/0004-6361/202037629}

\bibitem[{{Gaensler} {et~al.}(2011){Gaensler}, {Haverkorn}, {Burkhart},
  {Newton-McGee}, {Ekers}, {Lazarian}, {McClure-Griffiths}, {Robishaw},
  {Dickey}, \& {Green}}]{Gaensler:2011}
{Gaensler}, B.~M., {Haverkorn}, M., {Burkhart}, B., {et~al.} 2011, \nat, 478,
  214, \dodoi{10.1038/nature10446}

\bibitem[{{G{\'e}nova-Santos} {et~al.}(2017){G{\'e}nova-Santos},
  {Rubi{\~n}o-Mart{\'{\i}}n}, {Pel{\'a}ez-Santos}, {Poidevin}, {Rebolo},
  {Vignaga}, {Artal}, {Harper}, {Hoyland}, {Lasenby},
  {Mart{\'{\i}}nez-Gonz{\'a}lez}, {Piccirillo}, {Tramonte}, \&
  {Watson}}]{GenovaSantos_2017}
{G{\'e}nova-Santos}, R., {Rubi{\~n}o-Mart{\'{\i}}n}, J.~A.,
  {Pel{\'a}ez-Santos}, A., {et~al.} 2017, \mnras, 464, 4107,
  \dodoi{10.1093/mnras/stw2503}

\bibitem[{{Ghosh} {et~al.}(2017){Ghosh}, {Boulanger}, {Martin}, {Bracco},
  {Vansyngel}, {Aumont}, {Bock}, {Dor{\'e}}, {Haud}, {Kalberla}, \&
  {Serra}}]{Ghosh:2017}
{Ghosh}, T., {Boulanger}, F., {Martin}, P.~G., {et~al.} 2017, \aap, 601, A71,
  \dodoi{10.1051/0004-6361/201629829}

\bibitem[{{Girichidis} {et~al.}(2020){Girichidis}, {Offner}, {Kritsuk},
  {Klessen}, {Hennebelle}, {Kruijssen}, {Krause}, {Glover}, \&
  {Padovani}}]{Girichidis2020}
{Girichidis}, P., {Offner}, S. S.~R., {Kritsuk}, A.~G., {et~al.} 2020, \ssr,
  216, 68, \dodoi{10.1007/s11214-020-00693-8}

\bibitem[{{Gold} {et~al.}(2011){Gold}, {Odegard}, {Weiland}, {Hill}, {Kogut},
  {Bennett}, {Hinshaw}, {Chen}, {Dunkley}, {Halpern}, {Jarosik}, {Komatsu},
  {Larson}, {Limon}, {Meyer}, {Nolta}, {Page}, {Smith}, {Spergel}, {Tucker},
  {Wollack}, \& {Wright}}]{Gold:2011}
{Gold}, B., {Odegard}, N., {Weiland}, J.~L., {et~al.} 2011, \apjs, 192, 15,
  \dodoi{10.1088/0067-0049/192/2/15}

\bibitem[{Goldreich \& Kylafis(1981)}]{Goldreich1981}
Goldreich, P., \& Kylafis, N.~D. 1981, The Astrophysical Journal, 243, L75,
  \dodoi{10.1086/183446}

\bibitem[{{Goldreich} \& {Sridhar}(1995)}]{GS95}
{Goldreich}, P., \& {Sridhar}, S. 1995, \apj, 438, 763, \dodoi{10.1086/175121}

\bibitem[{{G{\'o}rski} {et~al.}(2005){G{\'o}rski}, {Hivon}, {Banday},
  {Wandelt}, {Hansen}, {Reinecke}, \& {Bartelmann}}]{Gorski2005}
{G{\'o}rski}, K.~M., {Hivon}, E., {Banday}, A.~J., {et~al.} 2005, \apj, 622,
  759, \dodoi{10.1086/427976}

\bibitem[{{Greaves} {et~al.}(2002){Greaves}, {Holland}, \&
  {Dent}}]{Greaves:2002}
{Greaves}, J.~S., {Holland}, W.~S., \& {Dent}, W.~R.~F. 2002, \apj, 578, 224,
  \dodoi{10.1086/342345}

\bibitem[{Greaves {et~al.}(1999)Greaves, Holland, Friberg, \&
  Dent}]{Greaves1999}
Greaves, J.~S., Holland, W.~S., Friberg, P., \& Dent, W. R.~F. 1999,
  Astrophysical Journal, 512, L139, \dodoi{10.1086/311888}

\bibitem[{{Guan} {et~al.}(2021){Guan}, {Clark}, {Hensley}, {Gallardo}, {Naess},
  {Duell}, {Aiola}, {Atkins}, {Calabrese}, {Choi}, {Cothard}, {Devlin},
  {Duivenvoorden}, {Dunkley}, {D{\"u}nner}, {Ferraro}, {Hasselfield}, {Hughes},
  {Koopman}, {Kosowsky}, {Madhavacheril}, {McMahon}, {Nati}, {Niemack}, {Page},
  {Salatino}, {Schaan}, {Sehgal}, {Sif{\'o}n}, {Staggs}, {Vavagiakis},
  {Wollack}, \& {Xu}}]{Guan:2021}
{Guan}, Y., {Clark}, S.~E., {Hensley}, B.~S., {et~al.} 2021, \apj, 920, 6,
  \dodoi{10.3847/1538-4357/ac133f}

\bibitem[{{Gudmundsson} {et~al.}(2021){Gudmundsson}, {Gallardo}, {Puddu},
  {Dicker}, {Adler}, {Ali}, {Bazarko}, {Chesmore}, {Coppi}, {Cothard},
  {Dachlythra}, {Devlin}, {D{\"u}nner}, {Fabbian}, {Galitzki}, {Golec}, {Patty
  Ho}, {Hargrave}, {Kofman}, {Lee}, {Limon}, {Matsuda}, {Mauskopf}, {Moodley},
  {Nati}, {Niemack}, {Orlowski-Scherer}, {Page}, {Partridge}, {Puglisi},
  {Reichardt}, {Sierra}, {Simon}, {Teply}, {Tucker}, {Wollack}, {Xu}, \&
  {Zhu}}]{SO_LAT2}
{Gudmundsson}, J.~E., {Gallardo}, P.~A., {Puddu}, R., {et~al.} 2021, \ao, 60,
  823, \dodoi{10.1364/AO.411533}

\bibitem[{{Guillet} {et~al.}(2018){Guillet}, {Fanciullo}, {Verstraete},
  {Boulanger}, {Jones}, {Miville-Desch{\^e}nes}, {Ysard}, {Levrier}, \&
  {Alves}}]{Guillet2018}
{Guillet}, V., {Fanciullo}, L., {Verstraete}, L., {et~al.} 2018, \aap, 610,
  A16, \dodoi{10.1051/0004-6361/201630271}

\bibitem[{{Guns} {et~al.}(2021){Guns}, {Foster}, {Daley}, {Rahlin},
  {Whitehorn}, {Ade}, {Ahmed}, {Anderes}, {Anderson}, {Archipley}, {Avva},
  {Aylor}, {Balkenhol}, {Barry}, {Basu Thakur}, {Benabed}, {Bender}, {Benson},
  {Bianchini}, {Bleem}, {Bouchet}, {Bryant}, {Byrum}, {Carlstrom}, {Carter},
  {Cecil}, {Chang}, {Chaubal}, {Chen}, {Cho}, {Chou}, {Cliche}, {Crawford},
  {Cukierman}, {de Haan}, {Denison}, {Dibert}, {Ding}, {Dobbs}, {Dutcher},
  {Everett}, {Feng}, {Ferguson}, {Fu}, {Galli}, {Gambrel}, {Gardner},
  {Goeckner-Wald}, {Gualtieri}, {Gupta}, {Guyser}, {Halverson},
  {Harke-Hosemann}, {Harrington}, {Henning}, {Hilton}, {Hivon}, {Holder},
  {Holzapfel}, {Hood}, {Howe}, {Huang}, {Irwin}, {Jeong}, {Jonas}, {Jones},
  {Khaire}, {Knox}, {Kofman}, {Korman}, {Kubik}, {Kuhlmann}, {Kuo}, {Lee},
  {Leitch}, {Lowitz}, {Lu}, {Marrone}, {Meyer}, {Michalik}, {Millea},
  {Montgomery}, {Nadolski}, {Natoli}, {Nguyen}, {Noble}, {Novosad}, {Omori},
  {Padin}, {Pan}, {Paschos}, {Pearson}, {Phadke}, {Posada}, {Prabhu}, {Quan},
  {Reichardt}, {Riebel}, {Riedel}, {Rouble}, {Ruhl}, {Sayre}, {Schiappucci},
  {Shirokoff}, {Smecher}, {Sobrin}, {Stark}, {Stephen}, {Story}, {Suzuki},
  {Thompson}, {Thorne}, {Tucker}, {Umilta}, {Vale}, {Vieira}, {Wang}, {Wu},
  {Yefremenko}, {Yoon}, {Young}, \& {Zhang}}]{Guns2021}
{Guns}, S., {Foster}, A., {Daley}, C., {et~al.} 2021, \apj, 916, 98,
  \dodoi{10.3847/1538-4357/ac06a3}

\bibitem[{{Hanany} {et~al.}(2019){Hanany}, {Alvarez}, {Artis}, {Ashton},
  {Aumont}, {Aurlien}, {Banerji}, {Barreiro}, {Bartlett}, {Basak}, {Battaglia},
  {Bock}, {Boddy}, {Bonato}, {Borrill}, {Bouchet}, {Boulanger}, {Burkhart},
  {Chluba}, {Chuss}, {Clark}, {Cooperrider}, {Crill}, {De Zotti},
  {Delabrouille}, {Di Valentino}, {Didier}, {Dor{\'e}}, {Eriksen}, {Errard},
  {Essinger-Hileman}, {Feeney}, {Filippini}, {Fissel}, {Flauger}, {Fuskeland},
  {Gluscevic}, {Gorski}, {Green}, {Hensley}, {Herranz}, {Hill}, {Hivon},
  {Hlo{\v{z}}ek}, {Hubmayr}, {Johnson}, {Jones}, {Jones}, {Knox}, {Kogut},
  {L{\'o}pez-Caniego}, {Lawrence}, {Lazarian}, {Li}, {Madhavacheril}, {Melin},
  {Meyers}, {Murray}, {Negrello}, {Novak}, {O'Brient}, {Paine}, {Pearson},
  {Pogosian}, {Pryke}, {Puglisi}, {Remazeilles}, {Rocha}, {Schmittfull},
  {Scott}, {Shirron}, {Stephens}, {Sutin}, {Tomasi}, {Trangsrud}, {van
  Engelen}, {Vansyngel}, {Wehus}, {Wen}, {Xu}, {Young}, \&
  {Zonca}}]{Hanany2019}
{Hanany}, S., {Alvarez}, M., {Artis}, E., {et~al.} 2019, arXiv e-prints,
  arXiv:1902.10541.
\newblock \doarXiv{1902.10541}

\bibitem[{{Hasselfield} {et~al.}(2013){Hasselfield}, {Moodley}, {Bond}, {Das},
  {Devlin}, {Dunkley}, {D{\"u}nner}, {Fowler}, {Gallardo}, {Gralla}, {Hajian},
  {Halpern}, {Hincks}, {Marriage}, {Marsden}, {Niemack}, {Nolta}, {Page},
  {Partridge}, {Schmitt}, {Sehgal}, {Sievers}, {Staggs}, {Swetz}, {Switzer}, \&
  {Wollack}}]{Hasselfield_2013}
{Hasselfield}, M., {Moodley}, K., {Bond}, J.~R., {et~al.} 2013, \apjs, 209, 17,
  \dodoi{10.1088/0067-0049/209/1/17}

\bibitem[{{Hazumi} {et~al.}(2020){Hazumi}, {Ade}, {Adler}, {Allys}, {Arnold},
  {Auguste}, {Aumont}, {Aurlien}, {Austermann}, {Baccigalupi}, {Banday},
  {Banjeri}, {Barreiro}, {Basak}, {Beall}, {Beck}, {Beckman}, {Bermejo}, {de
  Bernardis}, {Bersanelli}, {Bonis}, {Borrill}, {Boulanger}, {Bounissou},
  {Brilenkov}, {Brown}, {Bucher}, {Calabrese}, {Campeti}, {Carones}, {Casas},
  {Challinor}, {Chan}, {Cheung}, {Chinone}, {Cliche}, {Colombo}, {Columbro},
  {Cubas}, {Cukierman}, {Curtis}, {D'Alessandro}, {Dachlythra}, {De Petris},
  {Dickinson}, {Diego-Palazuelos}, {Dobbs}, {Dotani}, {Duband}, {Duff},
  {Duval}, {Ebisawa}, {Elleflot}, {Eriksen}, {Errard}, {Essinger-Hileman},
  {Finelli}, {Flauger}, {Franceschet}, {Fuskeland}, {Galloway}, {Ganga}, {Gao},
  {Genova-Santos}, {Gerbino}, {Gervasi}, {Ghigna}, {Gjerl{\o}w}, {Gradziel},
  {Grain}, {Grupp}, {Gruppuso}, {Gudmundsson}, {de Haan}, {Halverson},
  {Hargrave}, {Hasebe}, {Hasegawa}, {Hattori}, {Henrot-Versill{\'e}}, {Herman},
  {Herranz}, {Hill}, {Hilton}, {Hirota}, {Hivon}, {Hlozek}, {Hoshino}, {de la
  Hoz}, {Hubmayr}, {Ichiki}, {Iida}, {Imada}, {Ishimura}, {Ishino}, {Jaehnig},
  {Kaga}, {Kashima}, {Katayama}, {Kato}, {Kawasaki}, {Keskitalo}, {Kisner},
  {Kobayashi}, {Kogiso}, {Kogut}, {Kohri}, {Komatsu}, {Komatsu}, {Konishi},
  {Krachmalnicoff}, {Kreykenbohm}, {Kuo}, {Kushino}, {Lamagna}, {Lanen},
  {Lattanzi}, {Lee}, {Leloup}, {Levrier}, {Linder}, {Louis}, {Luzzi},
  {Maciaszek}, {Maffei}, {Maino}, {Maki}, {Mandelli}, {Martinez-Gonzalez},
  {Masi}, {Matsumura}, {Mennella}, {Migliaccio}, {Minami}, {Mitsuda},
  {Montgomery}, {Montier}, {Morgante}, {Mot}, {Murata}, {Murphy}, {Nagai},
  {Nagano}, {Nagasaki}, {Nagata}, {Nakamura}, {Namikawa}, {Natoli}, {Nerval},
  {Nishibori}, {Nishino}, {Noviello}, {O'Sullivan}, {Ogawa}, {Ogawa}, {Oguri},
  {Ohsaki}, {Ohta}, {Okada}, {Okada}, {Pagano}, {Paiella}, {Paoletti},
  {Patanchon}, {Peloton}, {Piacentini}, {Pisano}, {Polenta}, {Poletti},
  {Prouv{\'e}}, {Puglisi}, {Rambaud}, {Raum}, {Realini}, {Reinecke},
  {Remazeilles}, {Ritacco}, {Roudil}, {Rubino-Martin}, {Russell}, {Sakurai},
  {Sakurai}, {Sandri}, {Sasaki}, {Savini}, {Scott}, {Seibert}, {Sekimoto},
  {Sherwin}, {Shinozaki}, {Shiraishi}, {Shirron}, {Signorelli}, {Smecher},
  {Stever}, {Stompor}, {Sugai}, {Sugiyama}, {Suzuki}, {Suzuki}, {Svalheim},
  {Switzer}, {Takaku}, {Takakura}, {Takakura}, {Takase}, {Takeda}, {Tartari},
  {Taylor}, {Terao}, {Thommesen}, {Thompson}, {Thorne}, {Toda}, {Tomasi},
  {Tominaga}, {Trappe}, {Tristram}, {Tsuji}, {Tsujimoto}, {Tucker}, {Ullom},
  {Vermeulen}, {Vielva}, {Villa}, {Vissers}, {Vittorio}, {Wehus}, {Weller},
  {Westbrook}, {Wilms}, {Winter}, {Wollack}, {Yamasaki}, {Yoshida}, {Yumoto},
  {Zannoni}, \& {Zonca}}]{LiteBIRD2020}
{Hazumi}, M., {Ade}, P.~A.~R., {Adler}, A., {et~al.} 2020, in Society of
  Photo-Optical Instrumentation Engineers (SPIE) Conference Series, Vol. 11443,
  Society of Photo-Optical Instrumentation Engineers (SPIE) Conference Series,
  114432F, \dodoi{10.1117/12.2563050}

\bibitem[{{Hennebelle} \& {Inutsuka}(2019)}]{Hennebelle2019}
{Hennebelle}, P., \& {Inutsuka}, S.-i. 2019, Frontiers in Astronomy and Space
  Sciences, 6, 5, \dodoi{10.3389/fspas.2019.00005}

\bibitem[{{Hensley} \& {Draine}(2017)}]{Hensley_2017}
{Hensley}, B.~S., \& {Draine}, B.~T. 2017, \apj, 836, 179,
  \dodoi{10.3847/1538-4357/aa5c37}

\bibitem[{{Hensley} \& {Draine}(2021)}]{Hensley2021}
---. 2021, \apj, 906, 73, \dodoi{10.3847/1538-4357/abc8f1}

\bibitem[{{Herron} {et~al.}(2018){Herron}, {Burkhart}, {Gaensler}, {Lewis},
  {McClure-Griffiths}, {Bernardi}, {Carretti}, {Haverkorn}, {Kesteven},
  {Poppi}, \& {Staveley-Smith}}]{2018ApJ...855...29H}
{Herron}, C.~A., {Burkhart}, B., {Gaensler}, B.~M., {et~al.} 2018, \apj, 855,
  29, \dodoi{10.3847/1538-4357/aaafd0}

\bibitem[{{Herv{\'\i}as-Caimapo} \& {Huffenberger}(2021)}]{HerviasCaimapo2021}
{Herv{\'\i}as-Caimapo}, C., \& {Huffenberger}, K. 2021, arXiv e-prints,
  arXiv:2107.08317.
\newblock \doarXiv{2107.08317}

\bibitem[{{HI4PI Collaboration} {et~al.}(2016){HI4PI Collaboration}, {Ben
  Bekhti}, {Fl{\"o}er}, {Keller}, {Kerp}, {Lenz}, {Winkel}, {Bailin},
  {Calabretta}, {Dedes}, {Ford}, {Gibson}, {Haud}, {Janowiecki}, {Kalberla},
  {Lockman}, {McClure-Griffiths}, {Murphy}, {Nakanishi}, {Pisano}, \&
  {Staveley-Smith}}]{2016A&A...594A.116H}
{HI4PI Collaboration}, {Ben Bekhti}, N., {Fl{\"o}er}, L., {et~al.} 2016, \aap,
  594, A116, \dodoi{10.1051/0004-6361/201629178}

\bibitem[{{Hoang} {et~al.}(2013){Hoang}, {Lazarian}, \& {Martin}}]{Hoang2013}
{Hoang}, T., {Lazarian}, A., \& {Martin}, P.~G. 2013, \apj, 779, 152,
  \dodoi{10.1088/0004-637X/779/2/152}

\bibitem[{{Hoang} {et~al.}(2014){Hoang}, {Lazarian}, \& {Martin}}]{Hoang2014}
---. 2014, \apj, 790, 6, \dodoi{10.1088/0004-637X/790/1/6}

\bibitem[{{Huffenberger} {et~al.}(2020){Huffenberger}, {Rotti}, \&
  {Collins}}]{Huffenberger:2020}
{Huffenberger}, K.~M., {Rotti}, A., \& {Collins}, D.~C. 2020, \apj, 899, 31,
  \dodoi{10.3847/1538-4357/ab9df9}

\bibitem[{Hughes {et~al.}(2018)Hughes, Duchêne, \& Matthews}]{Hughes:2018}
Hughes, A.~M., Duchêne, G., \& Matthews, B.~C. 2018, Annual Review of
  Astronomy and Astrophysics, 56, 541–591,
  \dodoi{10.1146/annurev-astro-081817-052035}

\bibitem[{{Hunter}(2007)}]{Matplotlib}
{Hunter}, J.~D. 2007, Computing in Science and Engineering, 9, 90,
  \dodoi{10.1109/MCSE.2007.55}

\bibitem[{{Hurier}(2019)}]{hurier2019}
{Hurier}, G. 2019, arXiv e-prints.
\newblock \doarXiv{1903.11883}

\bibitem[{{Hurier} {et~al.}(2013){Hurier}, {Mac{\'\i}as-P{\'e}rez}, \&
  {Hildebrandt}}]{milca}
{Hurier}, G., {Mac{\'\i}as-P{\'e}rez}, J.~F., \& {Hildebrandt}, S. 2013, \aap,
  558, A118, \dodoi{10.1051/0004-6361/201321891}

\bibitem[{{Israel} {et~al.}(2010){Israel}, {Wall}, {Raban}, {Reach}, {Bot},
  {Oonk}, {Ysard}, \& {Bernard}}]{Israel2010}
{Israel}, F.~P., {Wall}, W.~F., {Raban}, D., {et~al.} 2010, \aap, 519, A67,
  \dodoi{10.1051/0004-6361/201014073}

\bibitem[{{Jansson} \& {Farrar}(2012)}]{Jansson:2012}
{Jansson}, R., \& {Farrar}, G.~R. 2012, \apj, 757, 14,
  \dodoi{10.1088/0004-637X/757/1/14}

\bibitem[{{Jones}(2009)}]{Jones_2009}
{Jones}, A.~P. 2009, \aap, 506, 797, \dodoi{10.1051/0004-6361/200810621}

\bibitem[{{Jones} {et~al.}(2017){Jones}, {K{\"o}hler}, {Ysard}, {Bocchio}, \&
  {Verstraete}}]{Jones2017}
{Jones}, A.~P., {K{\"o}hler}, M., {Ysard}, N., {Bocchio}, M., \& {Verstraete},
  L. 2017, \aap, 602, A46, \dodoi{10.1051/0004-6361/201630225}

\bibitem[{{Jones} {et~al.}(2018){Jones}, {Taylor}, {Aich}, {Copley}, {Chiang},
  {Davis}, {Dickinson}, {Grumitt}, {Hafez}, {Heilgendorff}, {Holler}, {Irfan},
  {Jew}, {John}, {Jonas}, {King}, {Leahy}, {Leech}, {Leitch}, {Muchovej},
  {Pearson}, {Peel}, {Readhead}, {Sievers}, {Stevenson}, \&
  {Zuntz}}]{Jones2018}
{Jones}, M.~E., {Taylor}, A.~C., {Aich}, M., {et~al.} 2018, \mnras, 480, 3224,
  \dodoi{10.1093/mnras/sty1956}

\bibitem[{{Jow} {et~al.}(2018){Jow}, {Hill}, {Scott}, {Soler}, {Martin},
  {Devlin}, {Fissel}, \& {Poidevin}}]{Jow:2018}
{Jow}, D.~L., {Hill}, R., {Scott}, D., {et~al.} 2018, \mnras, 474, 1018,
  \dodoi{10.1093/mnras/stx2736}

\bibitem[{{Kalberla} {et~al.}(2016){Kalberla}, {Kerp}, {Haud}, {Winkel}, {Ben
  Bekhti}, {Fl{\"o}er}, \& {Lenz}}]{Kalberla:2016}
{Kalberla}, P.~M.~W., {Kerp}, J., {Haud}, U., {et~al.} 2016, \apj, 821, 117,
  \dodoi{10.3847/0004-637X/821/2/117}

\bibitem[{{Kim} {et~al.}(2019){Kim}, {Choi}, \& {Flauger}}]{Kim:2019}
{Kim}, C.-G., {Choi}, S.~K., \& {Flauger}, R. 2019, \apj, 880, 106,
  \dodoi{10.3847/1538-4357/ab29f2}

\bibitem[{{Kim} \& {Martin}(1995)}]{Kim+Martin_1995}
{Kim}, S.-H., \& {Martin}, P.~G. 1995, \apj, 444, 293, \dodoi{10.1086/175604}

\bibitem[{{Kiuchi} {et~al.}(2020){Kiuchi}, {Adachi}, {Ali}, {Arnold}, {Ashton},
  {Austermann}, {Bazako}, {Beall}, {Chinone}, {Coppi}, {Crowley}, {Crowley},
  {Dicker}, {Dober}, {Duff}, {Fabbian}, {Galitzki}, {Golec}, {Gudmundsson},
  {Harrington}, {Hasegawa}, {Hattori}, {Hill}, {Ho}, {Hubmayr}, {Johnson},
  {Kaneko}, {Katayama}, {Keating}, {Kusaka}, {Lashner}, {Lee}, {Matsuda},
  {McCarrick}, {Murata}, {Nati}, {Nishinomiya}, {Page}, {Sathyanarayana Rao},
  {Reichardt}, {Sakaguri}, {Sakurai}, {Sibert}, {Spisak}, {Tajima}, {Teply},
  {Terasaki}, {Tsan}, {Walker}, {Wollack}, {Xu}, {Yamada}, {Zannoni}, \&
  {Zhu}}]{sat}
{Kiuchi}, K., {Adachi}, S., {Ali}, A.~M., {et~al.} 2020, in Society of
  Photo-Optical Instrumentation Engineers (SPIE) Conference Series, Vol. 11445,
  Society of Photo-Optical Instrumentation Engineers (SPIE) Conference Series,
  114457L, \dodoi{10.1117/12.2562016}

\bibitem[{{Knox}(1995)}]{Knox1995}
{Knox}, L. 1995, \prd, 52, 4307, \dodoi{10.1103/PhysRevD.52.4307}

\bibitem[{{Kogut}(2012)}]{Kogut2012}
{Kogut}, A. 2012, \apj, 753, 110, \dodoi{10.1088/0004-637X/753/2/110}

\bibitem[{{Kogut} {et~al.}(1996){Kogut}, {Banday}, {Bennett}, {Gorski},
  {Hinshaw}, \& {Reach}}]{Kogut_1996}
{Kogut}, A., {Banday}, A.~J., {Bennett}, C.~L., {et~al.} 1996, \apj, 460, 1,
  \dodoi{10.1086/176947}

\bibitem[{{Kogut} {et~al.}(2016){Kogut}, {Chluba}, {Fixsen}, {Meyer}, \&
  {Spergel}}]{PIXIE2016}
{Kogut}, A., {Chluba}, J., {Fixsen}, D.~J., {Meyer}, S., \& {Spergel}, D. 2016,
  in Society of Photo-Optical Instrumentation Engineers (SPIE) Conference
  Series, Vol. 9904, Space Telescopes and Instrumentation 2016: Optical,
  Infrared, and Millimeter Wave, ed. H.~A. {MacEwen}, G.~G. {Fazio},
  M.~{Lystrup}, N.~{Batalha}, N.~{Siegler}, \& E.~C. {Tong}, 99040W,
  \dodoi{10.1117/12.2231090}

\bibitem[{{Krachmalnicoff} {et~al.}(2016){Krachmalnicoff}, {Baccigalupi},
  {Aumont}, {Bersanelli}, \& {Mennella}}]{Krachmalnicoff2016}
{Krachmalnicoff}, N., {Baccigalupi}, C., {Aumont}, J., {Bersanelli}, M., \&
  {Mennella}, A. 2016, \aap, 588, A65, \dodoi{10.1051/0004-6361/201527678}

\bibitem[{{Krachmalnicoff} {et~al.}(2018){Krachmalnicoff}, {Carretti},
  {Baccigalupi}, {Bernardi}, {Brown}, {Gaensler}, {Haverkorn}, {Kesteven},
  {Perrotta}, {Poppi}, \& {Staveley-Smith}}]{Krachmalnicoff2018}
{Krachmalnicoff}, N., {Carretti}, E., {Baccigalupi}, C., {et~al.} 2018, \aap,
  618, A166, \dodoi{10.1051/0004-6361/201832768}

\bibitem[{{Krause} {et~al.}(2020){Krause}, {Offner}, {Charbonnel}, {Gieles},
  {Klessen}, {V{\'a}zquez-Semadeni}, {Ballesteros-Paredes}, {Girichidis},
  {Kruijssen}, {Ward}, \& {Zinnecker}}]{Krause2020}
{Krause}, M. G.~H., {Offner}, S. S.~R., {Charbonnel}, C., {et~al.} 2020, \ssr,
  216, 64, \dodoi{10.1007/s11214-020-00689-4}

\bibitem[{{Kritsuk} {et~al.}(2017){Kritsuk}, {Ustyugov}, \&
  {Norman}}]{Kritsuk_2017}
{Kritsuk}, A.~G., {Ustyugov}, S.~D., \& {Norman}, M.~L. 2017, New Journal of
  Physics, 19, 065003, \dodoi{10.1088/1367-2630/aa7156}

\bibitem[{{Krumholz}(2014)}]{krumreview2014}
{Krumholz}, M.~R. 2014, \physrep, 539, 49,
  \dodoi{10.1016/j.physrep.2014.02.001}

\bibitem[{{Krumholz} \& {Burkhart}(2016)}]{Krumholz_2016}
{Krumholz}, M.~R., \& {Burkhart}, B. 2016, \mnras, 458, 1671,
  \dodoi{10.1093/mnras/stw434}

\bibitem[{{Krumholz} {et~al.}(2019){Krumholz}, {McKee}, \&
  {Bland-Hawthorn}}]{Krumholz2019}
{Krumholz}, M.~R., {McKee}, C.~F., \& {Bland-Hawthorn}, J. 2019, \araa, 57,
  227, \dodoi{10.1146/annurev-astro-091918-104430}

\bibitem[{{Kusaka} {et~al.}(2014){Kusaka}, {Essinger-Hileman}, {Appel},
  {Gallardo}, {Irwin}, {Jarosik}, {Nolta}, {Page}, {Parker}, {Raghunathan},
  {Sievers}, {Simon}, {Staggs}, \& {Visnjic}}]{abs}
{Kusaka}, A., {Essinger-Hileman}, T., {Appel}, J.~W., {et~al.} 2014, Review of
  Scientific Instruments, 85, 024501, \dodoi{10.1063/1.4862058}

\bibitem[{{Kwon} {et~al.}(2006){Kwon}, {Looney}, {Crutcher}, \&
  {Kirk}}]{2006ApJ...653.1358K}
{Kwon}, W., {Looney}, L.~W., {Crutcher}, R.~M., \& {Kirk}, J.~M. 2006, \apj,
  653, 1358, \dodoi{10.1086/508920}

\bibitem[{{Lallement} {et~al.}(2019){Lallement}, {Babusiaux}, {Vergely},
  {Katz}, {Arenou}, {Valette}, {Hottier}, \& {Capitanio}}]{Lallement:2019}
{Lallement}, R., {Babusiaux}, C., {Vergely}, J.~L., {et~al.} 2019, \aap, 625,
  A135, \dodoi{10.1051/0004-6361/201834695}

\bibitem[{{Lawson} {et~al.}(1987){Lawson}, {Mayer}, {Osborne}, \&
  {Parkinson}}]{Lawson1987}
{Lawson}, K.~D., {Mayer}, C.~J., {Osborne}, J.~L., \& {Parkinson}, M.~L. 1987,
  \mnras, 225, 307, \dodoi{10.1093/mnras/225.2.307}

\bibitem[{{Leitch} {et~al.}(1997){Leitch}, {Readhead}, {Pearson}, \&
  {Myers}}]{Leitch:1997}
{Leitch}, E.~M., {Readhead}, A.~C.~S., {Pearson}, T.~J., \& {Myers}, S.~T.
  1997, \apjl, 486, L23, \dodoi{10.1086/310823}

\bibitem[{{Lenz} {et~al.}(2017){Lenz}, {Hensley}, \&
  {Dor{\'e}}}]{2017ApJ...846...38L}
{Lenz}, D., {Hensley}, B.~S., \& {Dor{\'e}}, O. 2017, \apj, 846, 38,
  \dodoi{10.3847/1538-4357/aa84af}

\bibitem[{{Lewis} {et~al.}(2000){Lewis}, {Challinor}, \&
  {Lasenby}}]{Lewis:1999bs}
{Lewis}, A., {Challinor}, A., \& {Lasenby}, A. 2000, \apj, 538, 473,
  \dodoi{10.1086/309179}

\bibitem[{{Lisenfeld} \& {V{\"o}lk}(2000)}]{Lisenfeld2000}
{Lisenfeld}, U., \& {V{\"o}lk}, H.~J. 2000, \aap, 354, 423.
\newblock \doarXiv{astro-ph/9912232}

\bibitem[{{Louis} {et~al.}(2017){Louis}, {Grace}, {Hasselfield}, {Lungu},
  {Maurin}, {Addison}, {Ade}, {Aiola}, {Allison}, {Amiri}, {Angile},
  {Battaglia}, {Beall}, {de Bernardis}, {Bond}, {Britton}, {Calabrese}, {Cho},
  {Choi}, {Coughlin}, {Crichton}, {Crowley}, {Datta}, {Devlin}, {Dicker},
  {Dunkley}, {D{\"u}nner}, {Ferraro}, {Fox}, {Gallardo}, {Gralla}, {Halpern},
  {Henderson}, {Hill}, {Hilton}, {Hilton}, {Hincks}, {Hlozek}, {Ho}, {Huang},
  {Hubmayr}, {Huffenberger}, {Hughes}, {Infante}, {Irwin}, {Muya Kasanda},
  {Klein}, {Koopman}, {Kosowsky}, {Li}, {Madhavacheril}, {Marriage}, {McMahon},
  {Menanteau}, {Moodley}, {Munson}, {Naess}, {Nati}, {Newburgh}, {Nibarger},
  {Niemack}, {Nolta}, {Nu{\~n}ez}, {Page}, {Pappas}, {Partridge}, {Rojas},
  {Schaan}, {Schmitt}, {Sehgal}, {Sherwin}, {Sievers}, {Simon}, {Spergel},
  {Staggs}, {Switzer}, {Thornton}, {Trac}, {Treu}, {Tucker}, {Van Engelen},
  {Ward}, \& {Wollack}}]{louis2017}
{Louis}, T., {Grace}, E., {Hasselfield}, M., {et~al.} 2017, \jcap, 2017, 031,
  \dodoi{10.1088/1475-7516/2017/06/031}

\bibitem[{{Lowe} {et~al.}(2020){Lowe}, {Coppi}, {Ade}, {Ashton}, {Austermann},
  {Beall}, {Clark}, {Cox}, {Devlin}, {Dicker}, {Dober}, {Fanfani}, {Fissel},
  {Galitzki}, {Gao}, {Hensley}, {Hubmayr}, {Li}, {Li}, {Lourie}, {Martin},
  {Mauskopf}, {Nati}, {Novak}, {Pisano}, {Romualdez}, {Sinclair}, {Soler},
  {Tucker}, {Vissers}, {Wheeler}, {Williams}, \& {Zannoni}}]{BLASTObs2020}
{Lowe}, I., {Coppi}, G., {Ade}, P. A.~R., {et~al.} 2020, in Society of
  Photo-Optical Instrumentation Engineers (SPIE) Conference Series, Vol. 11445,
  Society of Photo-Optical Instrumentation Engineers (SPIE) Conference Series,
  114457A, \dodoi{10.1117/12.2576146}

\bibitem[{{Macellari} {et~al.}(2011){Macellari}, {Pierpaoli}, {Dickinson}, \&
  {Vaillancourt}}]{Macellari_2011}
{Macellari}, N., {Pierpaoli}, E., {Dickinson}, C., \& {Vaillancourt}, J.~E.
  2011, \mnras, 418, 888, \dodoi{10.1111/j.1365-2966.2011.19542.x}

\bibitem[{{MacGregor} {et~al.}(2017){MacGregor}, {Matr{\`a}}, {Kalas},
  {Wilner}, {Pan}, {Kennedy}, {Wyatt}, {Duchene}, {Hughes}, {Rieke}, {Clampin},
  {Fitzgerald}, {Graham}, {Holland}, {Pani{\'c}}, {Shannon}, \&
  {Su}}]{2017ApJ...842....8M}
{MacGregor}, M.~A., {Matr{\`a}}, L., {Kalas}, P., {et~al.} 2017, \apj, 842, 8,
  \dodoi{10.3847/1538-4357/aa71ae}

\bibitem[{{MacLow} \& {Klessen}(2004)}]{MacLow2004}
{MacLow}, M.-M., \& {Klessen}, R.~S. 2004, Reviews of Modern Physics, 76, 125,
  \dodoi{10.1103/RevModPhys.76.125}

\bibitem[{{Malinen} {et~al.}(2016){Malinen}, {Montier}, {Montillaud}, {Juvela},
  {Ristorcelli}, {Clark}, {Bern{\'e}}, {Bernard}, {Pelkonen}, \&
  {Collins}}]{Malinen:2016}
{Malinen}, J., {Montier}, L., {Montillaud}, J., {et~al.} 2016, \mnras, 460,
  1934, \dodoi{10.1093/mnras/stw1061}

\bibitem[{{Marochnik} {et~al.}(1988){Marochnik}, {Mukhin}, \&
  {Sagdeev}}]{1988Sci...242..547M}
{Marochnik}, L.~S., {Mukhin}, L.~M., \& {Sagdeev}, R.~Z. 1988, Science, 242,
  547, \dodoi{10.1126/science.242.4878.547}

\bibitem[{{Martin} {et~al.}(2015){Martin}, {Blagrave}, {Lockman}, {Pinheiro
  Gon{\c{c}}alves}, {Boothroyd}, {Joncas}, {Miville-Desch{\^e}nes}, \&
  {Stephan}}]{Martin:2015}
{Martin}, P.~G., {Blagrave}, K.~P.~M., {Lockman}, F.~J., {et~al.} 2015, \apj,
  809, 153, \dodoi{10.1088/0004-637X/809/2/153}

\bibitem[{{Mathis} {et~al.}(1977){Mathis}, {Rumpl}, \&
  {Nordsieck}}]{Mathis1977}
{Mathis}, J.~S., {Rumpl}, W., \& {Nordsieck}, K.~H. 1977, \apj, 217, 425,
  \dodoi{10.1086/155591}

\bibitem[{{Matsuda} {et~al.}(2019){Matsuda}, {Lowry}, {Suzuki}, {Aguilar
  F{\'a}undez}, {Arnold}, {Barron}, {Bianchini}, {Cheung}, {Chinone},
  {Elleflot}, {Fabbian}, {Goeckner-Wald}, {Hasegawa}, {Kaneko}, {Katayama},
  {Keating}, {Lee}, {Navaroli}, {Nishino}, {Paar}, {Puglisi}, {Richards},
  {Seibert}, {Siritanasak}, {Tajima}, {Takatori}, {Tsai}, \&
  {Westbrook}}]{matsuda2019}
{Matsuda}, F., {Lowry}, L., {Suzuki}, A., {et~al.} 2019, Review of Scientific
  Instruments, 90, 115115, \dodoi{10.1063/1.5095160}

\bibitem[{{McCallum} {et~al.}(2021){McCallum}, {Thomas}, {Brown}, \&
  {Tessore}}]{2021MNRAS.501..802M}
{McCallum}, N., {Thomas}, D.~B., {Brown}, M.~L., \& {Tessore}, N. 2021, \mnras,
  501, 802, \dodoi{10.1093/mnras/staa3609}

\bibitem[{{McClure-Griffiths} {et~al.}(2018){McClure-Griffiths}, {D{\'e}nes},
  {Dickey}, {Stanimirovi{\'c}}, {}, {Staveley-Smith}, {Jameson}, {Di Teodoro},
  {Allison}, {Collier}, {Chippendale}, {Franzen}, {G{\"u}rkan}, {Heald},
  {Hotan}, {Kleiner}, {Lee-Waddell}, {McConnell}, {Popping}, {Rhee}, {Riseley},
  {Voronkov}, \& {Whiting}}]{McClure-Griffiths:2018}
{McClure-Griffiths}, N.~M., {D{\'e}nes}, H., {Dickey}, J.~M., {et~al.} 2018,
  Nature Astronomy, 2, 901, \dodoi{10.1038/s41550-018-0608-8}

\bibitem[{{McKee} \& {Ostriker}(2007)}]{McKee2007}
{McKee}, C.~F., \& {Ostriker}, E.~C. 2007, \araa, 45, 565,
  \dodoi{10.1146/annurev.astro.45.051806.110602}

\bibitem[{{Mirmelstein} {et~al.}(2021){Mirmelstein}, {Fabbian}, {Lewis}, \&
  {Peloton}}]{2021PhRvD.103l3540M}
{Mirmelstein}, M., {Fabbian}, G., {Lewis}, A., \& {Peloton}, J. 2021, \prd,
  103, 123540, \dodoi{10.1103/PhysRevD.103.123540}

\bibitem[{{Miville-Desch{\^e}nes} {et~al.}(2017){Miville-Desch{\^e}nes},
  {Murray}, \& {Lee}}]{Miville_2017}
{Miville-Desch{\^e}nes}, M.-A., {Murray}, N., \& {Lee}, E.~J. 2017, \apj, 834,
  57, \dodoi{10.3847/1538-4357/834/1/57}

\bibitem[{{Mouschovias} \& {Ciolek}(1999)}]{Mouschovias1999}
{Mouschovias}, T.~C., \& {Ciolek}, G.~E. 1999, in NATO Advanced Science
  Institutes (ASI) Series C, Vol. 540, NATO Advanced Science Institutes (ASI)
  Series C, ed. C.~J. {Lada} \& N.~D. {Kylafis}, 305

\bibitem[{{Naess} {et~al.}(2021{\natexlab{a}}){Naess}, {Aiola}, {Battaglia},
  {Bond}, {Calabrese}, {Choi}, {Cothard}, {Halpern}, {Hill}, {Koopman},
  {Devlin}, {McMahon}, {Dicker}, {Duivenvoorden}, {Dunkley}, {Van Engelen},
  {Fanfani}, {Ferraro}, {Gallardo}, {Guan}, {Han}, {Hasselfield}, {Hincks},
  {Huffenberger}, {Kosowsky}, {Louis}, {Macinnis}, {Madhavacheril}, {Nati},
  {Niemack}, {Page}, {Salatino}, {Schaan}, {Orlowski-Scherer}, {Schillaci},
  {Schmitt}, {Sehgal}, {Sif{\'o}n}, {Staggs}, \& {Wollack}}]{naess2021planet}
{Naess}, S., {Aiola}, S., {Battaglia}, N., {et~al.} 2021{\natexlab{a}}, arXiv
  e-prints, arXiv:2104.10264.
\newblock \doarXiv{2104.10264}

\bibitem[{{Naess} {et~al.}(2021{\natexlab{b}}){Naess}, {Battaglia}, {Richard
  Bond}, {Calabrese}, {Choi}, {Cothard}, {Devlin}, {Duell}, {Duivenvoorden},
  {Dunkley}, {D{\"u}nner}, {Gallardo}, {Gralla}, {Guan}, {Halpern}, {Colin
  Hill}, {Hilton}, {Huffenberger}, {Koopman}, {Kosowsky}, {Madhavacheril},
  {McMahon}, {Nati}, {Niemack}, {Page}, {Partridge}, {Salatino}, {Sehgal},
  {Spergel}, {Staggs}, {Wollack}, \& {Xu}}]{Naess2021transient}
{Naess}, S., {Battaglia}, N., {Richard Bond}, J., {et~al.} 2021{\natexlab{b}},
  \apj, 915, 14, \dodoi{10.3847/1538-4357/abfe6d}

\bibitem[{{N{\ae}ss} \& {Louis}(2013)}]{taylens}
{N{\ae}ss}, S.~K., \& {Louis}, T. 2013, \jcap, 2013, 001,
  \dodoi{10.1088/1475-7516/2013/09/001}

\bibitem[{{Nakamura} \& {Li}(2005)}]{Nakamura2005}
{Nakamura}, F., \& {Li}, Z.-Y. 2005, \apj, 631, 411, \dodoi{10.1086/432606}

\bibitem[{{Nashimoto} {et~al.}(2020){Nashimoto}, {Hattori}, {Poidevin}, \&
  {G{\'e}nova-Santos}}]{Nashimoto_2020}
{Nashimoto}, M., {Hattori}, M., {Poidevin}, F., \& {G{\'e}nova-Santos}, R.
  2020, \apjl, 900, L40, \dodoi{10.3847/2041-8213/abb29d}

\bibitem[{{Nederlander} {et~al.}(2021){Nederlander}, {Hughes}, {Fehr},
  {Flaherty}, {Su}, {Mo{\'o}r}, {Chiang}, {Andrews}, {Wilner}, \&
  {Marino}}]{2021ApJ...917....5N}
{Nederlander}, A., {Hughes}, A.~M., {Fehr}, A.~J., {et~al.} 2021, \apj, 917, 5,
  \dodoi{10.3847/1538-4357/abdd32}

\bibitem[{{Nibauer} {et~al.}(2020){Nibauer}, {Baxter}, \&
  {Jain}}]{Nibauer:2020}
{Nibauer}, J., {Baxter}, E., \& {Jain}, B. 2020, \aj, 159, 210,
  \dodoi{10.3847/1538-3881/ab8192}

\bibitem[{{Nibauer} {et~al.}(2021){Nibauer}, {Baxter}, {Jain}, {Van Saders},
  {Beaton}, \& {Teske}}]{2021ApJ...907..116N}
{Nibauer}, J., {Baxter}, E.~J., {Jain}, B., {et~al.} 2021, \apj, 907, 116,
  \dodoi{10.3847/1538-4357/abd0f1}

\bibitem[{{Oort}(1950)}]{Oort:1950}
{Oort}, J.~H. 1950, \bain, 11, 91

\bibitem[{{Padoan} \& {Nordlund}(1999)}]{Padoan1999}
{Padoan}, P., \& {Nordlund}, {\r{A}}. 1999, \apj, 526, 279,
  \dodoi{10.1086/307956}

\bibitem[{{Page} {et~al.}(2007){Page}, {Hinshaw}, {Komatsu}, {Nolta},
  {Spergel}, {Bennett}, {Barnes}, {Bean}, {Dor{\'e}}, {Dunkley}, {Halpern},
  {Hill}, {Jarosik}, {Kogut}, {Limon}, {Meyer}, {Odegard}, {Peiris}, {Tucker},
  {Verde}, {Weiland}, {Wollack}, \& {Wright}}]{Page2007}
{Page}, L., {Hinshaw}, G., {Komatsu}, E., {et~al.} 2007, \apjs, 170, 335,
  \dodoi{10.1086/513699}

\bibitem[{{Pan} \& {Sari}(2005)}]{2005Icar..173..342P}
{Pan}, M., \& {Sari}, R. 2005, \icarus, 173, 342,
  \dodoi{10.1016/j.icarus.2004.09.004}

\bibitem[{{Panopoulou} {et~al.}(2019){Panopoulou}, {Hensley}, {Skalidis},
  {Blinov}, \& {Tassis}}]{Panopoulou_2019}
{Panopoulou}, G.~V., {Hensley}, B.~S., {Skalidis}, R., {Blinov}, D., \&
  {Tassis}, K. 2019, \aap, 624, L8, \dodoi{10.1051/0004-6361/201935266}

\bibitem[{{Panopoulou} {et~al.}(2016){Panopoulou}, {Psaradaki}, \&
  {Tassis}}]{Panopoulou:2016}
{Panopoulou}, G.~V., {Psaradaki}, I., \& {Tassis}, K. 2016, \mnras, 462, 1517,
  \dodoi{10.1093/mnras/stw1678}

\bibitem[{{Paradis} {et~al.}(2011){Paradis}, {Bernard}, {M{\'e}ny}, \&
  {Gromov}}]{Paradis2011}
{Paradis}, D., {Bernard}, J.~P., {M{\'e}ny}, C., \& {Gromov}, V. 2011, \aap,
  534, A118, \dodoi{10.1051/0004-6361/201116862}

\bibitem[{{Pardo} {et~al.}(2001){Pardo}, {Cernicharo}, \&
  {Serabyn}}]{Pardo2001}
{Pardo}, J.~R., {Cernicharo}, J., \& {Serabyn}, E. 2001, IEEE Transactions on
  Antennas and Propagation, 49, 1683, \dodoi{10.1109/8.982447}

\bibitem[{{Pelgrims} {et~al.}(2021){Pelgrims}, {Clark}, {Hensley},
  {Panopoulou}, {Pavlidou}, {Tassis}, {Eriksen}, \& {Wehus}}]{Pelgrims:2021}
{Pelgrims}, V., {Clark}, S.~E., {Hensley}, B.~S., {et~al.} 2021, \aap, 647,
  A16, \dodoi{10.1051/0004-6361/202040218}

\bibitem[{{\sorthelp{Planck Collaboration 2011A}}{Planck Collaboration
  I}(2011)}]{Planck_Early_I}
{\sorthelp{Planck Collaboration 2011A}}{Planck Collaboration I}. 2011, \aap,
  536, A1, \dodoi{10.1051/0004-6361/201116464}

\bibitem[{{\sorthelp{Planck Collaboration 2011Q}}{Planck Collaboration
  XVII}(2011)}]{Planck_Early_XVII}
{\sorthelp{Planck Collaboration 2011Q}}{Planck Collaboration XVII}. 2011, \aap,
  536, A17, \dodoi{10.1051/0004-6361/201116473}

\bibitem[{{\sorthelp{Planck Collaboration 2014M}}{Planck Collaboration
  XIII}(2014)}]{Planck_2013_XIII}
{\sorthelp{Planck Collaboration 2014M}}{Planck Collaboration XIII}. 2014, \aap,
  571, A13, \dodoi{10.1051/0004-6361/201321553}

\bibitem[{{\sorthelp{Planck Collaboration 2015J}}{Planck Collaboration
  X}(2016)}]{Planck_2015_X}
{\sorthelp{Planck Collaboration 2015J}}{Planck Collaboration X}. 2016, \aap,
  594, A10, \dodoi{10.1051/0004-6361/201525967}

\bibitem[{{\sorthelp{Planck Collaboration 2015Y}}{Planck Collaboration
  XXV}(2016)}]{Planck_2015_XXV}
{\sorthelp{Planck Collaboration 2015Y}}{Planck Collaboration XXV}. 2016, \aap,
  594, A25, \dodoi{10.1051/0004-6361/201526803}

\bibitem[{{\sorthelp{Planck Collaboration 2015ZC}}{Planck Collaboration
  XXVIII}(2016)}]{Planck_2015_XXVIII}
{\sorthelp{Planck Collaboration 2015ZC}}{Planck Collaboration XXVIII}. 2016,
  \aap, 594, A28, \dodoi{10.1051/0004-6361/201525819}

\bibitem[{{\sorthelp{Planck Collaboration 2018D}}{Planck Collaboration
  IV}(2020)}]{Planck_2018_IV}
{\sorthelp{Planck Collaboration 2018D}}{Planck Collaboration IV}. 2020, \aap,
  641, A4, \dodoi{10.1051/0004-6361/201833881}

\bibitem[{{\sorthelp{Planck Collaboration 2018F}}{Planck Collaboration
  VI}(2020)}]{Planck_2018_VI}
{\sorthelp{Planck Collaboration 2018F}}{Planck Collaboration VI}. 2020, \aap,
  641, A6, \dodoi{10.1051/0004-6361/201833910}

\bibitem[{{\sorthelp{Planck Collaboration 2018K}}{Planck Collaboration
  XI}(2020)}]{Planck_2018_XI}
{\sorthelp{Planck Collaboration 2018K}}{Planck Collaboration XI}. 2020, \aap,
  641, A11, \dodoi{10.1051/0004-6361/201832618}

\bibitem[{{\sorthelp{Planck Collaboration 2018L}}{Planck Collaboration
  XII}(2020)}]{Planck_2018_XII}
{\sorthelp{Planck Collaboration 2018L}}{Planck Collaboration XII}. 2020, \aap,
  641, A12, \dodoi{10.1051/0004-6361/201833885}

\bibitem[{{\sorthelp{Planck Collaboration 2018}}{Planck Collaboration
  I}(2020)}]{Planck_2018_I}
{\sorthelp{Planck Collaboration 2018}}{Planck Collaboration I}. 2020, \aap,
  641, A1, \dodoi{10.1051/0004-6361/201833880}

\bibitem[{{\sorthelp{Planck Collaboration IntO}}{Planck Collaboration Int.
  XV}(2014)}]{Planck_Int_XV}
{\sorthelp{Planck Collaboration IntO}}{Planck Collaboration Int. XV}. 2014,
  \aap, 565, A103, \dodoi{10.1051/0004-6361/201322612}

\bibitem[{{\sorthelp{Planck Collaboration IntT}}{Planck Collaboration Int.
  XX}(2015)}]{Planck_Int_XX}
{\sorthelp{Planck Collaboration IntT}}{Planck Collaboration Int. XX}. 2015,
  \aap, 576, A105, \dodoi{10.1051/0004-6361/201424086}

\bibitem[{{\sorthelp{Planck Collaboration IntU}}{Planck Collaboration Int.
  XXI}(2015)}]{Planck_XXI}
{\sorthelp{Planck Collaboration IntU}}{Planck Collaboration Int. XXI}. 2015,
  \aap, 576, A106, \dodoi{10.1051/0004-6361/201424087}

\bibitem[{{\sorthelp{Planck Collaboration IntV}}{Planck Collaboration Int.
  XXII}(2015)}]{Planck_Int_XXII}
{\sorthelp{Planck Collaboration IntV}}{Planck Collaboration Int. XXII}. 2015,
  \aap, 576, A107, \dodoi{10.1051/0004-6361/201424088}

\bibitem[{{\sorthelp{Planck Collaboration IntZG}}{Planck Collaboration Int.
  XXXII}(2016)}]{Planck_Int_XXXII}
{\sorthelp{Planck Collaboration IntZG}}{Planck Collaboration Int. XXXII}. 2016,
  \aap, 586, A135, \dodoi{10.1051/0004-6361/201425044}

\bibitem[{{\sorthelp{Planck Collaboration IntZJ}}{Planck Collaboration Int.
  XXXV}(2016)}]{Planck_Int_XXXV}
{\sorthelp{Planck Collaboration IntZJ}}{Planck Collaboration Int. XXXV}. 2016,
  \aap, 586, A138, \dodoi{10.1051/0004-6361/201525896}

\bibitem[{{\sorthelp{Planck Collaboration IntZM}}{Planck Collaboration Int.
  XXXVIII}(2016)}]{Planck_Int_XXXVIII}
{\sorthelp{Planck Collaboration IntZM}}{Planck Collaboration Int. XXXVIII}.
  2016, \aap, 586, A141, \dodoi{10.1051/0004-6361/201526506}

\bibitem[{{\sorthelp{Planck Collaboration IntZQ}}{Planck Collaboration Int.
  XLII}(2016)}]{Planck_Int_XLII}
{\sorthelp{Planck Collaboration IntZQ}}{Planck Collaboration Int. XLII}. 2016,
  \aap, 596, A103, \dodoi{10.1051/0004-6361/201528033}

\bibitem[{{\sorthelp{Planck Collaboration IntZZB}}{Planck Collaboration Int.
  LII}(2017)}]{Planck_Int_LII}
{\sorthelp{Planck Collaboration IntZZB}}{Planck Collaboration Int. LII}. 2017,
  \aap, 607, A122, \dodoi{10.1051/0004-6361/201630311}

\bibitem[{{\sorthelp{Planck Collaboration IntZZG}}{Planck Collaboration Int.
  LVII}(2020)}]{NPIPE2020}
{\sorthelp{Planck Collaboration IntZZG}}{Planck Collaboration Int. LVII}. 2020,
  \aap, 643, 42, \dodoi{10.1051/0004-6361/202038073}

\bibitem[{{Presta} {et~al.}(2020){Presta}, {Ade}, {Battistelli}, {Castellano},
  {Colantoni}, {Columbro}, {Coppolecchia}, {D' Alessandro}, {de Bernardis},
  {Gordon}, {Lamagna}, {Masi}, {Mauskopf}, {Paiella}, {Pettinari},
  {Piacentini}, {Pisano}, \& {Tucker}}]{OLIMPO2020}
{Presta}, G., {Ade}, P.~A.~R., {Battistelli}, E.~S., {et~al.} 2020, in Journal
  of Physics Conference Series, Vol. 1548, Journal of Physics Conference
  Series, 012018, \dodoi{10.1088/1742-6596/1548/1/012018}

\bibitem[{{Puglisi} {et~al.}(2017){Puglisi}, {Fabbian}, \&
  {Baccigalupi}}]{Puglisi2016a}
{Puglisi}, G., {Fabbian}, G., \& {Baccigalupi}, C. 2017, \mnras, 469, 2982,
  \dodoi{10.1093/mnras/stx1029}

\bibitem[{{Regaldo-Saint Blancard} {et~al.}(2020){Regaldo-Saint Blancard},
  {Levrier}, {Allys}, {Bellomi}, \& {Boulanger}}]{Regaldo-SaintBlancard:2020}
{Regaldo-Saint Blancard}, B., {Levrier}, F., {Allys}, E., {Bellomi}, E., \&
  {Boulanger}, F. 2020, \aap, 642, A217, \dodoi{10.1051/0004-6361/202038044}

\bibitem[{{Robitaille} {et~al.}(2017){Robitaille}, {Scaife}, {Carretti},
  {Gaensler}, {McEwen}, {Leistedt}, {Haverkorn}, {Bernardi}, {Kesteven},
  {Poppi}, \& {Staveley-Smith}}]{Robitaille:2017}
{Robitaille}, J.~F., {Scaife}, A.~M.~M., {Carretti}, E., {et~al.} 2017, \mnras,
  468, 2957, \dodoi{10.1093/mnras/stx642}

\bibitem[{{Seifried} {et~al.}(2020){Seifried}, {Walch}, {Weis}, {Reissl},
  {Soler}, {Klessen}, \& {Joshi}}]{Seifried:2020}
{Seifried}, D., {Walch}, S., {Weis}, M., {et~al.} 2020, \mnras, 497, 4196,
  \dodoi{10.1093/mnras/staa2231}

\bibitem[{{Shu} {et~al.}(1987){Shu}, {Adams}, \& {Lizano}}]{Shu1987}
{Shu}, F.~H., {Adams}, F.~C., \& {Lizano}, S. 1987, \araa, 25, 23,
  \dodoi{10.1146/annurev.aa.25.090187.000323}

\bibitem[{{Siebenmorgen} {et~al.}(2014){Siebenmorgen}, {Voshchinnikov}, \&
  {Bagnulo}}]{Siebenmorgen2014}
{Siebenmorgen}, R., {Voshchinnikov}, N.~V., \& {Bagnulo}, S. 2014, \aap, 561,
  A82, \dodoi{10.1051/0004-6361/201321716}

\bibitem[{{Simons Observatory Collaboration}(2019)}]{SO_2019}
{Simons Observatory Collaboration}. 2019, \jcap, 2019, 056,
  \dodoi{10.1088/1475-7516/2019/02/056}

\bibitem[{{Skalidis} \& {Pelgrims}(2019)}]{Skalidis2019}
{Skalidis}, R., \& {Pelgrims}, V. 2019, \aap, 631, L11,
  \dodoi{10.1051/0004-6361/201936547}

\bibitem[{{Soler} \& {Hennebelle}(2017)}]{Soler:2017}
{Soler}, J.~D., \& {Hennebelle}, P. 2017, \aap, 607, A2,
  \dodoi{10.1051/0004-6361/201731049}

\bibitem[{{Soler} {et~al.}(2013){Soler}, {Hennebelle}, {Martin},
  {Miville-Desch{\^e}nes}, {Netterfield}, \& {Fissel}}]{Soler:2013}
{Soler}, J.~D., {Hennebelle}, P., {Martin}, P.~G., {et~al.} 2013, \apj, 774,
  128, \dodoi{10.1088/0004-637X/774/2/128}

\bibitem[{{Stein} {et~al.}(2019){Stein}, {Alvarez}, \&
  {Bond}}]{2019MNRAS.483.2236S}
{Stein}, G., {Alvarez}, M.~A., \& {Bond}, J.~R. 2019, \mnras, 483, 2236,
  \dodoi{10.1093/mnras/sty3226}

\bibitem[{{Stein} {et~al.}(2020){Stein}, {Alvarez}, {Bond}, {van Engelen}, \&
  {Battaglia}}]{2020JCAP...10..012S}
{Stein}, G., {Alvarez}, M.~A., {Bond}, J.~R., {van Engelen}, A., \&
  {Battaglia}, N. 2020, \jcap, 2020, 012, \dodoi{10.1088/1475-7516/2020/10/012}

\bibitem[{{Stern} {et~al.}(1991){Stern}, {Stocke}, \& {Weissman}}]{Stern:1991}
{Stern}, S.~A., {Stocke}, J., \& {Weissman}, P.~R. 1991, \icarus, 91, 65,
  \dodoi{10.1016/0019-1035(91)90126-E}

\bibitem[{{Stevens} {et~al.}(2018){Stevens}, {Goeckner-Wald}, {Keskitalo},
  {McCallum}, {Ali}, {Borrill}, {Brown}, {Chinone}, {Gallardo}, {Kusaka},
  {Lee}, {McMahon}, {Niemack}, {Page}, {Puglisi}, {Salatino}, {Mak}, {Teply},
  {Thomas}, {Vavagiakis}, {Wollack}, {Xu}, \& {Zhu}}]{Stevens2018}
{Stevens}, J.~R., {Goeckner-Wald}, N., {Keskitalo}, R., {et~al.} 2018, in
  Society of Photo-Optical Instrumentation Engineers (SPIE) Conference Series,
  Vol. 10708, Millimeter, Submillimeter, and Far-Infrared Detectors and
  Instrumentation for Astronomy IX, ed. J.~{Zmuidzinas} \& J.-R. {Gao},
  1070841, \dodoi{10.1117/12.2313898}

\bibitem[{{Stompor} {et~al.}(2009){Stompor}, {Leach}, {Stivoli}, \&
  {Baccigalupi}}]{stompor2009}
{Stompor}, R., {Leach}, S., {Stivoli}, F., \& {Baccigalupi}, C. 2009, \mnras,
  392, 216, \dodoi{10.1111/j.1365-2966.2008.14023.x}

\bibitem[{{Sullivan} {et~al.}(2021){Sullivan}, {Fissel}, {King}, {Chen}, {Li},
  \& {Soler}}]{Sullivan:2021}
{Sullivan}, C.~H., {Fissel}, L.~M., {King}, P.~K., {et~al.} 2021, \mnras, 503,
  5006, \dodoi{10.1093/mnras/stab596}

\bibitem[{{Sutin} {et~al.}(2018){Sutin}, {Alvarez}, {Battaglia}, {Bock},
  {Bonato}, {Borrill}, {Chuss}, {Cooperrider}, {Crill}, {Delabrouille},
  {Devlin}, {Essinger-Hileman}, {Fissel}, {Flauger}, {Gorski}, {Green},
  {Hanany}, {Hubmayr}, {Johnson}, {Jones}, {Knox}, {Kogut}, {Lawrence},
  {McMahon}, {Matsumura}, {Negrello}, {O'Brient}, {Paine}, {Pryke}, {Shirron},
  {Trangsrud}, {Wen}, {Young}, \& {de Zotti}}]{PICO2018}
{Sutin}, B.~M., {Alvarez}, M., {Battaglia}, N., {et~al.} 2018, in Society of
  Photo-Optical Instrumentation Engineers (SPIE) Conference Series, Vol. 10698,
  Space Telescopes and Instrumentation 2018: Optical, Infrared, and Millimeter
  Wave, ed. M.~{Lystrup}, H.~A. {MacEwen}, G.~G. {Fazio}, N.~{Batalha},
  N.~{Siegler}, \& E.~C. {Tong}, 106984F, \dodoi{10.1117/12.2311326}

\bibitem[{{Svalheim} {et~al.}(2020){Svalheim}, {Andersen}, {Aurlien},
  {Banerji}, {Bersanelli}, {Bertocco}, {Brilenkov}, {Carbone}, {Colombo},
  {Eriksen}, {Foss}, {Franceschet}, {Fuskeland}, {Galeotta}, {Galloway},
  {Gerakakis}, {Gjerl{\o}w}, {Hensley}, {Herman}, {Iacobellis}, {Ieronymaki},
  {Ihle}, {Jewell}, {Karakci}, {Keih{\"a}nen}, {Keskitalo}, {Maggio}, {Maino},
  {Maris}, {Paradiso}, {Partridge}, {Reinecke}, {Suur-Uski}, {Tavagnacco},
  {Thommesen}, {Watts}, {Wehus}, \& {Zacchei}}]{Svalheim2020}
{Svalheim}, T.~L., {Andersen}, K.~J., {Aurlien}, R., {et~al.} 2020, arXiv
  e-prints, arXiv:2011.08503.
\newblock \doarXiv{2011.08503}

\bibitem[{{Tassis} {et~al.}(2018){Tassis}, {Ramaprakash}, {Readhead}, {Potter},
  {Wehus}, {Panopoulou}, {Blinov}, {Eriksen}, {Hensley}, {Karakci},
  {Kypriotakis}, {Maharana}, {Ntormousi}, {Pavlidou}, {Pearson}, \&
  {Skalidis}}]{PASIPHAE}
{Tassis}, K., {Ramaprakash}, A.~N., {Readhead}, A. C.~S., {et~al.} 2018, arXiv
  e-prints, arXiv:1810.05652.
\newblock \doarXiv{1810.05652}

\bibitem[{{Teague} {et~al.}(2021){Teague}, {Hull}, {Guilloteau}, {Bergin},
  {Dutrey}, {Henning}, {Kuiper}, {Semenov}, {Stephens}, \&
  {Vlemmings}}]{Teague2021}
{Teague}, R., {Hull}, C. L.~H., {Guilloteau}, S., {et~al.} 2021, arXiv
  e-prints, arXiv:2109.09247.
\newblock \doarXiv{2109.09247}

\bibitem[{{The CMB-S4 Collaboration} {et~al.}(2020){The CMB-S4 Collaboration},
  {:}, {Abazajian}, {Addison}, {Adshead}, {Ahmed}, {Akerib}, {Ali}, {Allen},
  {Alonso}, {Alvarez}, {Amin}, {Anderson}, {Arnold}, {Ashton}, {Baccigalupi},
  {Bard}, {Barkats}, {Barron}, {Barry}, {Bartlett}, {Basu Thakur}, {Battaglia},
  {Bean}, {Bebek}, {Bender}, {Benson}, {Bianchini}, {Bischoff}, {Bleem},
  {Bock}, {Bocquet}, {Boddy}, {Bond}, {Borrill}, {Bouchet}, {Brinckmann},
  {Brown}, {Bryan}, {Buza}, {Byrum}, {Hervias Caimapo}, {Calabrese}, {Calafut},
  {Caldwell}, {Carlstrom}, {Carron}, {Cecil}, {Challinor}, {Chang}, {Chinone},
  {Cho}, {Cooray}, {Coulton}, {Crawford}, {Crites}, {Cukierman}, {Cyr-Racine},
  {de Haan}, {Delabrouille}, {Devlin}, {Di Valentino}, {Dierickx}, {Dobbs},
  {Duff}, {Dunkley}, {Dvorkin}, {Eimer}, {Elleflot}, {Errard},
  {Essinger-Hileman}, {Fabbian}, {Feng}, {Ferraro}, {Filippini}, {Flauger},
  {Flaugher}, {Fraisse}, {Frolov}, {Galitzki}, {Gallardo}, {Galli}, {Ganga},
  {Gerbino}, {Gluscevic}, {Goeckner-Wald}, {Green}, {Grin}, {Grohs},
  {Gualtieri}, {Gudmundsson}, {Gullett}, {Gupta}, {Habib}, {Halpern},
  {Halverson}, {Hanany}, {Harrington}, {Hasegawa}, {Hasselfield}, {Hazumi},
  {Heitmann}, {Henderson}, {Hensley}, {Hill}, {Hill}, {Hlozek}, {Ho}, {Hoang},
  {Holder}, {Holzapfel}, {Hood}, {Hubmayr}, {Huffenberger}, {Hui}, {Irwin},
  {Jeong}, {Johnson}, {Jones}, {Kang}, {Karkare}, {Katayama}, {Keskitalo},
  {Kisner}, {Knox}, {Koopman}, {Kosowsky}, {Kovac}, {Kovetz}, {Kuhlmann},
  {Kuo}, {Kusaka}, {L{\"a}hteenm{\"a}ki}, {Lawrence}, {Lee}, {Lewis}, {Li},
  {Linder}, {Loverde}, {Lowitz}, {Lubin}, {Madhavacheril}, {Mantz}, {Marques},
  {Matsuda}, {Mauskopf}, {McCarrick}, {McMahon}, {Meerburg}, {Melin},
  {Menanteau}, {Meyers}, {Millea}, {Mohr}, {Moncelsi}, {Monzani},
  {Mroczkowski}, {Mukherjee}, {Nagy}, {Namikawa}, {Nati}, {Natoli}, {Newburgh},
  {Niemack}, {Nishino}, {Nord}, {Novosad}, {O'Brient}, {Padin}, {Palladino},
  {Partridge}, {Petravick}, {Pierpaoli}, {Pogosian}, {Prabhu}, {Pryke},
  {Puglisi}, {Racine}, {Rahlin}, {Sathyanarayana Rao}, {Raveri}, {Reichardt},
  {Remazeilles}, {Rocha}, {Roe}, {Roy}, {Ruhl}, {Salatino}, {Saliwanchik},
  {Schaan}, {Schillaci}, {Schmitt}, {Schmittfull}, {Scott}, {Sehgal},
  {Shandera}, {Sherwin}, {Shirokoff}, {Simon}, {Slosar}, {Spergel}, {St.
  Germaine}, {Staggs}, {Stark}, {Starkman}, {Stompor}, {Stoughton}, {Suzuki},
  {Tajima}, {Teply}, {Thompson}, {Thorne}, {Timbie}, {Tomasi}, {Tristram},
  {Tucker}, {Umilt{\`a}}, {van Engelen}, {Vavagiakis}, {Vieira}, {Vieregg},
  {Wagoner}, {Wallisch}, {Wang}, {Watson}, {Westbrook}, {Whitehorn}, {Wollack},
  {Kimmy Wu}, {Xu}, {Yang}, {Yasini}, {Yefremenko}, {Yoon}, {Young}, {Yu}, \&
  {Zonca}}]{CMBS42020}
{The CMB-S4 Collaboration}, {:}, {Abazajian}, K., {et~al.} 2020, arXiv
  e-prints, arXiv:2008.12619.
\newblock \doarXiv{2008.12619}

\bibitem[{{Thorne} {et~al.}(2017){Thorne}, {Dunkley}, {Alonso}, \&
  {N{\ae}ss}}]{Thorne2017}
{Thorne}, B., {Dunkley}, J., {Alonso}, D., \& {N{\ae}ss}, S. 2017, \mnras, 469,
  2821, \dodoi{10.1093/mnras/stx949}

\bibitem[{{Unger} \& {Farrar}(2017)}]{Unger:2017}
{Unger}, M., \& {Farrar}, G.~R. 2017, in International Cosmic Ray Conference,
  Vol. 301, 35th International Cosmic Ray Conference (ICRC2017), 558.
\newblock \doarXiv{1707.02339}

\bibitem[{{van der Walt} {et~al.}(2011){van der Walt}, {Colbert}, \&
  {Varoquaux}}]{NumPy}
{van der Walt}, S., {Colbert}, S.~C., \& {Varoquaux}, G. 2011, Computing in
  Science and Engineering, 13, 22, \dodoi{10.1109/MCSE.2011.37}

\bibitem[{{V{\'a}zquez-Semadeni} {et~al.}(2011){V{\'a}zquez-Semadeni},
  {Banerjee}, {G{\'o}mez}, {Hennebelle}, {Duffin}, \& {Klessen}}]{Vazquez2011}
{V{\'a}zquez-Semadeni}, E., {Banerjee}, R., {G{\'o}mez}, G.~C., {et~al.} 2011,
  \mnras, 414, 2511, \dodoi{10.1111/j.1365-2966.2011.18569.x}

\bibitem[{{Virtanen} {et~al.}(2020){Virtanen}, {Gommers}, {Oliphant},
  {Haberland}, {Reddy}, {Cournapeau}, {Burovski}, {Peterson}, {Weckesser},
  {Bright}, {van der Walt}, {Brett}, {Wilson}, {Millman}, {Mayorov}, {Nelson},
  {Jones}, {Kern}, {Larson}, {Carey}, {Polat}, {Feng}, {Moore}, {Vand erPlas},
  {Laxalde}, {Perktold}, {Cimrman}, {Henriksen}, {Quintero}, {Harris},
  {Archibald}, {Ribeiro}, {Pedregosa}, {van Mulbregt}, \& {SciPy 1. 0
  Contributors}}]{SciPy}
{Virtanen}, P., {Gommers}, R., {Oliphant}, T.~E., {et~al.} 2020, Nature
  Methods, 17, 261, \dodoi{10.1038/s41592-019-0686-2}

\bibitem[{{Weiland} {et~al.}(2020){Weiland}, {Addison}, {Bennett}, {Halpern},
  \& {Hinshaw}}]{Weiland:2020}
{Weiland}, J.~L., {Addison}, G.~E., {Bennett}, C.~L., {Halpern}, M., \&
  {Hinshaw}, G. 2020, \apj, 893, 119, \dodoi{10.3847/1538-4357/ab7ea6}

\bibitem[{{Weiland} {et~al.}(2011){Weiland}, {Odegard}, {Hill}, {Wollack},
  {Hinshaw}, {Greason}, {Jarosik}, {Page}, {Bennett}, {Dunkley}, {Gold},
  {Halpern}, {Kogut}, {Komatsu}, {Larson}, {Limon}, {Meyer}, {Nolta}, {Smith},
  {Spergel}, {Tucker}, \& {Wright}}]{Weiland_2011}
{Weiland}, J.~L., {Odegard}, N., {Hill}, R.~S., {et~al.} 2011, \apjs, 192, 19,
  \dodoi{10.1088/0067-0049/192/2/19}

\bibitem[{{Wittenmyer} {et~al.}(2020){Wittenmyer}, {Wang}, {Horner}, {Butler},
  {Tinney}, {Carter}, {Wright}, {Jones}, {Bailey}, {O'Toole}, \&
  {Johns}}]{2020MNRAS.492..377W}
{Wittenmyer}, R.~A., {Wang}, S., {Horner}, J., {et~al.} 2020, \mnras, 492, 377,
  \dodoi{10.1093/mnras/stz3436}

\bibitem[{{Wolff} {et~al.}(1997){Wolff}, {Clayton}, {Kim}, {Martin}, \&
  {Anderson}}]{Wolff_1997}
{Wolff}, M.~J., {Clayton}, G.~C., {Kim}, S.-H., {Martin}, P.~G., \& {Anderson},
  C.~M. 1997, \apj, 478, 395, \dodoi{10.1086/303789}

\bibitem[{{Xu} {et~al.}(2021){Xu}, {Adachi}, {Ade}, {Beall}, {Bhandarkar},
  {Bond}, {Chesmore}, {Chinone}, {Choi}, {Connors}, {Coppi}, {Cothard},
  {Crowley}, {Devlin}, {Dicker}, {Dober}, {Duff}, {Galitzki}, {Gallardo},
  {Golec}, {Gudmundsson}, {Haridas}, {Harrington}, {Hervias-Caimapo}, {Patty
  Ho}, {Huber}, {Hubmayr}, {Iuliano}, {Kaneko}, {Kofman}, {Koopman}, {Lashner},
  {Limon}, {Link}, {Lucas}, {Matsuda}, {McCarrick}, {Nati}, {Niemack},
  {Orlowski-Scherer}, {Piccirillo}, {Sarmiento}, {Schaan}, {Silva-Feaver},
  {Sonka}, {Sutariya}, {Tajima}, {Teply}, {Terasaki}, {Thornton}, {Tucker},
  {Ullom}, {Vavagiakis}, {Vissers}, {Walker}, {Whipps}, {Wollack}, {Zannoni},
  {Zhu}, {Zonca}, \& {Simons Observatory Collaboration}}]{SO_LAT}
{Xu}, Z., {Adachi}, S., {Ade}, P., {et~al.} 2021, Research Notes of the
  American Astronomical Society, 5, 100, \dodoi{10.3847/2515-5172/abf9ab}

\bibitem[{{Zhang} {et~al.}(2017){Zhang}, {Telesco}, {Hoang}, {Li}, {Pantin},
  {Wright}, {Li}, \& {Barnes}}]{Zhang_2017}
{Zhang}, H., {Telesco}, C.~M., {Hoang}, T., {et~al.} 2017, \apj, 844, 6,
  \dodoi{10.3847/1538-4357/aa77ff}

\bibitem[{{Zhang} {et~al.}(2019){Zhang}, {Lazarian}, {Ho}, {Yuen}, {Yang}, \&
  {Hu}}]{2019MNRAS.486.4813Z}
{Zhang}, J.-F., {Lazarian}, A., {Ho}, K.~W., {et~al.} 2019, \mnras, 486, 4813,
  \dodoi{10.1093/mnras/stz1176}

\bibitem[{Zonca {et~al.}(2019)Zonca, Singer, Lenz, Reinecke, Rosset, Hivon, \&
  Gorski}]{Zonca2019}
Zonca, A., Singer, L., Lenz, D., {et~al.} 2019, Journal of Open Source
  Software, 4, 1298, \dodoi{10.21105/joss.01298}

\bibitem[{{Zonca} {et~al.}(2021){Zonca}, {Thorne}, {Krachmalnicoff}, \&
  {Borrill}}]{pysm3}
{Zonca}, A., {Thorne}, B., {Krachmalnicoff}, N., \& {Borrill}, J. 2021, arXiv
  e-prints, arXiv:2108.01444.
\newblock \doarXiv{2108.01444}

\bibitem[{{Zubko} {et~al.}(2004){Zubko}, {Dwek}, \& {Arendt}}]{Zubko2004}
{Zubko}, V., {Dwek}, E., \& {Arendt}, R.~G. 2004, \apjs, 152, 211,
  \dodoi{10.1086/382351}

\end{thebibliography}

\end{document}